\documentclass[11pt,draftcls,onecolumn]{IEEEtran}
\usepackage{graphicx}
\usepackage{amstext}
\usepackage{amsmath}
\usepackage{amsfonts,amssymb}
\usepackage{amsmath,tabu}
\usepackage{amssymb}
\usepackage[dvips]{epsfig}
\usepackage{epsfig}
\usepackage[dvips]{graphicx}
\usepackage{multirow}
\usepackage{multirow} 
\usepackage[utf8]{inputenc}
\usepackage[english]{babel}
\usepackage{caption}
\usepackage{subcaption}
\usepackage{float}
\usepackage{ltablex}
\usepackage{algorithmic}
\usepackage{setspace}
\usepackage[nospace,noadjust]{cite}
\usepackage{titlesec}
\titlespacing*{\section}{0pt}{1.1\baselineskip}{\baselineskip}
\titlespacing{\subsection}{0pt}{*0}{*0}
\titlespacing{\subsubsection}{0pt}{*0}{*0}
\newtheorem{theorem}{Theorem}[section]

\newtheorem{lemma}[theorem]{Lemma}

\def\bTh{{\boldsymbol{\Theta}}}

\def\bPhi{{\boldsymbol{\Phi}}}

\def\bpsi{{\boldsymbol{\Psi}}}

\def\b1{{\boldsymbol{1}}}
\def\c1{{\textcircled{a}}}
\def\ba{{\boldsymbol{a}}}

\def\bd{{\boldsymbol{d}}}

\def\bh{{\boldsymbol{h}}}

\def\bp{{\mathbf{p}}}
\def\bq{{\boldsymbol{q}}}
\def\br{{\mathbf{r}}}
\def\bs{{\boldsymbol{s}}}

\def\bu{{\boldsymbol{u}}}
\def\bv{{\boldsymbol{v}}}

\def\bx{{\boldsymbol{x}}}
\def\by{{\mathbf{y}}}

\def\bA{{\mathbf{A}}}
\def\bB{{\boldsymbol{B}}}

\def\bD{{\boldsymbol{D}}}
\def\bE{{\boldsymbol{E}}}

\def\bG{{\boldsymbol{G}}}

\def\bI{{\mathbf{I}}}

\def\bM{{\boldsymbol{M}}}

\def\bS{{\boldsymbol{S}}}

\def\bU{{\boldsymbol{U}}}

\def\bX{{\boldsymbol{X}}}
\def\bY{{\boldsymbol{Y}}}

\def\bzero{{\boldsymbol{0}}}

\providecommand{\keywords}[1]{\textbf{\textit{Index terms---}} #1}
\begin{document}
\title{TELET: A Monotonic Algorithm to Design Large Dimensional Equiangular Tight Frames for Applications in Compressed Sensing}
\author{R.~Jyothi and P.~Babu}
\maketitle
\begin{abstract}
An Equiangular tight frame (ETF) - also known as the Welch-bound-equality sequences - consists of a sequence of unit norm vectors whose absolute inner product is identical and minimal. Due to this unique property, these frames are preferred in different applications such as in constructing sensing matrices for compressed sensing systems, robust transmission and quantum computing. Construction of ETFs involves solving a challenging non-convex minimax optimization problem, and only a few methods were successful in constructing them, albeit only for smaller dimensions. In this paper, we propose an iterative algorithm named \textbf{TE}chnique to devise \textbf{L}arge dimensional \textbf{E}quiangular \textbf{T}ight-frames (\textbf{TELET}-frames) based on the majorization minimization (MM) procedure - in which we design and minimize a tight upper bound for the ETF cost function at every iteration. Since TELET is designed using the MM approach, it inherits useful properties of MM such as monotonicity and guaranteed convergence to a stationary point. Subsequently, we use the derived frames to construct optimized sensing matrix for compressed sensing systems. In the numerical simulations, we show that the proposed algorithm can generate complex and real frames (in the order of hundreds) with very low mutual coherence value when compared to the state-of-the-art algorithm, with a slight increase in computational cost. Experiments using synthetic data and real images reveal that the optimized sensing matrix obtained through the frames constructed by TELET performs better, in terms of image reconstruction accuracy, than the sensing matrix constructed using state-of-the-art methods.
\end{abstract}
\section{Introduction}
\vspace{-2.3mm}
Orthonormal basis is a powerful tool and has applications in diverse fields such as in computer vision, signal and image processing \cite{fourier}. Sometimes, instead of orthonormal basis, one may prefer the use of an overcomplete spanning set of vectors. An example where such a situation occurs is in Gabor analysis, wherein orthonormal Gabor basis with good time-frequency localization do not exist, while it is not difficult to find overcomplete Gabor system with excellent time-frequency localization \cite{gabor}. The overcomplete spanning set of vectors in finite dimensions is called a \emph{Frame}. Formally, a frame is defined as a sequence of vectors $\{\bx_{i}\}_{i=1}^{N}$ drawn from $\mathbb{C}^{d}$ ($d \leq N$) that satisfies the generalized Parseval condition: 
\begin{equation}\label{Parseval}
\begin{array}{ll}
a\|\bv\|_{2}^{2} \leq \displaystyle\sum_{n=1}^{N}|\bv^{H}\bx_{i}|^{2} \leq b\|\bv\|_{2}^{2} \:\: {\textrm{for all}} \:\:\bv \in \mathbb{C}^{d}
\end{array}
\end{equation} 
where $\|\cdot\|_{2}$ is the Euclidian norm, $(\cdot)^{H}$ is the Hermitian operator, $a$ and $b$ are the lower and upper frame bounds, respectively. If $a=b$ in (\ref{Parseval}), then the frame is called $a$-tight frame, and if $a=b=1$, it is a Parseval frame. When the frame is $a$-tight, the condition in (\ref{Parseval}) is equivalent to the statement that $\bX\bX^{H} = a\,\bI_{d}$, where $\bX = [\bx_{1},\bx_{2}, \cdots, \bx_{N}] \in \mathbb{C}^{d \times N}$ is the frame synthesis operator.  This implies that the $d$ non-zero singular values of $\bX$ are equal and is equal to $\sqrt{a}$. Hence, being a tight frame is a spectral constraint on the matrix $\bX$, which as discussed later in this section is used by the state-of-the-art algorithms to construct $a$-tight frames. When each $\bx_{i}$ of an $a$-tight frame has unit norm i.e. if $\|\bx_{i}\|_{2}=1$, then such an $a$-tight frame is called as a Unit Norm Tight Frame (UNTF). These frames are commonly used in the construction of signature sequences of CDMA systems \cite{comm}.  In applications such as sparse approximation \cite{sparse2}, robust transmission \cite{rt}, and quantum
computing \cite{qt}, it is required that the vectors comprising the UNTFs be maximally uncorrelated. The maximum correlation among the pair of vectors in a frame can be estimated using the mutual coherence metric $\mu(\bX)$ which is defined as:
\begin{equation}
\begin{array}{ll}
\mu(\bX)=\underset{i, j = 1, 2, \cdots N \: (i \neq j)}{\rm maximize}\: \dfrac{\left| \bx_{i}^{H}\bx_{j}\right| }{\|\bx_{i}\|_{2}\|\bx_{j}\|_{2}}
\end{array}
\end{equation} 
Frames with small mutual coherence are known as incoherent frames. A lower bound (known as the Welch bound) on the minimum achievable correlation for any arbitrary frame is given as: 
\begin{equation}\label{bound}
\begin{array}{ll}
\mu(\bX) \geq \sqrt{\dfrac{N-d}{d(N-1)}}
\end{array}
\end{equation}
Unit norm tight frames which achieve equality in the Welch bound are known as Equiangular Tight Frame (ETF). Geometrically, it corresponds to a frame in which each pair of distinct unit norm vectors meet at the same angle $\theta = {\rm{cos}}^{-1}\left(\sqrt{\dfrac{N-d}{d(N-1)}}\right)$.
Note that, all orthonormal bases are ETFs with $\mu(\bX)$ equal to zero. Hence, ETFs can be perceived as a generalization of orthonormal bases. However, ETFs do not exist for all values of $d$ and $N$. A necessary condition for the existence of a real ETF ($\bx_{i}$'s taking only real values) is given by  $N \leq \dfrac{1}{2} d(d+1)$  while a complex ETF ($\bx_{i}$'s taking complex values) can only exist when $N \leq d^{2}$ \cite{upperbound}.
Construction of an ETF is challenging as it involves solving a challenging non-convex minimax optimization problem. The existing methods are based on the alternating projection framework and they focus on the construction of the Gram matrix $\bG = \bX^{H}\bX$ of the ETF rather than directly constructing the frame itself \cite{tropp,xiong,agelos}. These methods consists of two steps, where in Step. I, the algorithms \cite{tropp,xiong,agelos} finds a Gram matrix $\bar{\bG}^{t}$ with the desired structural constraint using the shrinkage operator defined as:
\begin{equation} \label{SO}
\begin{array}{ll}
\bar{\bG}^{t}(i,j)=\left\{ \begin{array}{ll}
G^{t}(i,j)&  i \neq j, |G^{t}(i,j)| \leq \eta\\
\eta e^{i{\rm{arg}}G^{t}(i,j)}&  i \neq j, |G^{t}(i,j)| > \eta\\
1 & i=j\\
\end{array}\right..
\end{array}
\end{equation}
In Step. II, the nearest $\alpha$-tight matrix ${\bG^{t}_{\alpha}}$ to $\bar{\bG}^{t}$ is obtained, according to ${\bG^{t}_{\alpha}} = \alpha\bU
\bU^{*}$, where $\bU$ is a $N \times d$ matrix whose columns are the eigenvectors corresponding to the $d$ largest  eigenvalues of $\bar{\bG}^{t}$. Hence at every iteration, the algorithms in \cite{tropp,xiong,agelos} alternates between finding a Gram matrix with the desired structural and spectral constraint. After performing alternating projection, the tight frames are obtained from the output Gram matrix using either eigenvalue decomposition (EVD) or rank-revealing QR factorization. The algorithms in \cite{tropp,xiong,agelos} differs essentially in the choice of the threshold $\eta$ used in the shrinkage operator in (\ref{SO}) and the tightness constraint $\alpha$. The algorithms in \cite{tropp}, \cite{xiong} uses the threshold parameter $\eta$ as $\sqrt{1/d}$ and \cite{agelos} computes $\eta$ as $\sqrt{{(N-d)}/{d(N-1)}}$. The tightness constraint $\alpha$ in the case of  \cite{tropp} and \cite{agelos} is chosen as $\sqrt{{N}/{d}}$ while \cite{xiong} computes $\alpha=\displaystyle\sum_{k=1}^{d}\lambda_{k}/{d}$ where $\lambda_{k}$'s are the eigenvalues of $\bar{\bG}^{t}$.
One of the drawbacks of the alternating projection based methods is that  they cannot construct ETFs for large values of $N$ and $d$ \cite{tropp}. This could be because of the high complexity of the algorithm due to the computation of EVD at every iteration. \cite{alternatingprojection}. Different from the alternating projection based methods, the authors in \cite{tahir,codebook,preclic,rusu,bcasc} proposed algorithms which constructs real equiangular frames by directly modifying the frame synthesis operator $\bX$ and not its Gram matrix $\bG$. The authors in \cite{preclic} and \cite{rusu} proposed an iterative algorithm wherein at every iteration the frame synthesis operator is constructed column-by-column and each update involves solving a convex optimization problem using interior point method - which makes the algorithm computationally expensive. We next discuss an application of ETFs  to design the optimal sensing matrix in the compressed sensing (CS) systems.  
\subsection{Application to compressed sensing}
Based on the theoretical findings of \cite{ub}, ETFs can be employed in the CS framework.  In the following we briefly discuss the CS model and the need for optimal sensing matrix in CS systems. The CS framework involves reconstructing the high-dimensional $K$-sparse signal vector $\bu \in \mathbb{C}^{N \times 1}$ from its corresponding low-dimensional measurement vector $\by \in \mathbb{C}^{d \times 1} $ ($d << N$) via a carefully constructed sensing matrix $\bTh \in \mathbb{C}^{d \times N}$ using the relation $\by = \bTh\bu$. 
The vector $\bu$ is $K$-sparse, since it can be represented using $K$ non-zero coefficients over a fixed dictionary $\bpsi \in \mathbb{C}^{N \times N}$ i.e. $\bu = \bpsi\bs$ (where $\bs \in \mathbb{C}^{N}$ has $K$ non-zero entries i.e. $\|\bs\|_{0} = |\{i: s_{i}\neq 0\}|= K$). Hence, the low-dimensional measurement vector $\by$ can be expressed as $\by = \bTh\bpsi\bs$. 
Now, given the measurement vector $\by$, the dictionary $\bpsi$ and the sensing matrix $\bTh$, the sparse signal vector $\bs$ can be reconstructed by solving the following $\ell_{0}$ minimization problem:
\begin{equation}\label{CS-4}
\begin{array}{ll}
\underset{\bs}{\rm minimize} \: \|\bs\|_{0} \:\:\:\: {\rm{s.t.}}\:\: \by =\bTh\bpsi\bs
\end{array}
\end{equation}
The problem in (\ref{CS-4}) is NP-hard and can be solved using greedy algorithms such as orthogonal matching pursuit (OMP) \cite{r2,o1} or Basis Pursuit (BP) algorithm \cite{bp}.
As shown in \cite{ub}, the $K$-sparse signal $\bs$ can be uniquely recovered from the measurements $\by =\bTh\bpsi\bs$ as a solution to (\ref{CS-4}) as long as: 
\begin{equation}\label{CS-5}
\begin{array}{ll}
\|\bs\|_{0} <\dfrac{1}{2}\left[ 1+ \dfrac{1}{\mu(\bX_{\rm{ED}})}\right]
\end{array}
\end{equation}
where $\bX_{\rm{ED}}=\bTh\bpsi$ is called the equivalent dictionary. Note that the above upper bound is dictated by mutual coherence of $\mu(\bX_{\rm{ED}})$, smaller the value of $\mu(\bX_{\rm{ED}})$, larger the sparsity of signal $\bs$ which can be successfully recovered. Also, as reported in \cite{r2}, the incoherence property of $\bX_{\rm{ED}}$ plays an important role in the performance of the OMP and BP algorithm. Hence, given the dictionary $\bpsi$, we need to find the optimal sensing matrix $\bTh$ such that  $\bX_{\rm{ED}}$ is as close as possible to an ETF. \\
Elad in \cite{elad} was the first to come up with the idea of optimizing the sensing matrix $\bTh$ such that the equivalent dictionary is incoherent. However, instead of decreasing the mutual coherence of the equivalent dictionary, they worked on decreasing its average mutual coherence. Since then many different approaches have appeared. Majority of the approaches \cite{sapiro,Li,cao,gradient} designed the sensing matrix $\bTh$ such that the Gram matrix of  the equivalent dictionary $\bX_{\rm{ED}}$ is close to a target Gram matrix $\bG_{{\rm{Target}}}$.
The authors  in \cite{agelos}, obtained the optimal sensing matrix by first constructing the incoherent dictionary $\bX_{\rm{ED}}$ as an ETF using the alternating projection method (as discussed earlier) and then found the optimal sensing matrix $\bTh$ over the fixed dictionary by solving the following least squares problem:
\begin{equation}\label{LS-1}
\begin{array}{ll}
\underset{\bTh}{\rm minimize}  \|\bX_{\rm{ED}}- \bTh\bpsi\|_{F}^{2} 
 \end{array}
 \end{equation}
Sometimes, even though the optimal sensing matrix is constructed such that the equivalent dictionary is as close as possible to an ETF, it may not perform well. This could happen when the sparse representation error (SRE)  defined as $\bE = \bU -  \bpsi\bS$, where $\bU = [\bu_{1},\bu_{2}, \cdots \bu_{R}]$ is the original signal set and $\bS$ is its corresponding sparse coefficient matrix $\bS = [\bs_{1},\bs_{2}, \cdots \bs_{R}]$ is not taken into account. Projecting SRE to the measurement domain through the  sensing matrix $\bTh$ that is not properly designed can affect the performance of the CS systems. Hence, the optimal sensing matrix must be constructed by also taking the SRE into consideration. Huang et.al. in \cite{xiong} found the optimal sensing matrix such that equivalent dictionary is close to a target frame and also by taking the SRE into account:
\begin{equation}
\begin{array}{ll}
\underset{\bTh,\bX_{\rm{ED-Target}}}{\rm minimize}\:\:\omega\|\bX_{{\rm{ED-Target}}}- \bTh\bpsi\|_{F}^{2} + (1-\omega)\|\bTh(\bU -  \bpsi\bS)\|_{F}^{2}
\end{array}
\end{equation}
where $\bX_{\rm{ED-Target}}$ is a certain target equivalent dictionary and $\omega$ is the trade-off factor which takes the value from $[0,1]$. 
\subsection{Contributions and outline}
 In this paper, we propose an algorithm named \textbf{TE}chnique to devise \textbf{L}arge dimensional \textbf{E}quiangular \textbf{T}ight-frames ({TELET-frames}) based on Majorization Minimization (MM) procedure to construct ETFs. We also use TELET to construct optimal sensing matrix for the CS systems. Since our method does not involve computing EVD and requires only computing some matrix vector product, it is scalable to construct medium to large ETFs. As the proposed algorithm is based on MM, it enjoys nice properties such as monotonicity and guaranteed convergence to a stationary point of the problem. The major contributions of the paper are as follows:
\begin{enumerate}
\item{A MM based iterative algorithm TELET is proposed to construct  the ETFs. The computational complexity of TELET is dictated only by matrix vector products and does not involve computing EVD or matrix inversion at any iteration. We also show a computational efficient way of implementing TELET. Subsequently, we  show that the frames constructed using TELET can be used to construct optimal sensing matrix  for the CS system using the model proposed in \cite{xiong}.}
\item{We also propose a way to accelerate the convergence of the TELET algorithm. The monotonic convergence of the TELET algorithm to the stationary point of the frame design problem is proved. }
\item{We show through numerical simulations that when compared to the state-of-the-art algorithms, TELET can construct frames (especially of large dimensions, of order $N=1000$) with low mutual coherence but with higher run time. Various numerical simulations using synthetic data and real images are also conducted to compare the performance of the sensing matrix constructed via TELET with the other state-of-the-art algorithms used in the CS systems.}
\end{enumerate}
The paper is organized as follows. In Sec. \ref{sec:2}, we formulate the ETF design problem and briefly discuss the MM procedure. Next, in Sec. \ref{sec:4} we develop the proposed algorithm TELET using the MM procedure. In Sec. \ref{sec:5}, we compare the proposed algorithm with the state-of-the-art algorithm through numerical simulations and conclude the paper in Sec. \ref{sec:6}. 
\vspace{-2mm}
\section{ETF Problem Formulation and Preliminaries}\label{sec:2}
\vspace{-2mm}
A unit norm ETF is constructed by minimizing the maximum absolute inner product between pairs of distinct vectors in a frame: 
\begin{equation}\label{Eql}
\begin{aligned}
&\underset{\bx_{1},\cdots, \bx_{N}}{\rm minimize}& & \underset{i, j = 1, 2, \cdots N \: (i \neq j)}{\rm maximize}\:  \left| \bx_{i}^{H}\bx_{j}\right| \\
 & \text{} &  & \text{subject to}\: \: \|\bx_{i}\|_{2}= 1  \quad i=1,2, \cdots N
\end{aligned}
\end{equation}
Suppose $\bx = [\bx_{1}^{T}, \bx_{2}^{T}, \cdots \bx_{N}^{T}]^{T} \in \mathbb{C}^{Nd \times 1}$ is formed by stacking $\{\bx_{i}\}_{i=1}^{N}$ and $\bS_{l}$ be a $d \times dN$ selection matrix defined as $\bS_{l} = [\bzero_{d \times (l-1)d}, \bI_{d}, \bzero_{d \times (N-l)d}]$, then we can obtain $\bx_{l}$ using the following relation:
\begin{equation} 
\begin{array}{ll}
\bx_{l} = \bS_{l}\bx,\: l= 1,2, \cdots, N
\end{array}
\end{equation}
Using the above relation, we rewrite the absolute inner product $|\bx_{i}^{H}\bx_{j}|$ in (\ref{Eql}) as: 
\begin{equation}\label{ip}
\begin{array}{ll}
|\bx_{i}^{H}\bx_{j}|= |\bx^{H} \bS_{j}^{H}\bS_{i} \bx | = |\bx^{H}\bA_{ij} \bx |
\end{array}
\end{equation}
Then using (\ref{ip}), the problem in (\ref{Eql}) becomes:  
\begin{equation}  \label{eq:eqlre}
\begin{aligned}
& \underset{\boldsymbol{\boldsymbol{\boldsymbol{x}}}}{\text{\text{minimize}}} &  & \text{\ensuremath{\underset{i, j = 1, 2, \cdots N \: (i \neq j)}{\text{\text{maximum}}}}} &  & 
\bar{f}_{ij}(\bx) \\
 & \text{} &  & \text{subject to} &  &\|\bx_{i}\|_{2}= 1  \quad i=1,2, \cdots N
\end{aligned}
\end{equation}
where $\bar{f}_{ij}(\bx)=2\left(|{{{\bx}}}^{H}{{{{\bA}_{ij}}}}{{{\bx}}}|^{2}\right)$. Please note that we have squared the objective function (which will not change the optimum) and scaled by two for our convenience. The problem in (\ref{eq:eqlre}) can be further rewritten as: 
\begin{equation}\label{eq:vecprob}
\begin{aligned} & \underset{\boldsymbol{\boldsymbol{{\bY}}},\boldsymbol{\boldsymbol{\boldsymbol{x}}}}{\text{\text{minimize}}} &  & \text{\ensuremath{\underset{i, j = 1, 2, \cdots N \: (i \neq j)}{\text{\text{maximum}}}}} &  & \text{vec}^{H}(\boldsymbol{\boldsymbol{\boldsymbol{Y}}}){\boldsymbol{\boldsymbol{\boldsymbol{\Phi}}}_{ij}}\text{vec}(\boldsymbol{\boldsymbol{\boldsymbol{Y}}})\\
 & \text{} &  & \text{subject to} &  & \|\bx_{i}\|_{2}= 1  \quad i=1,2, \cdots N\\
 &  &  &  &  & \boldsymbol{\boldsymbol{\boldsymbol{Y}}}=\boldsymbol{\boldsymbol{x}}\boldsymbol{\boldsymbol{x}}^{H},
\end{aligned}
\end{equation}
where ${\boldsymbol{\boldsymbol{\boldsymbol{\Phi}}}_{ij}}=\text{vec}\left({{\bA}_{ij}}\right)\text{vec}^{H}\left(\left({{\bA}_{ij}}\right)^{H}\right)+ \text{vec}\left(\left({{\bA}_{ij}}\right)^{H}\right)\text{vec}^{H}\left({{\bA}_{ij}}\right)$. We solve the above non-convex minimax problem using the MM procedure which will be explained in the next subsection. 
\subsection{MM for the minimax problem}
Consider the following optimization problem: 
\begin{equation}\label{minimax}
\begin{array}{ll}
\underset{\bx \in \chi}{\rm minimize} \: f(\bx)
\end{array}
\end{equation}
where the objective function $f(\bx)$ is equal to the maximum value of the $i^{th}$ sub-function $\bar{f}_{i}(\bx)$ i.e. $f(\bx) = \underset{i =1,2, \cdots, I}{\rm maximize}\:\bar{f}_{i}(\bx)$ .  The MM based algorithms solves the above problem by first constructing a surrogate function $g(\bx|\bx^{t})$ which majorizes the objective function $f(\bx)$ at the current iterate $\bx^{t}$ and in the next step, the surrogate function is minimized to get the next iterate.
\begin{equation}  \label{eq:mmc}
\bx^{t+1} \in \underset{\bx \in \chi}{\rm arg\:min} \: g\left(\bx|\bx^{t}\right)
\end{equation}
Majorization combines the tangency and the upper bound condition, i.e.,  a function qualifies to be a surrogate function if it satisfies the following conditions: 
\begin{equation}  \label{eq:mma}
g\left(\bx^{t}|\bx^{t}\right) = f\left(\bx^{t}\right) 
\end{equation}
\begin{equation}\label{eq:mmb}
g\left(\bx|\bx^{t}\right) \geq f\left(\bx\right) 
\end{equation} 
Constructing $g(\bx|\bx^{t})$ for the minimax problem in (\ref{minimax}) does not look obvious, but can be worked out as follows, let: 
\begin{equation}\label{surrogateminmax}
\begin{array}{ll}
g(\bx|\bx^{t})= \underset{i =1,2, \cdots, I}{\rm maximize}\: \bar{g}_{i}(\bx|\bx^{t}) 
\end{array}
\end{equation}
where each $\bar{g}_{i}(\bx)$ is a tight upper bound on $\bar{f}_{i}(\bx)$ satisfying the following conditions: 
\begin{equation}\label{cond1}
\begin{array}{ll}
\bar{g}_{i}(\bx^{t}|\bx^{t}) = \bar{f}_{i}(\bx^{t})
\end{array}
\end{equation}
\begin{equation}\label{cond2}
\begin{array}{ll}
\bar{g}_{i}(\bx|\bx^{t}) \geq \bar{f}_{i}(\bx)
\end{array}
\end{equation}
We now show that the surrogate function ${g}(\bx|\bx^{t})$  defined in (\ref{surrogateminmax}) satisfies condition (\ref{eq:mma}) and (\ref{eq:mmb}):\\
\emph{Proof of $g(\bx|\bx^{t})$ satisfying condition (\ref{eq:mma})}:
\begin{equation}\label{mmd}
\begin{array}{ll}
g(\bx^{t}|\bx^{t}) = \underset{i =1,2, \cdots, I}{\rm maximize}\: \bar{g}_{i}(\bx^{t}|\bx^{t})  = \underset{i =1,2, \cdots, I}{\rm maximize}\: \bar{f}_{i}(\bx^{t})  = f(\bx^{t})
\end{array}
\end{equation}
where the first and second equality is by (\ref{surrogateminmax}) and (\ref{cond1}) and the last equality is by the definition of $f(\bx)$. \\
\emph{Proof of $g(\bx|\bx^{t})$ satisfying condition (\ref{eq:mmb})}:
\begin{equation}\label{mme}
\begin{array}{ll}
g(\bx|\bx^{t})= \underset{i =1,2, \cdots, I}{\rm maximize}\: \bar{g}_{i}(\bx|\bx^{t}) \geq \underset{i =1,2, \cdots, I}{\rm maximize}\: \bar{f}_{i}(\bx^{t})  = f(\bx^{t})
\end{array}
\end{equation}
where the first equality is given by (\ref{surrogateminmax}), second inequality is by (\ref{cond2}) and the last equality is given by the definition of $f(\bx)$.
Using (\ref{eq:mmc}), (\ref{eq:mma}) and (\ref{eq:mmb}) it can be shown that the sequence of points $\{\bx^{t}\}$ generated by the MM procedure monotonically decrease the objective function. Summary of different techniques used to construct the surrogate function can be found in (\cite{hunter}, \cite{sir}). Further, since the proposed algorithm is developed using MM, it can be shown using Theorem 1 in (\cite{convergence}) that TELET will converge to a stationary point of the problem in (\ref{Eql}). Due to lack of space, we do not include the proof of convergence here. 
\vspace{-0.5cm}
\section{Algorithm for constructing large dimensional equiangular tight frames}\label{sec:4}
In this section, we develop  \textbf{TE}chnique to devise \textbf{L}arge dimensional \textbf{E}quiangular \textbf{T}ight-frames construction algorithm (\textbf{TELET}) using the MM procedure. For better readability, we present our MM based algorithm in two stages: majorization function construction and minimization. We also propose a way to accelerate the convergence of TELET algorithm using SQUAREM acceleration scheme. 
\subsection{Majorization Function construction}
In this subsection, we construct the surrogate function for the objective function in (\ref{eq:vecprob}): 
\begin{equation}\label{eq:vecprobrep}
\begin{aligned} & \underset{\boldsymbol{\boldsymbol{{\bY}}},\boldsymbol{\boldsymbol{\boldsymbol{x}}}}{\text{\text{minimize}}} &  & \text{\ensuremath{\underset{i, j = 1, 2, \cdots N \: (i \neq j)}{\text{\text{maximum}}}}} &  & \bar{f}_{ij}(\bY)\\
 & \text{} &  & \text{subject to} &  & \|\bx_{i}\|_{2}= 1  \quad i=1,2, \cdots N\\
 &  &  &  &  & \boldsymbol{\boldsymbol{\boldsymbol{Y}}}=\boldsymbol{\boldsymbol{x}}\boldsymbol{\boldsymbol{x}}^{H},
\end{aligned}
\end{equation}
where $ \bar{f}_{ij}(\bY)=\text{vec}^{H}(\boldsymbol{\boldsymbol{\boldsymbol{Y}}}){\boldsymbol{\boldsymbol{\boldsymbol{\Phi}}}_{ij}}\text{vec}(\boldsymbol{\boldsymbol{\boldsymbol{Y}}})$. As discussed in Sec. \ref{sec:2}.A, constructing majorizing function for $\text{\ensuremath{\underset{i, j = 1, 2, \cdots N \: (i \neq j)}{\text{\text{maximum}}}}} \bar{f}_{ij}(\bY)$ boils down to finding a majorizing function one for each $\bar{f}_{ij}(\bY)$. Thus, we can focus on $\bar{f}_{ij}(\bY)$ only. Next, we discuss the following lemmas which will be used to find a tight
upper bound for $\bar{f}_{ij}(\bY)$.
\begin{lemma} \label{lemma 1}
Let $f(\bx) \in \mathbb{C}^{md \times 1} \rightarrow \mathbb{R} $ be a continuous twice differentiable function in $\bx$ and if $f(\bx)$ has a bounded curvature, then there exists a matrix $\bM$ such that $\boldsymbol{\boldsymbol{M}}\succeq\nabla^{2}f(\boldsymbol{\boldsymbol{\boldsymbol{x}}})$ and $f(\bx)$ can be upper bounded as: 
\begin{equation}
\begin{array}{ll}
f(\bx) \leq f(\bx^{t}) + {\rm{Re}}\left(\nabla f(\bx^{t})^{H}(\bx - \bx^{t})\right)+\dfrac{1}{2}(\bx - \bx^{t})^{H}\bM( \bx - \bx^{t})
\end{array}
\end{equation}
where $\bx = \bx^{t}$ is the value taken by $\bx$ at the $t^{th}$ iteration. The upper bound for $f(\bx)$ is quadratic and differentiable in $\bx$. 
\end{lemma}
\begin{IEEEproof}
Suppose there exists a matrix $\bM$ such that $\boldsymbol{\boldsymbol{M}}\succeq\nabla^{2}f(\boldsymbol{\boldsymbol{\boldsymbol{x}}})$, then we have the following inequality by second order
Taylor expansion:
\begin{equation}
\begin{array}{ll}
f(\bx) \leq f(\bx^{t}) + {\rm{Re}}\left(\nabla f(\bx^{t})^{H}(\bx - \bx^{t})\right)+\dfrac{1}{2}(\bx - \bx^{t})^{H}\bM( \bx - \bx^{t})
\end{array}
\end{equation}
and equality is achieved at $\bx = \bx^{t}$. 
\end{IEEEproof}
\begin{lemma} \label{lemma 2}
Given any ${\bx= \bx^{t}}$, a concave function $f\left(\bx\right) \in \mathbb{C}^{md \times 1} \rightarrow \mathbb{R}$  can be upper  bounded as:
\begin{equation}\label{eq:41}
\begin{array}{ll}
{f}\left(\bx\right) \leq  {f}\left(\bx^{t}\right) +  {\rm{Re}}\left(\left(\nabla f \left(\bx^{t}\right) \right)^{H}\left(\bx-\bx^{t}\right)\right)
\end{array}
\end{equation}
The upper bound for $f\left(\bx\right)$ is linear in $\bx$.
\end{lemma}
\begin{IEEEproof}
Since $f(\bx)$ is concave, linearizing it around ${\bx^{t}}$ using the first order Taylor series gives the above inequality.
\end{IEEEproof}
Since $\bar{f}_{ij}(\bY)$ in (\ref{eq:vecprobrep}) is quadratic and differentiable in ${\textrm{vec}}(\bY)$, we apply Lemma \ref{lemma 1} with ${\boldsymbol{\boldsymbol{M}}}_{ij}=\lambda_{\text{max}}({\boldsymbol{\boldsymbol{\Phi}}}_{ij})\boldsymbol{\boldsymbol{I}}_{N^{2}d^{2}} = d{\boldsymbol{I}}_{N^{2}d^{2}}$ (refer Appendix 1 for proof) and use the relation $\text{vec}^{H}(\boldsymbol{\boldsymbol{\boldsymbol{Y}}})\text{vec}(\boldsymbol{\boldsymbol{\boldsymbol{Y}}})=N^{2}$ to get the following surrogate function: 
\begin{equation}
\begin{array}{ll} \label{eq:vecfinal}
\tilde{g}_{ij}\left(\bY|\bY^{t}\right) =  -\text{vec}^{H}(\boldsymbol{\boldsymbol{\boldsymbol{Y}}}^{t})\boldsymbol{\boldsymbol{\boldsymbol{\Phi}}}_{ij}\text{vec}(\boldsymbol{\boldsymbol{\boldsymbol{Y}}}^{t})+2N^{2}d+
2\text{Re}(\text{vec}^{H}(\boldsymbol{\boldsymbol{\boldsymbol{Y}}}^{t})\boldsymbol{\boldsymbol{\boldsymbol{\Phi}}}_{ij}\text{vec}(\boldsymbol{\boldsymbol{\boldsymbol{Y}}}))-2d\text{Re}(\text{vec}^{H}(\boldsymbol{\boldsymbol{\boldsymbol{Y}}}^{t})\text{vec}(\boldsymbol{\boldsymbol{\boldsymbol{Y}}})).
\end{array}
\end{equation}
Next, we rewrite $\tilde{g}_{ij}\left(\bY|\bY^{t}\right)$ in terms of $\bx$ by using $\bY = \bx\bx^{H}$:
\begin{equation}\label{s}
\begin{aligned}
\tilde{g}_{ij}\left(\bx|\bx^{t}\right) = 
 -2|(\boldsymbol{\boldsymbol{\boldsymbol{x}}}^{t})^{H}{{\bA}_{ij}}\boldsymbol{\boldsymbol{\boldsymbol{x}}}^{t}|^{2}+2(\boldsymbol{\boldsymbol{\boldsymbol{x}}}^{H}{\bB}^{t}_{ij}\boldsymbol{\boldsymbol{\boldsymbol{x}}})-2d(\boldsymbol{\boldsymbol{\boldsymbol{x}}}^{H}\boldsymbol{\boldsymbol{\boldsymbol{x}}}^{t}(\boldsymbol{\boldsymbol{\boldsymbol{x}}}^{t})^{H}\boldsymbol{\boldsymbol{\boldsymbol{x}}})+2N^{2}d
\end{aligned}
\end{equation}
where $\boldsymbol{B}^{t}_{ij}=\boldsymbol{\boldsymbol{\boldsymbol{A}}}_{ij}((\boldsymbol{\boldsymbol{\boldsymbol{x}}}^{t})^{H}\boldsymbol{\boldsymbol{\boldsymbol{A}}}_{ij}^{H}\boldsymbol{\boldsymbol{\boldsymbol{x}}}^{t})+\boldsymbol{\boldsymbol{\boldsymbol{A}}}_{ij}^{H}((\boldsymbol{\boldsymbol{\boldsymbol{x}}}^{t})^{H}\boldsymbol{\boldsymbol{\boldsymbol{A}}}_{ij}\boldsymbol{\boldsymbol{\boldsymbol{x}}}^{t})$. Using (\ref{ip}), ${\bB}^{t}_{ij}$ can be written compactly as ${\bB}^{t}_{ij}= \bA_{ij} \left(\left(\bx_{j}^{t}\right)^{H}\bx_{i}^{t}\right)+\bA_{ij}^{H}\left(\left(\bx_{i}^{t}\right)^{H}\bx_{j}^{t}\right)$. Note that the above surrogate function $\tilde{g}_{ij}\left(\bx|\bx^{t}\right)$ has a  dominant convex quadratic term in $\bx$: $2(\boldsymbol{\boldsymbol{\boldsymbol{x}}}^{H}{\bB}^{t}_{ij}\boldsymbol{\boldsymbol{\boldsymbol{x}}})$ which makes the resulting surrogate minimization problem intractable under the unit norm constraint. To overcome this, we first add and subtract $\lambda_{\text{max}}({\bB}^{t}_{ij})$ to $\tilde{g}_{ij}\left(\bx|\bx^{t}\right)$ and obtain the surrogate function:  
\begin{equation}
\begin{array}{ll}
\tilde{g}_{ij}\left(\bx|\bx^{t}\right)= 
 -2|(\boldsymbol{\boldsymbol{\boldsymbol{x}}}^{t})^{H}{{\bA}_{ij}}\boldsymbol{\boldsymbol{\boldsymbol{x}}}^{t}|^{2}+2(\boldsymbol{\boldsymbol{\boldsymbol{x}}}^{H}\hat{\bB}^{t}_{ij}\boldsymbol{\boldsymbol{\boldsymbol{x}}})+
 2\lambda_{\text{max}}({\bB}^{t}_{ij})N -2d(\boldsymbol{\boldsymbol{\boldsymbol{x}}}^{H}\boldsymbol{\boldsymbol{\boldsymbol{x}}}^{t}(\boldsymbol{\boldsymbol{\boldsymbol{x}}}^{t})^{H}\boldsymbol{\boldsymbol{\boldsymbol{x}}})+2N^{2}d
\end{array}
\label{eq:backx2}
\end{equation}
where $\hat{\bB}^{t}_{ij}={\bB}^{t}_{ij}-(\lambda_{\text{max}}({\bB}^{t}_{ij})\boldsymbol{\boldsymbol{I}}_{Nd}) ={\bB}^{t}_{ij}-|\left({\bx_{i}^{t}}\right)^{H}{{\bx}_{j}^{t}}|\boldsymbol{\boldsymbol{I}}_{Nd}$ (see Appendix 2 for proof of $\lambda_{\text{max}}({\bB}^{t}_{ij})=|\left({\bx_{i}^{t}}\right)^{H}{{\bx}_{j}^{t}}|$). Next, since the surrogate function $\tilde{g}_{ij}\left(\bx|\bx^{t}\right)$ is now concave in $\bx$, we majorize it once again using Lemma \ref{lemma 2}: 
\begin{equation}
\begin{array}{ll}\label{eq:backxfinal}
\bar{g}_{ij}(\bx|\bx^{t}) = -2|(\boldsymbol{\boldsymbol{\boldsymbol{x}}}^{t})^{H}{{\bA}_{ij}}\boldsymbol{\boldsymbol{\boldsymbol{x}}}^{t}|^{2}+
2\left(-(\boldsymbol{\boldsymbol{\boldsymbol{x}}}^{t})^{H}\hat{\bB}^{t}_{ij}\boldsymbol{\boldsymbol{\boldsymbol{x}}}^{t}+2\text{Re}\left((\boldsymbol{\boldsymbol{\boldsymbol{x}}}^{t})^{H}{\hat{\bB}^{t}_{ij}}\boldsymbol{\boldsymbol{\boldsymbol{x}}}\right)\right)+2\lambda_{\text{max}}({\bB}^{t}_{ij})N -\\
2d \left(-N^{2}+2\text{Re}\left(N\boldsymbol{\boldsymbol{\boldsymbol{x}}}^{H}\boldsymbol{\boldsymbol{\boldsymbol{x}}}^{t}\right)\right)+2N^{2}d.\vspace{-3.5mm}
\end{array}
\end{equation}
The above surrogate function $\bar{g}_{ij}(\bx|\bx^{t})$ is linear in $\bx$ and hence can now be minimized. Note that although we do double majorization i.e. find surrogate of the surrogate function, since $\bar{g}_{ij}(\bx|\bx^{t})$ is a tighter surrogate for $\tilde{g}_{ij}\left(\bx|\bx^{t}\right)$, it can be viewed as a direct surrogate to the objective function $\bar{f}_{ij}(\bx)$ in (\ref{eq:eqlre}). The surrogate function in (\ref{eq:backxfinal}) can be rewritten compactly as: 
\begin{equation} \label{surrogate}
\begin{aligned}
\bar{g}_{ij}(\bx|\bx^{t})  = 4\text{Re}(\boldsymbol{\boldsymbol{x}}{}^{H}{{\bd^{t}_{ij}}})+s^{t}_{ij}
\end{aligned}
\end{equation}
where $s^{t}_{ij} = - 6|(\bx^{t}_{i})^{H}\bx^{t}_{j}|^{2} +4N|(\bx^{t}_{i})^{H}\bx^{t}_{j}|+4N^{2}d$ and $\bd_{ij}^{t} = \left(\bA_{ij}\bx^{t}\right) ((\bx^{t}_{j})^{H}\bx^{t}_{i}) + \left(\bA_{ij}^{H}\bx^{t}\right)((\bx^{t}_{i})^{H}\bx^{t}_{j}) - \left(|(\bx^{t}_{i})^{H}\bx^{t}_{j})|\bI_{Nd}\right)\bx^{t} - Nd\bx^{t}$. The superscript $t$ in $s^{t}_{ij}$ and $\bd_{ij}^{t}$ denotes that the computation of these quantities depends on the previous iterate value $\bx^{t}$. The only computational complexity involved in computing $\bar{g}_{ij}(\bx|\bx^{t})$ is the matrix vector product $\bA_{ij}\bx^{t}$ and $\bA_{ij}^{H}\bx^{t}$. Exploiting the sparse structure of  $\bA_{ij}$, the matrix vector product  can be carried out efficiently as: $\bA_{ij}\bx^{t} = [\bzero^{T}_{d(j-1) \times 1},({\bx_{i}^{t}})^{T},\bzero^{T}_{d(N-j)\times 1}]^{T}$ and $\bA_{ij}^{H}\bx^{t} = [\bzero^{T}_{d(i-1) \times 1},({\bx_{j}^{t}})^{T},\bzero^{T}_{d(N-i)\times 1}]^{T}$
Then the majorizing function for $\text{\ensuremath{\underset{i, j = 1, 2, \cdots N \: (i \neq j)}{\text{\text{maximum}}}}} \bar{f}_{ij}(\bx)$  is: 
\begin{equation}\label{surrogatefinal}
\begin{array}{ll}
g(\bx|\bx^{t}) = \text{\ensuremath{\underset{i, j = 1, 2, \cdots N \: (i \neq j)}{\text{\text{maximum}}}}} \bar{g}_{ij}(\bx|\bx^{t}) 
\end{array}
\end{equation}
Hence at any iteration, given $\bx^{t}$, the surrogate minimization problem becomes: 
\begin{equation}
\begin{aligned} & \underset{\boldsymbol{\boldsymbol{\boldsymbol{x}}}}{\text{\text{minimize}}} &  & \text{\ensuremath{\underset{i, j = 1, 2, \cdots N \: (i \neq j)}{\text{\text{maximum}}}}} &  & 4\text{Re}(\boldsymbol{\boldsymbol{x}}{}^{H}{{\bd^{t}_{ij}}})+s^{t}_{ij}\\
 & \text{} &  & \text{subject to} &  & \|\bx_{i}\|_{2} = 1,\,i = 1,2,\cdots ,N
\end{aligned}
\label{eq:Maj1}
\end{equation}
\vspace{-6mm} 
\subsection{Solution to the Minimization Problem}
\vspace{-1mm}
The surrogate minimization problem in (\ref{eq:Maj1}) is non-convex in $\bx$ because of the equality constraint. However, 
the equality constraint can be relaxed with the inequality constraint as the optimal solution of the relaxed problem would lie only on the boundary of the relaxed constraint set \cite{boyd}. The relaxed problem is: 
\begin{equation}
\begin{aligned} & \underset{\bx}{\text{\text{minimize}}} &  & \text{\ensuremath{\underset{i,j = 1,2, \cdots, N (i\neq j)}{\text{\text{maximum}}}}} &  & 4\text{Re}(\boldsymbol{\boldsymbol{x}}{}^{H}\boldsymbol{\boldsymbol{d}}^{t}_{ij})+s^{t}_{ij}\\
 & \text{} &  & \text{subject to} &  & \|\bx_{i}\|_{2} \leq 1,\,i = 1,2,\cdots ,N\ 
\end{aligned}
\label{eq:epiundo}
\end{equation}
Next, we rewrite the inner maximization problem over the indices as maximization over a simplex variable:  
\begin{equation}
\begin{aligned} & \underset{\bx}{\text{\text{minimize}}} &  & \text{\ensuremath{\underset{\bq \geq 0, \b1^{T}\bq=1}{\text{\text{maximum}}}}} &  &  4\text{Re}\left(\left({\bD}^{t}\bq\right)^{H}\bx\right)+ \bq^{T}\bs^{t}\\
 & \text{} &  & \text{subject to} &  &   \|\bx_{i}\|_{2} \leq 1,\,i = 1,2,\cdots ,N\ 
\end{aligned}
\end{equation}
where $\bs^{t} = [s^{t}_{12} ,\, s^{t}_{13} \cdots s^{t}_{N(N-1)}]^{T} \in \mathbb{R}^{N(N-1) \times 1}$ and ${\bD}^{t}= [{\bd}^{t}_{12}, {\bd}^{t}_{13}, \cdots, {\bd}^{t}_{N(N-1)}] \in \mathbb{R}^{2Nd \times N(N-1)}$.
The objective function in the above problem is bilinear in $\bq$ and $\bx$ and the constraints are compact convex sets, then by minimax theorem \cite{minmax}, we can swap minimax to maximin without altering the solution: 
\begin{equation}
\begin{aligned} &  \text{\ensuremath{\underset{\bq \geq 0, \b1^{T}\bq=1}{\text{\text{maximum}}}}} &  & \underset{\bx}{\text{\text{minimize}}} &  & 4\text{Re}\left(\left({\bD}^{t}\bq\right)^{H}\bx\right)+ \bq^{T}\bs^{t}\\
 & \text{} &  & \text{subject to} &  &   \|\bx_{i}\|_{2} \leq 1,\,i = 1,2,\cdots ,N\ 
\end{aligned}
\end{equation}
which is equivalent to 
\begin{equation}
\begin{aligned} & \text{\ensuremath{\underset{\bq \geq 0, \b1^{T}\bq=1}{\text{\text{maximize}}}}}  &  & h(\bq) 
\end{aligned}
\label{max}
\end{equation}
where 
\begin{equation}\label{min}
\begin{array}{ll}
h(\bq)= \underset{\bx}{{\textrm{minimize}}} \:  4\text{Re}\left(\left({\bD}^{t}\bq\right)^{H}\bx\right)+ \bq^{T}\bs^{t}\\
\:\:\:\:\:\:\:\:\:\: \:\:\:{\textrm{subject to}} \:  \|\bx_{i}\|_{2} \leq 1,\,i = 1,2,\cdots ,N\ 
\end{array}
\end{equation}
The above problem has a closed-form solution and is given as 
\begin{equation}\label{solz}
\bx^{t+1}_{i}= -\dfrac{{{\ba}^{t}_{i}}}{\|{{\ba}^{t}_{i}}\|_{2}} ,\: i=1, 2, \cdots, N
\end{equation}
where $\ba^{t}$ is made of blocks $\ba^{t}_{1}, \ba^{t}_{2}, \cdots, \ba^{t}_{N}$ and $\ba^{t} = {\bD}^{t}\bq$ where $\bq$ is obtained by solving the problem in (\ref{max}). However, the problem in (\ref{max}) does not have a closed-form solution and hence we solve it iteratively using Mirror Descent Algorithm (MDA) whose update step is given as:  
\begin{equation}\label{mdaupdate}
\begin{aligned}
\bq^{k+1}  = {\underset{\bq \geq 0, \b1^{T}\bq=1}{\text{\text{arg\:max}}}}  & {(\bh^{k})}^{T}\bq + \dfrac{1}{\gamma_{k}} \bB_{\psi} \left(\bq,\bq^{k}\right)\\
\end{aligned}
\end{equation}
where $\bq^{k}$ represents the value taken by $\bq$ at the $k^{th}$ iteration of the MDA algorithm (this should not be confused with superscript $t$ which is used as the iteration index in the outer loop), $\bh^{k} \in \partial h(\bq^{k})$ is the subgradient of $h(\bq)$, $\gamma_{k} > 0$ is the step size given by $\gamma_{k}= \dfrac{\eta}{\sqrt{k}}$ where $\eta$ is some constant and $\bB_{\psi}\left(\bq,\bq^{k}\right)$ is Bregman-like distance generated by $\psi$ and is given by $\bB_{\psi}\left(\bq,\bq^{k}\right) = \psi(\bq) - \psi(\bq^{k}) - \nabla^{T}\psi\left(\bq^{k}\right)\left(\bq -\bq^{k}\right)$. Since the constraint in the maximization problem is a unit simplex, similar to \cite{MDA1}, we choose $\psi(\bq)$ as: 
\begin{equation} 
\begin{array}{ll}
\psi(\bq)=\left\{ \begin{array}{ll}
\displaystyle\sum_{i=1}^{N(N-1)}q_{i}\textrm{log}\:q_{i} & \bq \in \mathcal{Q}\\
+\infty & \rm{otherwise}\\
\end{array}\right..
\end{array}
\end{equation}
\vspace{-1mm}
where $\mathcal{Q} = \{\bq \in \mathbf{R}^{N(N-1)} | \b1^{T}\bq =1, \bq \geq 0 \}$. Hence, the update step in (\ref{mdaupdate}) is simplified to:
\vspace{-1mm}
\begin{equation}\label{simplified}
\begin{array}{ll}
q_{j}^{k+1} = \dfrac{q_{j}^{k}\textrm{exp} \left(-\gamma_{k}h_{j}^{k}\right)}{\displaystyle\sum_{j=1}^{N(N-1)}q_{j}^{k}\textrm{exp} \left(-\gamma_{k}h_{j}^{k}\right)}\: j=1,2,\cdots N(N-1)
\end{array}
\end{equation} 
The subgradient $\bh^{k}$ is given as:
\begin{equation}
\begin{array}{ll}
\bh^{k} = 4\text{Re}\left(({\bD}^{t})^{H}\by^{k}\right) +{\bs}^{t}
\end{array}
\end{equation}
where 
\begin{equation}
\begin{array}{ll}
\by^{k} = \underset{\by}{\rm{min}}\: 4\text{Re}\left(\left(\left({\bD}\right)^{t}\bq^{k}\right)^{H}\by\right)\\
\hspace{5mm} \:{\textrm{subject to}} \:  \|\by_{i}\|_{2} \leq 1,\,i = 1,2,\cdots ,N\ 
\end{array}
\end{equation}
The above problem has closed-form solution and is given by $\by^{k}_{i}= -\dfrac{{{\ba}^{k,t}_{i}}}{\|{{\ba}^{k,t}_{i}}\|_{2}}$ for $i=1, 2 \cdots N$ and $\ba^{k,t} = {\bD}^{t}\bq^{k}$ (we have used the superscript $k,t$ for the vector $\ba$ here since it depends on the $k^{th}$ iteration value of $\bq$ of the MDA algorithm and also on the $t^{th}$ iteration value of $\bD$). The update step in (\ref{simplified}) is repeated until convergence and once the optimal $\bq$ is found, $\bx$ can be calculated using (\ref{solz}). 
The pseudocode of the proposed algorithm is shown in Algorithm 1. 
\vspace{-3mm}
\begin{center}
\begin{tabular}{ p{15cm} }
\hline
\hline
\bf{Algorithm 1: Pseudocode of the proposed algorithm} \\
\hline
\hline
{\bf{Input}}: Number of vectors in frame $N$ and the dimension of each vector $d$.  \\
{\bf{Initialize}}: Set $t=0$. Initialize ${{\bx}^{0}}$  \\       
{\bf{Repeat}}: \\
1)\, Compute the following for $i, j= 1, 2, \cdots, N$ and $i \neq j$\\
a) Compute $\bd^{t}_{ij} =  [\bzero^{T}_{d(j-1) \times 1},({\bx_{i}^{t}})^{T},\bzero^{T}_{d(N-j)\times 1}]^{T}((\bx^{t}_{j})^{H}\bx^{t}_{i})+  [\bzero^{T}_{d(i-1) \times 1},({\bx_{j}^{t}})^{T},\bzero^{T}_{d(N-i)\times 1}]^{T}((\bx^{t}_{i})^{H}\bx^{t}_{j})   - (|(\bx^{t}_{i})^{H}\bx^{t}_{j})|+ Nd)\bx^{t}$.\\
b) Compute $s^{t}_{ij} =  - 6|(\bx^{t}_{i})^{H}\bx^{t}_{j}|^{2} +4N|(\bx^{t}_{i})^{H}\bx^{t}_{j}|+4N^{2}d$. \\
2)\, Compute ${\bD}^{t} = [{\bd^{t}_{12}}, {\bd^{t}_{13}}, \cdots, {\bd^{t}}_{(N(N-1))}]$ and $\ba^{t} = {\bD^{t}}\bq$ where $\bq$ is obtained by solving the maximization problem in (\ref{max}) using the MDA algorithm as discussed in Subsection. \ref{sec:4}-B. \\ 
3)\, Compute $\bx^{t+1}_{i}=-\dfrac{{{\ba}^{t}_{i}}}{\|{{\ba}^{t}_{i}}\|_{2}}$ for $i=1, 2, \cdots, N$.\\
$t \leftarrow t+1$, \textbf{until convergence}\\
\hline
\hline
\end{tabular}
\end{center}
\vspace{-2mm}
Before we end this subsection, we will here discuss the computational complexity of TELET. As shown in the algorithmic development, TELET consists of an inner loop which implements MDA to solve the surrogate minimization problem and an outer loop which updates each element of $\bd^{t}_{ij}$ and $s^{t}_{ij}$. We first discuss the complexity of MDA - which is mostly dictated by the matrix vector product $({\bD}^{t})^{H}\by^{k}$ and $({\bD}^{t})\bq^{k}$. The columns of the matrix ${\bD^{t}}$ denoted as ${\bd^{t}_{ij}}$ has a special structure which we utilize to carry out this matrix vector operation efficiently. Note that each $\bd^{t}_{ij}$  is made by adding a sparse vector:  $[\bzero^{T}_{d(j-1) \times 1},({\bx_{i}^{t}})^{T},\bzero^{T}_{d(N-j)\times 1}]^{T}((\bx^{t}_{j})^{H}\bx^{t}_{i})+  [\bzero^{T}_{d(i-1) \times 1},({\bx_{j}^{t}})^{T},\bzero^{T}_{d(N-i)\times 1}]^{T}((\bx^{t}_{i})^{H}\bx^{t}_{j})$- which consists of $2d$ non-zero elements and a non-sparse vector: $(|(\bx^{t}_{i})^{H}\bx^{t}_{j})|+ Nd)\bx^{t}$ - which is nothing but a scalar vector product. Exploiting this structure, the matrix vector products can be carried out efficiently with complexity of $\mathcal{O}\left(N^{2}d\right)$ per iteration. In the case of outer loop, the major computations are that of computing the inner products $(\bx^{t}_{i})^{H}\bx^{t}_{j}$ which requires only $\mathcal{O}\left(Nd\right)$ computation per iteration. Hence, after neglecting the lower terms, the computational complexity of TELET is equal to $\mathcal{O}\left(N^{2}d\right)$. 
\subsection{Proof of convergence}
We denote the objective function of the problem in (\ref{Eql}) as $f_{_{ETF}}(\bx)$. Then, from the derivation of the proposed algorithm, we know that $f_{_{ETF}}(\bx)$ is majorized by $g(\bx|\bx^{t})$ in (\ref{surrogatefinal}) at $\bx^{t}$ over the constraint  $\|\bx_{i}\|_{2} \leq 1$. Using (\ref{eq:mmc}), (\ref{mmd}) and (\ref{mme}) it can be shown that the sequence of points $\{\bx^{t}\}$ generated by the MM procedure monotonically decrease the objective function:
\begin{equation}\label{conv}
\begin{array}{ll}
f_{_{ETF}}(\bx^{t+1}) \leq g(\bx^{t+1}|\bx^{t}) \leq g(\bx^{t}|\bx^{t})  = f_{_{ETF}}(\bx^{t})
\end{array}
\end{equation}
where the first inequality and the last equality are obtained by using (\ref{mmd}) and (\ref{mme}). The second inequality is by (\ref{eq:mmc}). Since $f_{_{ETF}}(\bx)$ in (\ref{Eql}) is bounded below by zero and the points $\{\bx^{t}\}$ monotonically decrease the objective function, it is guaranteed that the sequence $\{f_{_{ETF}}(\bx^{t})\}$ will converge to a finite value.\\
In the following, we show that the sequence $\{\bx^{t}\}$ converges to the stationary point of the problem in (\ref{Eql}). Assume that there is a subsequence $\tilde{\bx}^{r_{j}}$ converging to a limit point ${\bp}$. Then from (\ref{mmd}), (\ref{mme}) and  (\ref{conv}) we get:
\begin{equation}
\begin{array}{ll}
g(\tilde{\bx}^{r_{j+1}}|\tilde{\bx}^{r_{j+1}}) = f_{_{\textrm{ETF}}}({\tilde{\bx}^{r_{j+1}}}) \leq f_{_{\textrm{ETF}}}({\tilde{\bx}^{r_{j}+1}})\\
 \leq g(\tilde{\bx}^{r_{j}+1}|\tilde{\bx}^{r_{j}}) \leq g(\tilde{\bx}|\tilde{\bx}^{r_{j}})
\end{array}
\end{equation}
where $g(.)$ is the surrogate function as defined in (\ref{surrogatefinal}). Then, letting $j \rightarrow \infty$, we get:
\begin{equation}
\begin{array}{ll}
g({\bp}|{\bp})  \leq g(\tilde{\bx}|{\bp})
\end{array}
\end{equation}
which implies $g'({\bp}|{\bp}) \geq 0$. Since the first order behavior of surrogate function is same as function $f_{_{\textrm{ETF}}}({\tilde{\bx}})$ (\cite{convergence}), $g'({\bp}|{\bp}) \geq 0$  implies $f'_{_{\textrm{ETF}}}({{\bp}}) \geq 0$. Hence, ${\bp}$ is the stationary point of $f_{_{\textrm{ETF}}}({\tilde{\bx}})$ and therefore the proposed algorithm converges to the stationary point of the ETF problem.
\subsection{SQUAREM acceleration scheme}
As discussed in Sec. \ref{sec:2}, the choice of surrogate function will dictate the convergence speed of the MM based algorithm. In our case, since the surrogate function is constructed through double majorization of the objective function, it may lead to slower convergence of the proposed algorithm. Squared Iterative Method (SQUAREM) \cite{squarem} is based on Cauchy-Brazilai-Brownein method, to accelerate the convergence rate.
Let $F_{\rm{{MM}}}(\cdot)$ denote the nonlinear fixed-point iteration map of the proposed algorithm. Then the MM updating scheme can be expressed as: 
\begin{equation}
\bx^{t+1} = F_{\rm{{MM}}}(\bx^{t})
\end{equation}
The pseudocode of the proposed algorithm accelerated using SQUAREM acceleration scheme is shown in Algorithm. 2. 
\begin{center}
\begin{tabular}{ p{15cm} }
\hline
\hline
\bf{Algorithm 2: Proposed algorithm acceleration scheme via SQUAREM} \\
\hline
\hline
{\bf{Input}}: Number of vectors in frame $N$ and the dimension of each vector $d$.  \\
{\bf{Initialize}}: Set $t=0$. Initialize ${{\bx}^{0}}$  \\       
{\bf{Repeat}}: \\
1)\, $\bar{\bx}^{1} = F_{MM}(\bx^{t})$ , $\bar{\bx}^{2} = F_{MM}(\bar{\bx}^{1})$ \\
2)\, $\br = \bar{\bx}^{1} -\bx^{t}$, $\bv = \bar{\bx}^{2} - \bar{\bx}^{1} - \br$\\
3)\, Compute step length $\alpha = -\dfrac{\|\br\|_{2}}{\|\bv\|_{2}}$\\
4)\, Compute $\bx^{t+1} = \bx^{t} - 2\alpha\br + \alpha^{2}\bv$ and normalize the frame. \\
5)\, Modify the step length $\alpha$ if $\underset{i, j = 1, 2, \cdots N \: (i \neq j)}{\rm maximize}\:  \bar{f}_{ij}(\bx^{t})>\underset{i, j = 1, 2, \cdots N \: (i \neq j)}{\rm maximize}\:  \bar{f}_{ij}(\bx^{t+1}) $ as shown in section 6 in \cite{squarem}. Update $\bx^{t+1} = \bx^{t} - 2\alpha\br + \alpha^{2}\bv$ and normalize the frame. \\
$t \leftarrow t+1$, \textbf{until convergence}\\
\hline
\hline
\end{tabular}
\end{center}
\section{Numerical Simulations and CS Application}\label{sec:5}
\vspace{-2mm}
In this section we first compare TELET (whose convergence speed was accelerated using the scheme discussed in the Subsection. \ref{sec:4}.D) with the state-of-the-art algorithms used to construct the ETFs. When compared to the other state-of-the-art algorithm, we show that TELET is monotonic and is able to construct new ETFs of medium to large dimensions. Next, using synthetic data and real images we compare the performance of optimal sensing matrix constructed using TELET with the state-of-the-art algorithms used in the CS systems. All the algorithms were implemented in MATLAB. The simulations were carried out in a PC with $2.40$ GHz processor with $16$ GB RAM. 
\subsection{Comparison of TELET with the state-of-the-art algorithms}
\begin{figure}[h]
\begin{subfigure}{0.49\textwidth}
\centering
\captionsetup{justification=centering}

\includegraphics[height=3.1in,width=3.4in]{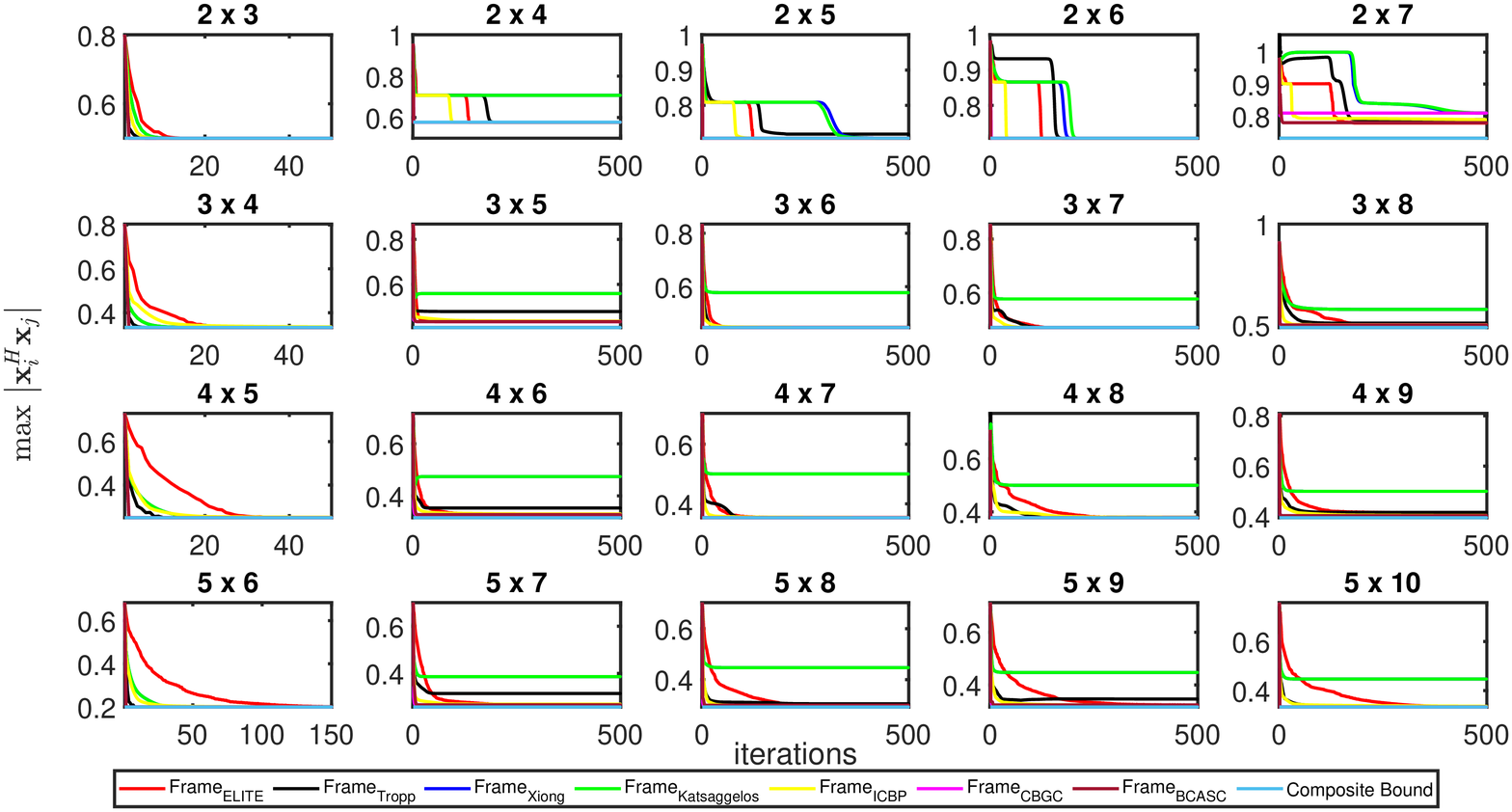}
\caption{Convergence plot: ${\rm max}\:  \left| \bx_{i}^{H}\bx_{j}\right|$ vs. iteration for complex ETFs of different dimensions.}
\end{subfigure}
\centering
\begin{subfigure}{0.49\textwidth}
\centering
\captionsetup{justification=centering}
\includegraphics[height=3.1in,width=3.4in]{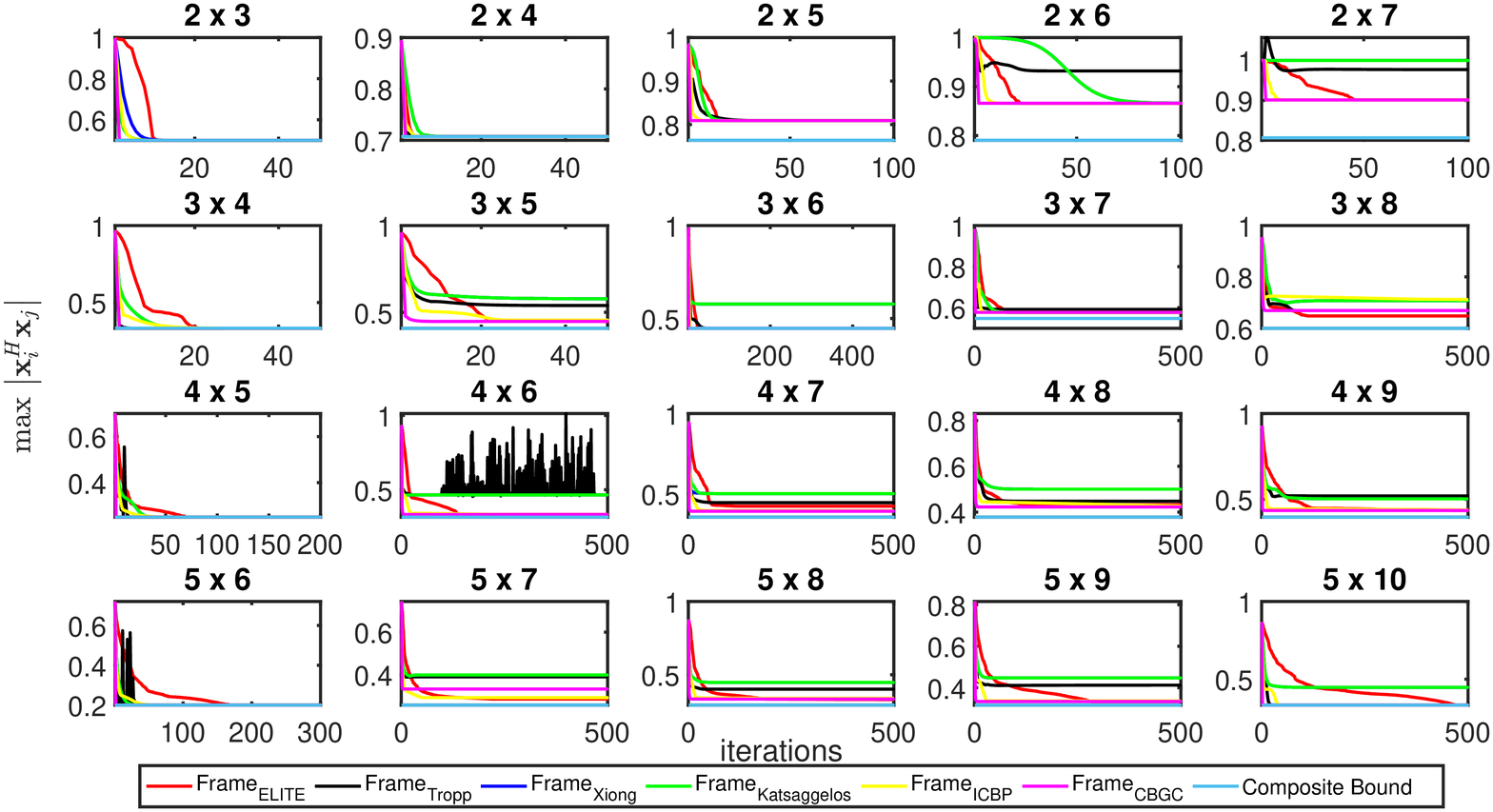}
\caption{Convergence plot: ${\rm max}\:  \left| \bx_{i}^{H}\bx_{j}\right|$ vs. iteration for real ETFs of different dimensions.}
\end{subfigure}
\caption{Convergence plot for different values of $d$ and $N$ for real and complex ETF.}\vspace{-1.5mm}
\label{iterboth}
\end{figure}

\begin{figure} [!h]
\centering
\begin{tabular}{ccccccc}
\includegraphics[height=0.15\textwidth, width=0.15\textwidth]{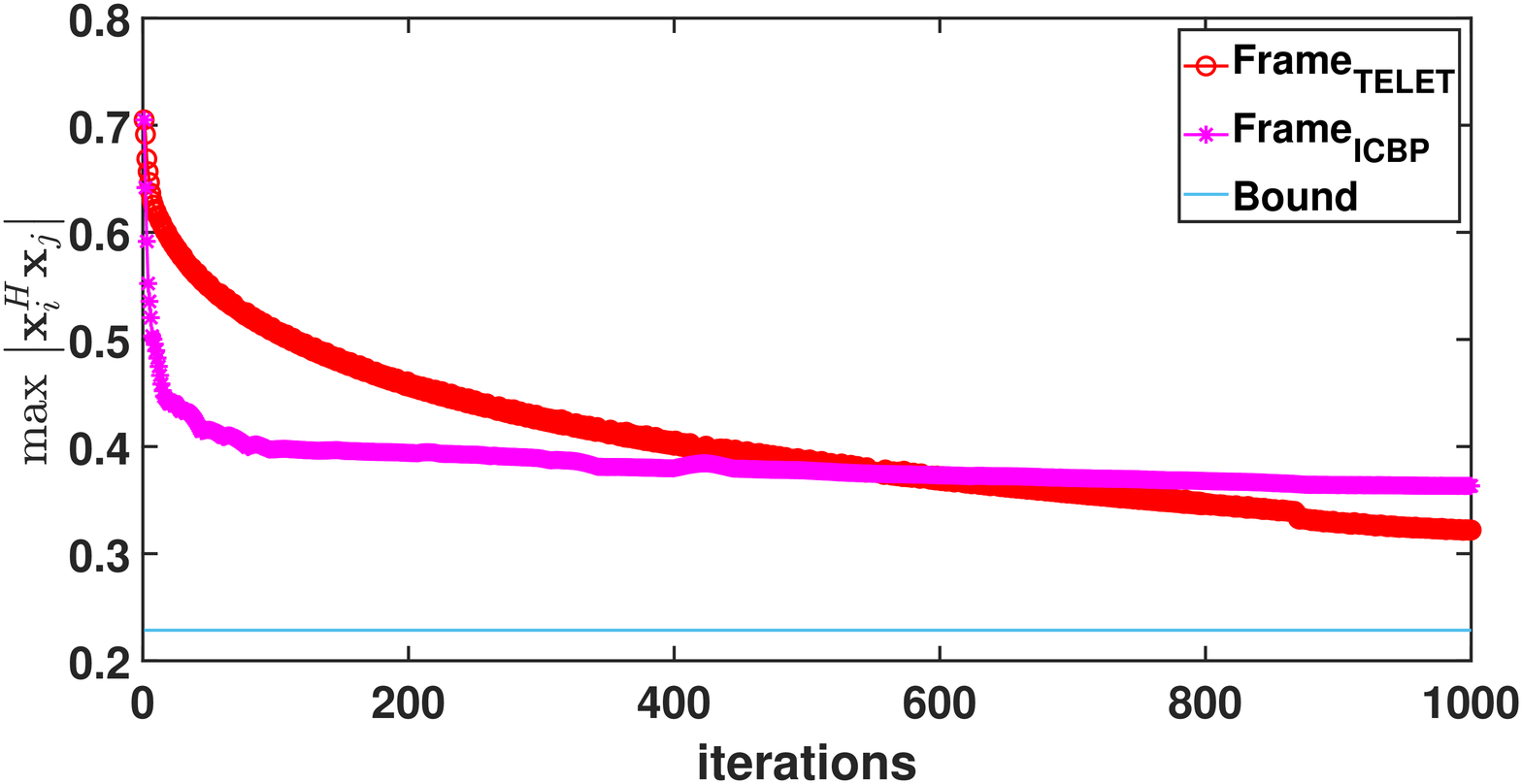} &
\includegraphics[height=0.15\textwidth,width=0.15\textwidth]{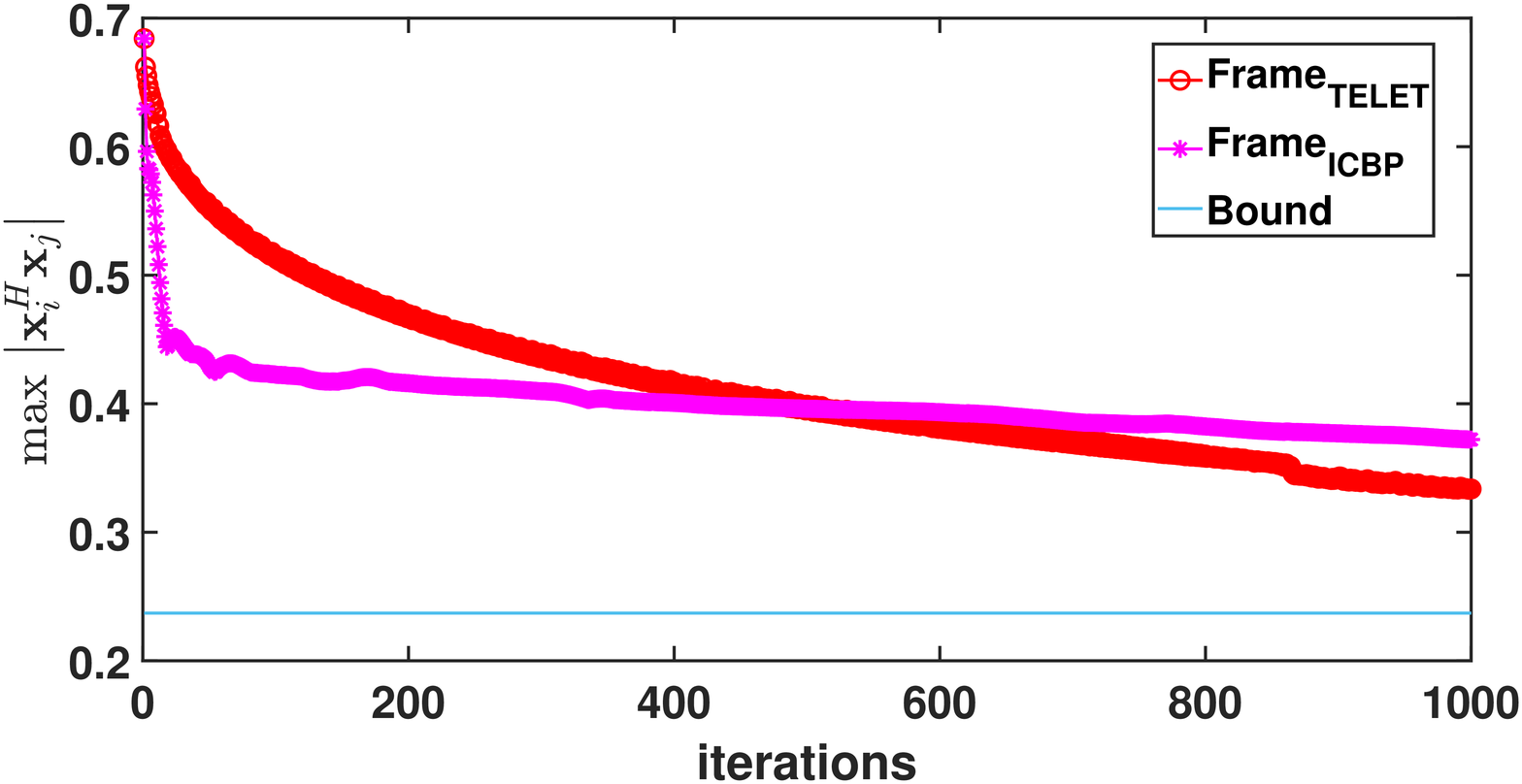} &
\includegraphics[height=0.15\textwidth,width=0.15\textwidth]{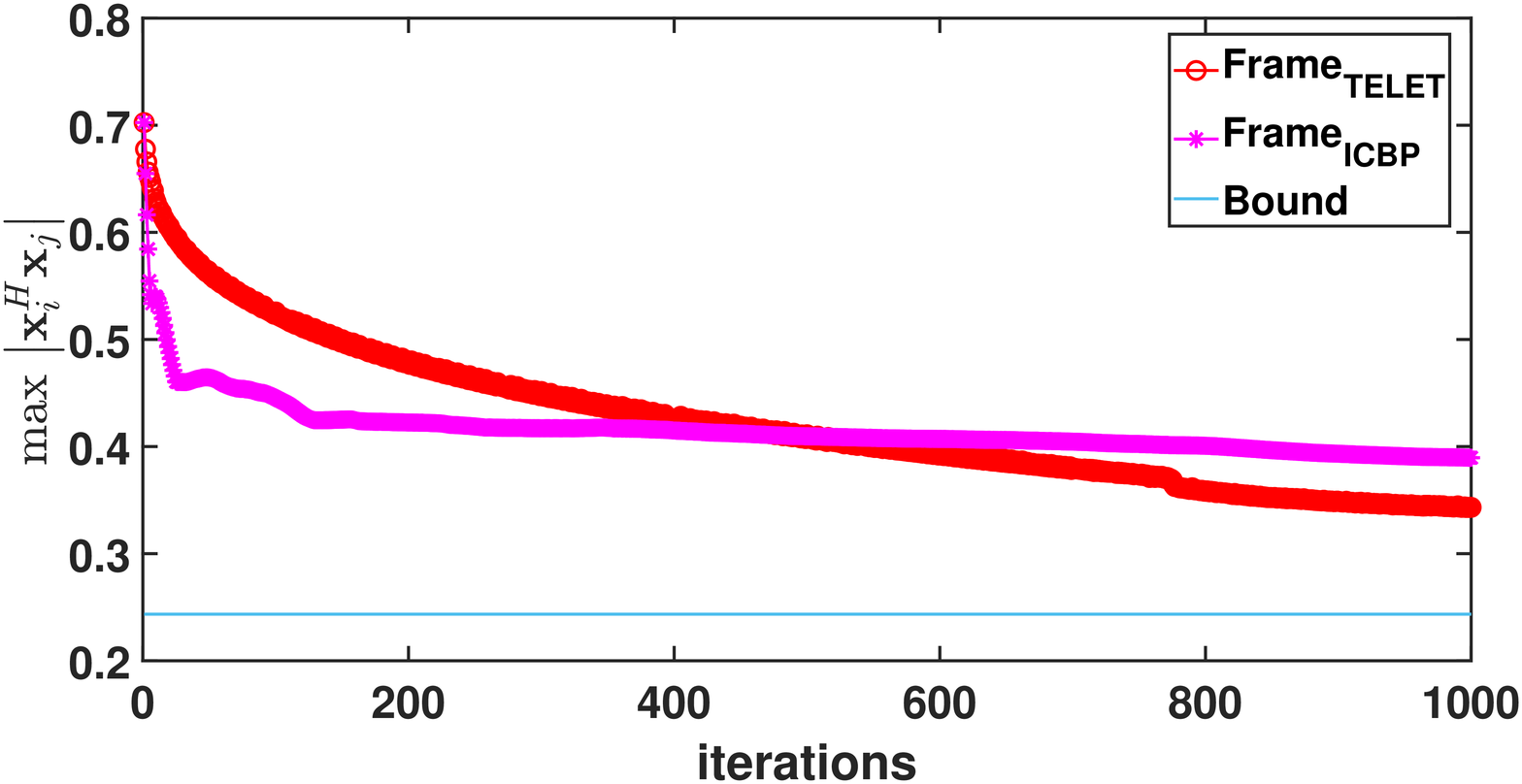} &
\includegraphics[height=0.15\textwidth, width=0.15\textwidth]{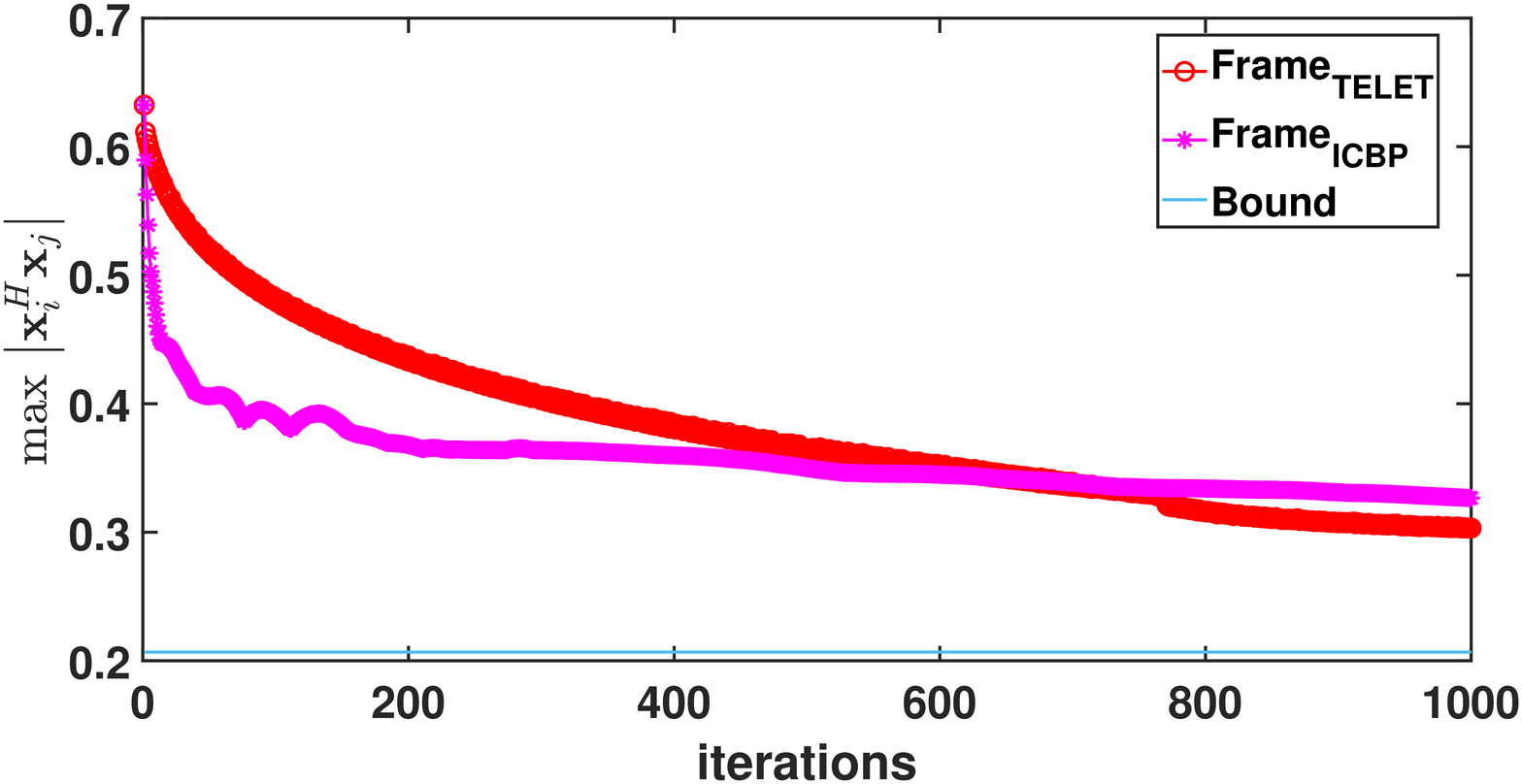} &
\includegraphics[height=0.15\textwidth, width=0.15\textwidth]{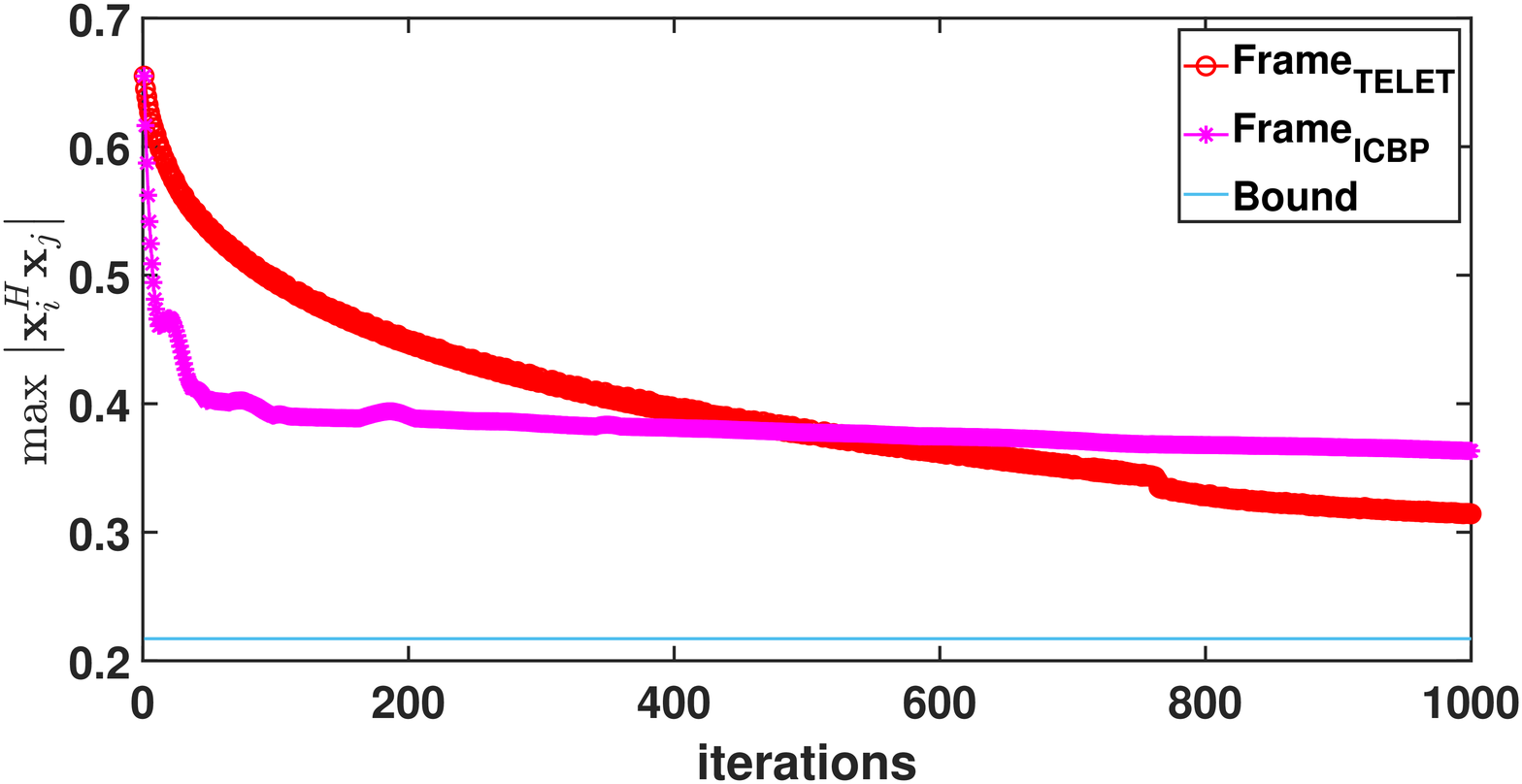} &
\includegraphics[height=0.15\textwidth, width=0.15\textwidth]{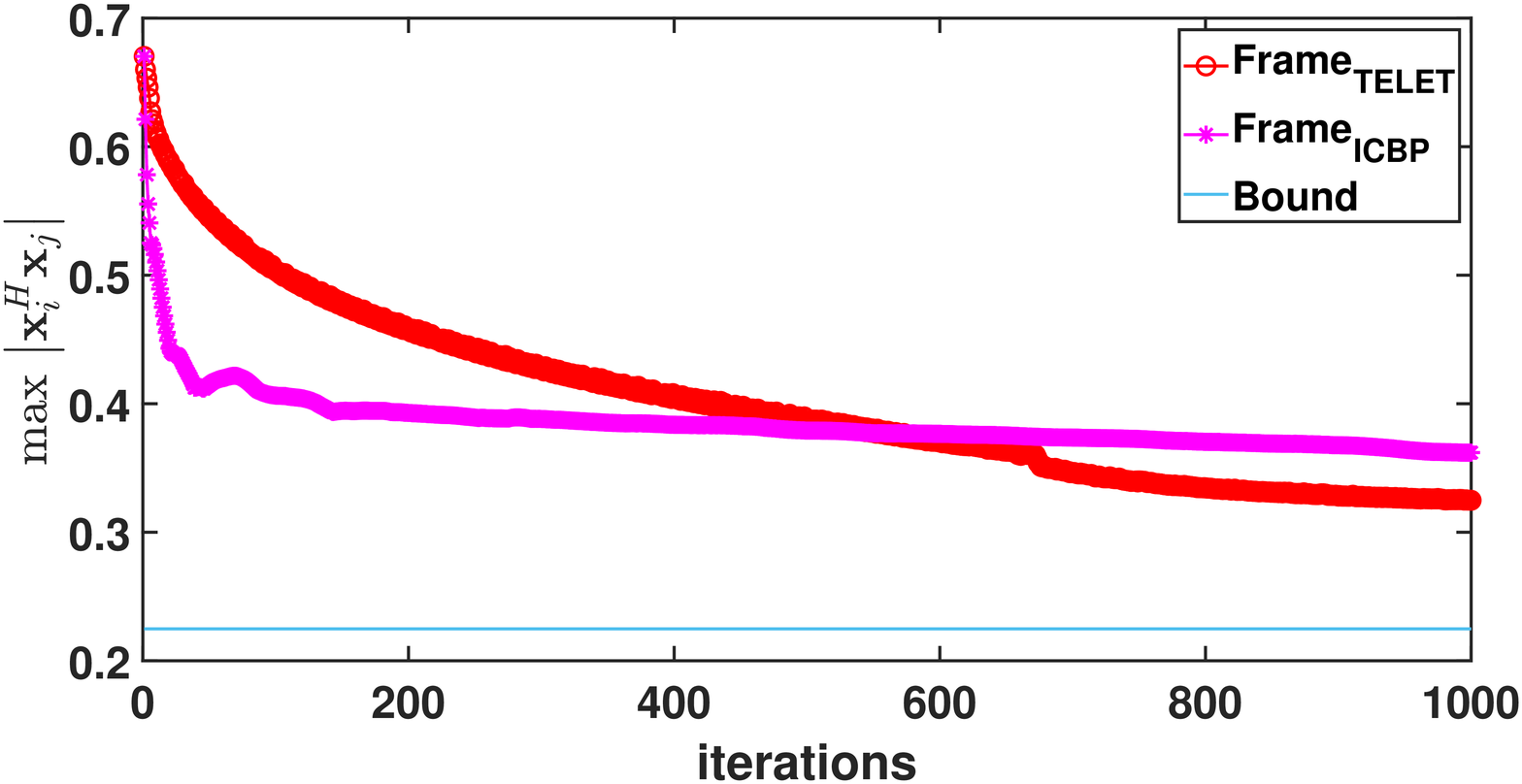} \\
\boldmath{$23\times 700$}  & \boldmath{$23\times800$} & \boldmath{$23\times 900$}&\boldmath{$25\times 700$}  & \boldmath{$25\times800$} & \boldmath{$25\times 900$} \\[6pt]
\end{tabular}
\begin{tabular}{ccccccc}
\includegraphics[height=0.15\textwidth, width=0.15\textwidth]{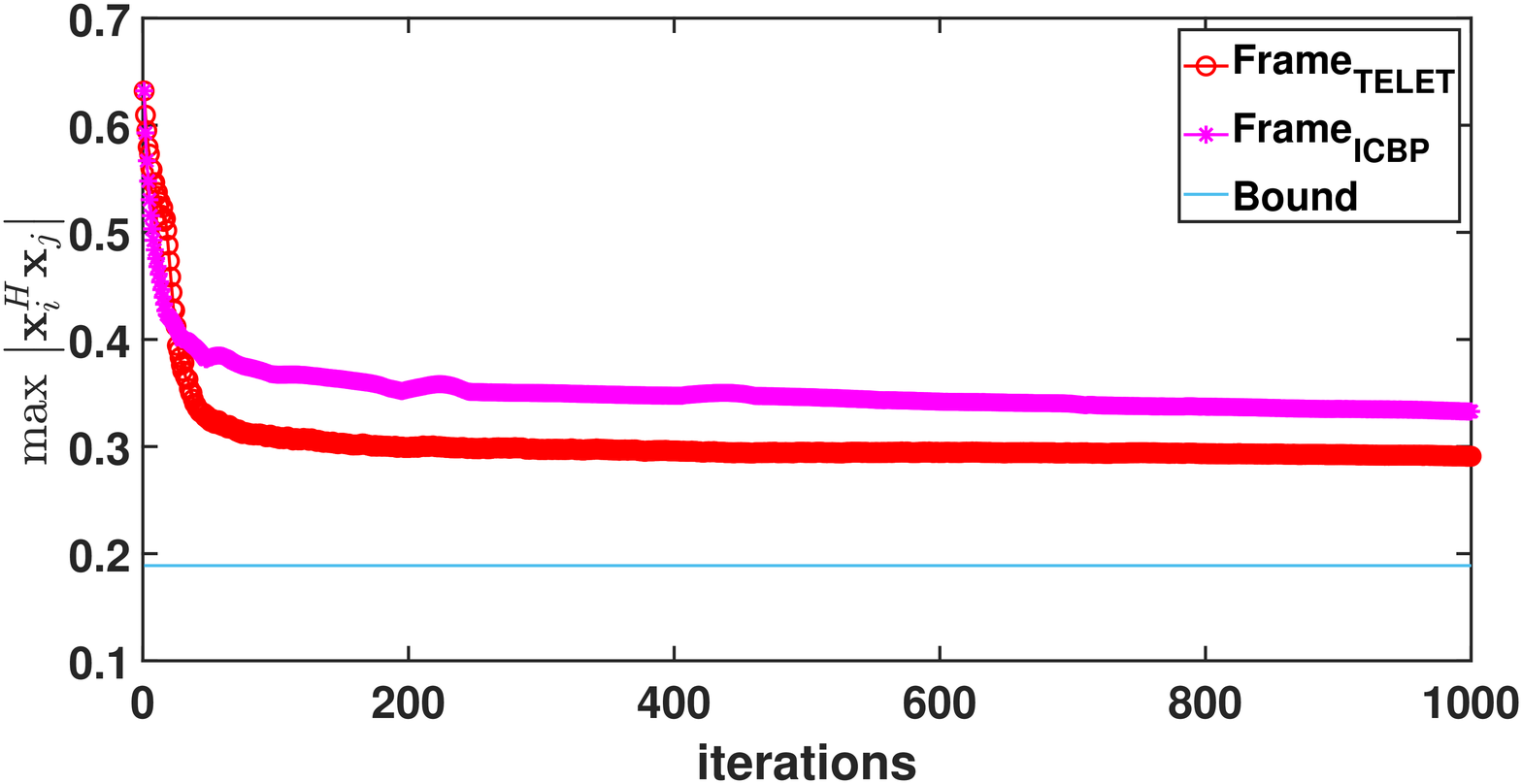} &
\includegraphics[height=0.15\textwidth, width=0.15\textwidth]{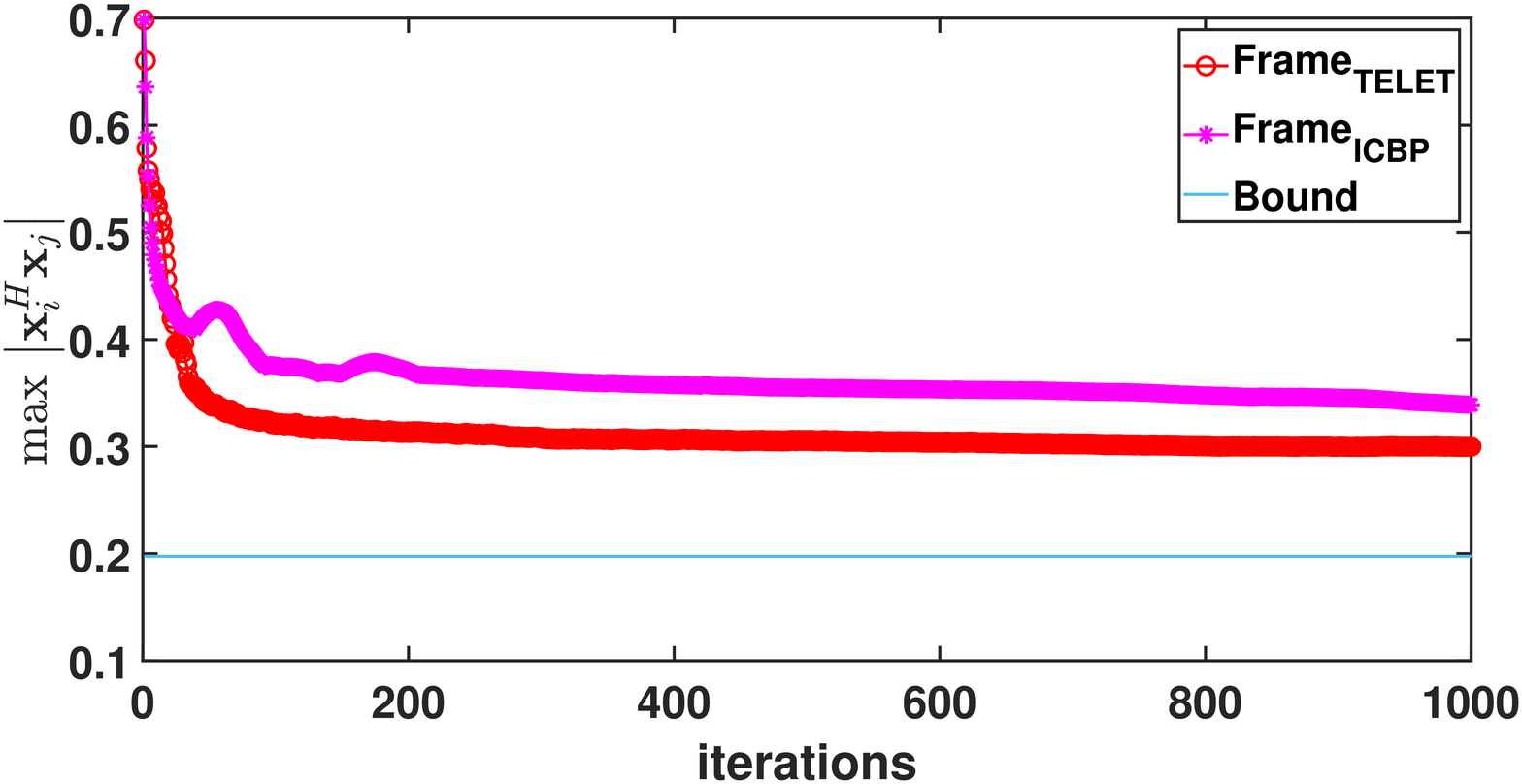} &
\includegraphics[height=0.15\textwidth, width=0.15\textwidth]{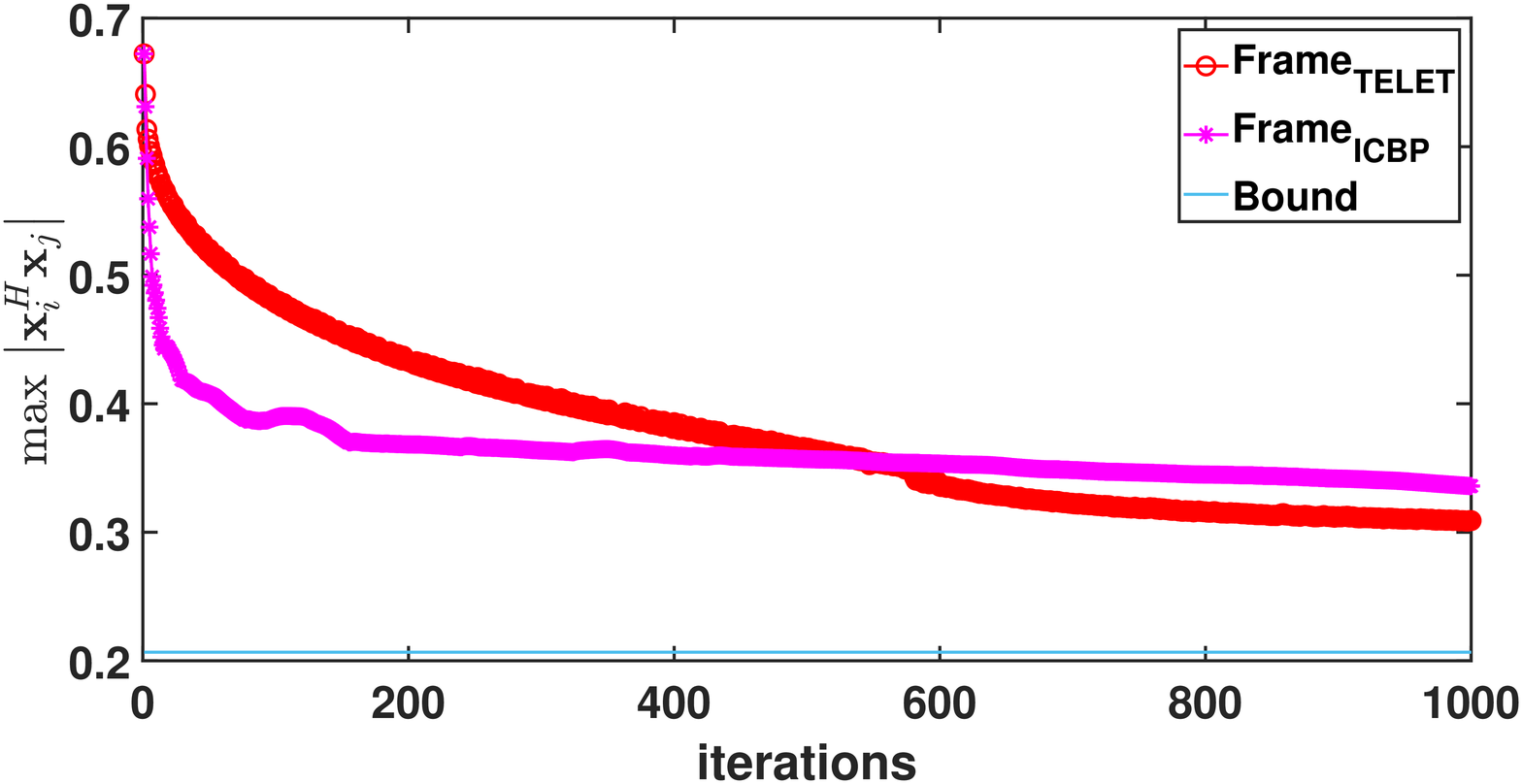} &
\includegraphics[height=0.15\textwidth, width=0.15\textwidth]{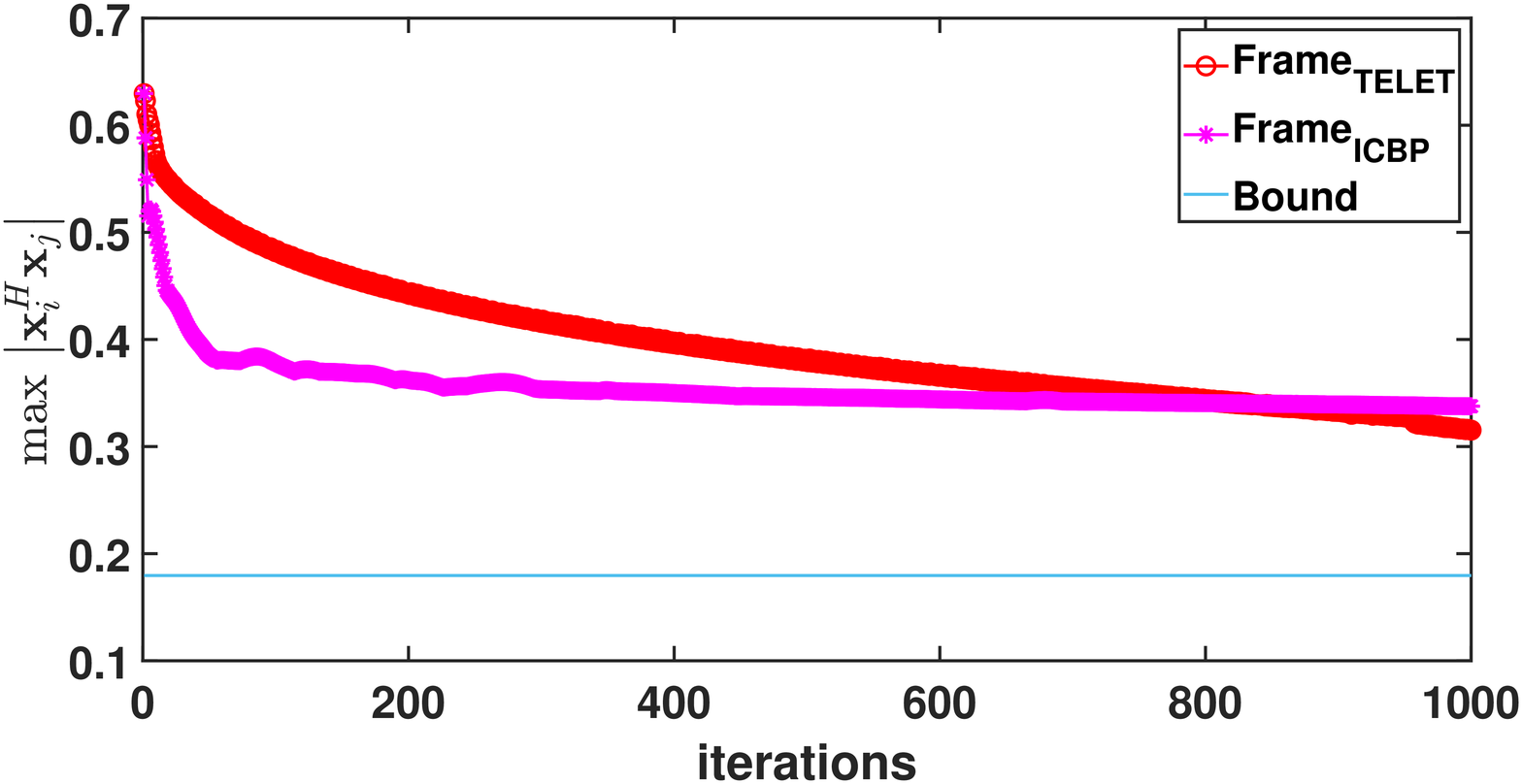} &
\includegraphics[height=0.15\textwidth, width=0.15\textwidth]{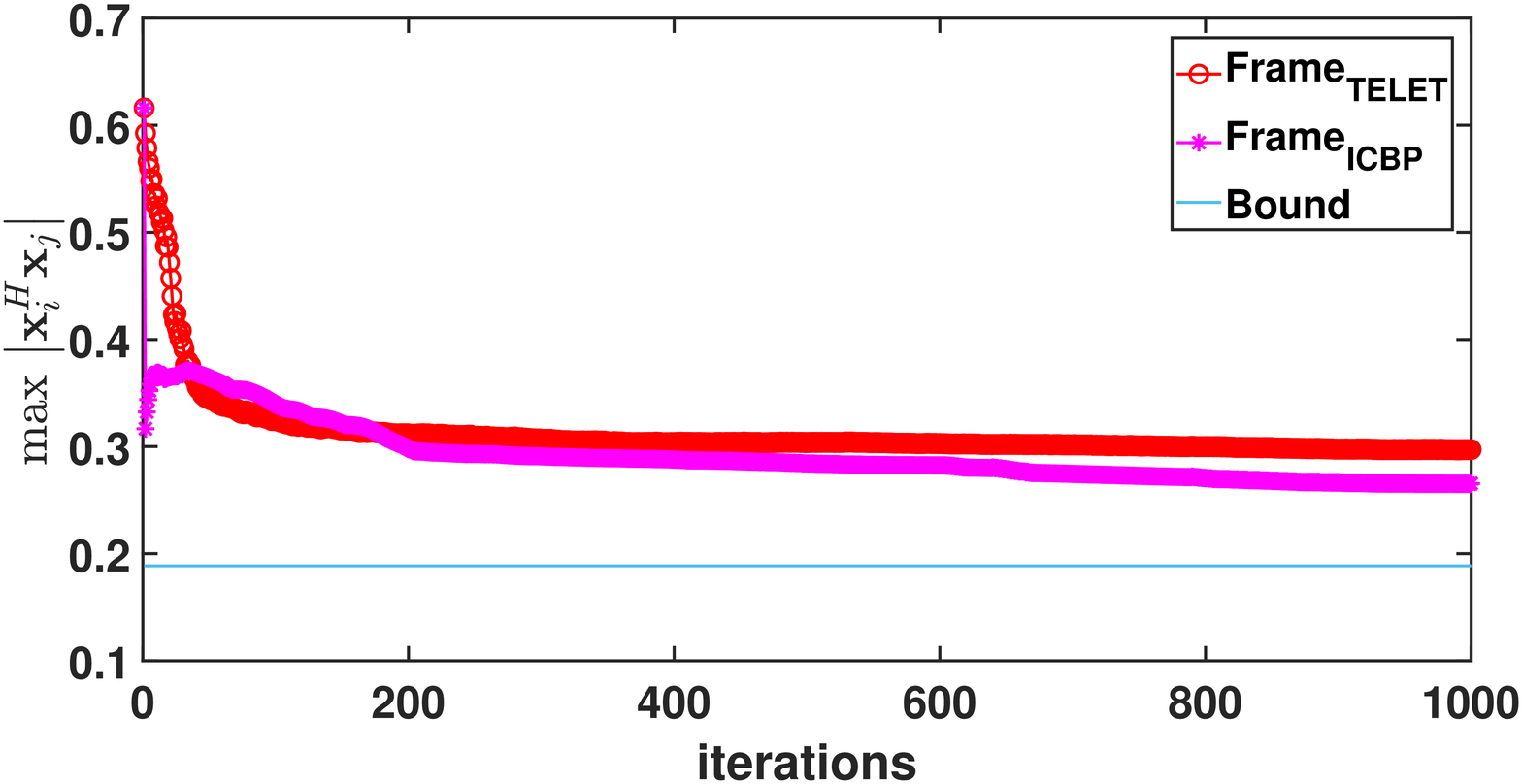} &
\includegraphics[height=0.15\textwidth, width=0.15\textwidth]{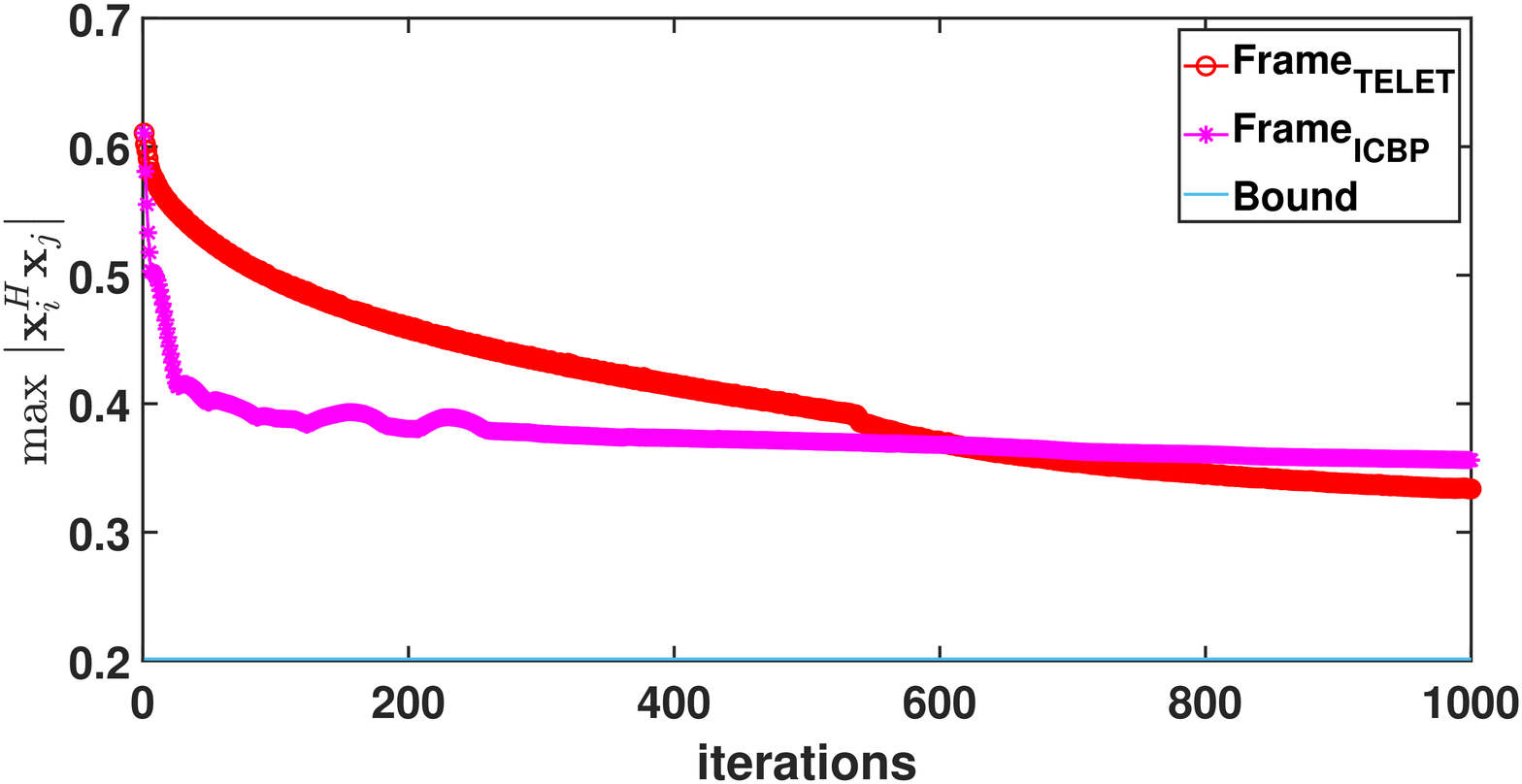} \\
\boldmath{$27\times 700$}  & \boldmath{$27\times800$} & \boldmath{$27\times 900$}&\boldmath{$30\times 900$}  & \boldmath{$30\times1000$} & \boldmath{$30\times 1200$}  \\[6pt]
\end{tabular}
\caption{Convergence plot: ${\rm max}\:  \left| \textbf{x}_{i}^{H}\textbf{x}_{j}\right|$ vs. iteration of complex frames of large dimensions. Red line - converge of ${\textrm{Frame}}_{\textrm{TELET}}$; Magenta line - convergence of ${\textrm{Frame}}_{\textrm{ICBP}}$ and Blue line - composite bound.}\vspace{-1mm}
\label{complexbig}
\end{figure}

1. In this simulation, we compare the convergence plot
of the proposed algorithm and that of the state-of-the-art algorithms for real and complex ETFs. In the case of complex ETFs, we compare our algorithm with \cite{tropp,xiong,agelos} and \cite{codebook,tahir,bcasc} and in the case of real ETFs, we compare our algorithm with the algorithms in \cite{tropp, xiong, agelos} and \cite{tahir,codebook}. We denote the frames created by our method and others as ${\textrm{Frame}}_{\textrm{TELET}}$, ${\textrm{Frame}}_{\textrm{Tropp}}$ \cite{tropp}, ${\textrm{Frame}}_{\textrm{Xiong}}$ \cite{xiong}, ${\textrm{Frame}}_{\textrm{Katsaggelos}}$ \cite{agelos}, ${\textrm{Frame}}_{\textrm{CBGC}}$\cite{codebook}, ${\textrm{Frame}}_{\textrm{ICBP}}$\footnote{The code for ICBP is obtained from https://www.nt.tuwien.ac.at/christian-doppler-laboratory/cd-download/} \cite{tahir} and ${\textrm{Frame}}_{\textrm{BCASC}}$\footnote{The codes for CBGC and BCASC are obtained from https://codeocean.com/capsule/3494920/tree/v1} \cite{bcasc}. The algorithm were initialized using the following technique: Generate a large collection of random vectors ($>>N$) $\bx_{i}$'s whose $j^{th}$ element was generated using:
\begin{equation}\label{gen}
\begin{array}{ll}
x_{j} = {\rm{exp}}(j2 \pi \phi_{j})
\end{array}
\end{equation}
where $\phi_{j}$ is randomly generated from a uniform distribution from $[0,1]$. Remove the vectors having the largest inner product. Repeat this process until the desired number of vectors $N$ are reached.  All the algorithms were made to run until the following criteria was met:
\begin{equation}\label{cond}
|\mu(\textbf{X}^{k})- \mu_{\textrm{CB}}| < 10^{-5}
\end{equation}
where $\mu_{\textrm{CB}}$ is the composite bound (which is a tighter bound when compared to the Welch bound in the region $N >d^{2}$), or until certain maximum number of iterations was met. The composite bound in the case of complex frames  is defined as:
{\small{\[
\mu_{\textrm{CB}}= 
\begin{cases}
\text{if } N \leq d^{2}:&\\
    \sqrt{\dfrac{N-d}{d(N-1)}},&\\
\text{if } d^{2}<N\leq2(d^{2}-1):&\\
{\rm{max}}\left(\sqrt{\dfrac{1}{d}}, \sqrt{\dfrac{2N-d^{2}-d}{(d+1)(N-d)}}, 1- 2N^{-\dfrac{1}{d-1}}\right)& \\
\text{if } N>2(d^{2}-1):&\\
{\rm{max}}\left(\sqrt{\dfrac{2N-d^{2}-d}{(d+1)(N-d)}}, 1- 2N^{-\dfrac{1}{d-1}}\right)& 
\end{cases}
\]}}
In the case of real frames, composite bound is defined as:
{\small{ \begin{equation*}
 \mu_{\textrm{CB}}= {\rm{max}}\left(\sqrt{\dfrac{N-d}{d(N-1)}}, \sqrt{\dfrac{3N-d^{2}-2d}{(d+2)(N-d)}}\right)
 \end{equation*}}}
Fig. \ref{iterboth}.a and Fig. \ref{iterboth}.b compares the convergence plot of the frame design algorithms for complex and real ETF, respectively. From Fig. \ref{iterboth}.a it can be seen that for some frame dimensions such as $3 \times 6$ and  $4 \times7$, the algorithms $\textrm{Frame}_{\textrm{Kastaggelos}}$ and $\textrm{Frame}_{\textrm{Xiong}}$ converges to a higher mutual coherence value when compared to the other state-of-the-art algorithms. Also, from Fig. \ref{iterboth}.a, it can be seen that the remaining algorithms converges closer to the composite bound, except for frame dimensions $3 \times 5$, $4 \times 6$ and $5 \times 7$,  where ${\textrm{Frame}}_{\textrm{Tropp}}$ algorithm converges to a larger coherence value. Also, from Fig. \ref{iterboth}.b it can be seen that almost always ${\textrm{Frame}}_{\textrm{ICBP}}$, ${\textrm{Frame}}_{\textrm{CBGC}}$  and the proposed algorithm converge to similar coherence value for the frame dimensions considered in Fig. \ref{iterboth}.b.  \\
\\
2. We now compare the algorithms with respect to the following metrics: mutual coherence and run time for frames of small, medium and large dimensions. The maximum number of iterations was set equal to $10^{4}$ in the case of small and medium frame dimensions and $10^{3}$ in the case of large frame dimensions (due to larger computational complexity of the methods in comparison we have reduced the number of iterations for large frame dimensions). Table. \ref{coh} - Table. \ref{timec} compares the mutual coherence value and run time of the algorithms for frames of various dimensions, respectively. In the case of CBGC and BCASC we report the cumulative sum of run time which was required to solve each sub-problem (problem (6) in \cite{codebook}). For the other algorithms, we report the run time either until the condition in (\ref{cond}) was met or the run time corresponding to the minimum mutual coherence value achieved by the algorithms.  In Table. \ref{coh} - Table. \ref{timec} we have highlighted the algorithms which performs the best with respect to mutual coherence value and run time, respectively. From Table. \ref{coh} it can be seen that when compared to the other algorithms ${\textrm{Frame}}_{\textrm{Tropp}}$ and ${\textrm{Frame}}_{\textrm{Xiong}}$ have higher mutual coherence value for small frame dimensions. Hence, we did not include them in the comparison for medium and large frame dimensions. From Table. \ref{time} it can be seen that ${\textrm{Frame}}_{\textrm{CBGC}}$ and ${\textrm{Frame}}_{\textrm{BCASC}}$ takes more time to converge when compared to the other algorithms. This could be because unlike the other algorithms, ${\textrm{Frame}}_{\textrm{CBGC}}$ and ${\textrm{Frame}}_{\textrm{BCASC}}$ requires solving sub-problems for various values of $p$ to obtain frames with minimum coherence. Hence, we did not include them for comparison for large frame dimensions. Also, from Table. \ref{coh} and Table. \ref{cohc} it can be observed that out of the $54$ frame dimensions considered, the proposed algorithm achieves the lowest coherence value $42$ times. Especially, for large frame dimensions, it can be observed that the proposed algorithm almost always achieves lower mutual coherence value when compared to the state-of-the-art algorithm. From Table. \ref{timec} it can be seen that for all the medium and large frame dimensions considered, ${\textrm{Frame}}_{\textrm{ICBP}}$ always takes lesser time when compared to the proposed and state-of-the-art algorithms. In Fig. \ref{complexbig} we also show the convergence plot for some of the large frame dimensions considered in Table. \ref{cohc}. From Fig. \ref{complexbig} it can be seen that for some of the frame dimensions such as $25 \times 800$ and $30 \times 1200$, the proposed algorithm even though achieves lower coherence value, has not converged and had to be stopped because of the maximum iteration limit.
\begin{table*}[!htbp]
\small
\begin{center}
\caption{Comparison of Mutual coherence of complex frames of small and medium dimensions.}
\label{coh}
\begin{tabular}{|p{1.3cm}|p{1.5cm}|p{1.7cm}|p{1.7cm}|p{1.7cm}|p{1.7cm}|p{1.7cm}|p{1.7cm}|}
\hline
$(d,N)$  &$\mu_{\textrm{CB}}$&${\textrm{Frame}}_{\textrm{TELET}}$&${\textrm{Frame}}_{\textrm{ICBP}}$&${\textrm{Frame}}_{\textrm{CBGC}}$&${\textrm{Frame}}_{\textrm{BCASC}}$&${\textrm{Frame}}_{\textrm{Xiong}}$&${\textrm{Frame}}_{\textrm{Tropp}}$\\       
 \hline
$(2,8)$   & $0.75$&$\textbf{0.7941}$& $\textbf{0.7941}$ &$0.7942$&$0.7950$&$0.8110$&$0.9270$ \\
$(3,16)$ &$0.6202$&$\textbf{0.6483}$&$0.6547$&$0.6491$&$0.6499$&$0.6870$&0.6970\\
(4,5)&0.25&\textbf{0.25}&\textbf{0.25}&\textbf{0.25}&\textbf{0.25}&\textbf{0.25}&\textbf{0.25}\\
(4,6)&0.3162&\textbf{0.3273}&0.3274&0.3276&0.3282&0.4425&0.3536\\
(4,7)&0.3536&\textbf{0.3536}&\textbf{0.3536}&\textbf{0.3536}&\textbf{0.3536}&0.5&\textbf{0.3536}\\
(4,8)&0.3780&\textbf{0.3780}&\textbf{0.3780}&\textbf{0.3780}&\textbf{0.3780}&0.5&\textbf{0.3780}\\
(4,9)&0.3853&\textbf{0.4019}&0.4131&0.4022&0.4025&0.5&0.416\\
(4,10)&0.4082&\textbf{0.4110}&0.4112&0.4118&0.4124&0.5&0.414\\
(4,19)&0.5&\textbf{0.5}&0.5188&\textbf{0.5}&\textbf{0.5}&0.5330&0.5313\\
(4,20)&0.5&0.5272&0.5371&0.5278&\textbf{0.5264}&0.5398&0.5932\\
(5,6)&0.2&\textbf{0.2}&\textbf{0.2}&\textbf{0.2}&\textbf{0.2}&\textbf{0.2}&\textbf{0.2}\\
(5,7)&0.2582&\textbf{0.2664}&0.2665&0.2667&0.2676&0.3658&0.3154\\
(5,8)&0.2928&0.2953&\textbf{0.2952}&0.2954&0.2957&0.4472&0.3055\\
(5,9)&0.3162&\textbf{0.3203}&0.3216&0.3205&0.3212&0.4472&0.3331\\
(5,10)&0.3333&\textbf{0.3333}&\textbf{0.3333}&\textbf{0.3333}&\textbf{0.3333}&0.4472&0.3334\\
(5,16)&0.3830&\textbf{0.3892}&0.3941&0.3895&0.3905&0.4472&0.4041\\
(5,26)&0.4472&0.4486&0.4511&0.4495&\textbf{0.4479}&0.4564&0.4671\\
(6,37)&0.4082&0.4180&0.4176&\textbf{0.4147}&0.4164&0.4258&0.4357\\
(20,30)&0.1313&0.1315&0.1317&\textbf{0.1314}&0.1317&-&-\\
(20,50)&0.1750&0.1753&0.1753&\textbf{0.1751}&0.1754&-&-\\
(20,100)&0.2010&0.2152&0.2118&\textbf{0.2109}&-&-&-\\
(30,40)&0.0925&0.0930&0.0930&\textbf{0.0927}&-&-&-\\
(30,50)&0.1166&0.1168&0.1169&\textbf{0.1167}&-&-&-\\
(30,100)&0.1535&0.1606&0.1553&\textbf{0.1549}&-&-&-\\
\hline
\end{tabular}
\end{center}
\end{table*}

\begin{table}[!h]
\small
\begin{center}
\caption{Comparison of Mutual coherence of large dimension complex frames constructed by different algorithms}
\label{cohc}
\begin{tabular}{|p{1.3cm}|p{1.5cm}|p{1.7cm}|p{1.7cm}|p{1.7cm}|p{1.7cm}|p{1.7cm}|p{1.7cm}|}
\hline
$(d,N)$  &$\mu_{\textrm{CB}}$&${\textrm{Frame}}_{\textrm{TELET}}$&${\textrm{Frame}}_{\textrm{ICBP}}$\\       
 \hline
(23,500)&0.2039&\textbf{0.2178}&0.3257\\
(23,600)&0.2163&\textbf{0.3195}&0.3401\\
(23,700)&0.2285&\textbf{0.3218}&0.3632\\
(23,800)&0.2371&\textbf{0.3334}&0.3721\\
(23,900)&0.2435&\textbf{0.3431}&0.3895\\
(23,1000)&0.2485&\textbf{0.3511}&{0.3758}\\
(23,1200)&0.2558&\textbf{0.3580}&0.3978\\
 (25,500)&0.1951&\textbf{0.2807}&0.3047\\
(25,600)&0.1960&\textbf{0.2069}&0.3391\\
(25,700)&0.2067&\textbf{0.3032}&0.3265\\
(25,800)&0.2171&\textbf{0.3143}&0.3632\\
(25,900)&0.2248&\textbf{0.3249}&0.3619\\
(25,1000)&0.2308&\textbf{0.3292}&0.3702\\
(25,1200)&0.2393&\textbf{0.3424}&0.3428\\
(27,500)&0.1874&\textbf{0.2718}&0.2913\\
(27,600)&0.1882&\textbf{0.2791}&0.3011\\
(27,700)&0.1888&\textbf{0.2909}&0.3327\\
(27,800)&0.1975&\textbf{0.2999}&0.3388\\
(27,900)&0.2067&\textbf{0.3086}&0.3358\\
(27,1000)&0.2137&\textbf{0.3144}&0.3517\\
 (27,1200)&0.2237&\textbf{0.3307}&0.3632\\
 (30,800)&0.1792&0.2823&\textbf{0.2443}\\
(30,900)&0.1796&\textbf{0.3153}&0.3376\\
(30,1000)&0.1886&0.2962&\textbf{0.2650}\\
(30,1200)&0.2013&\textbf{0.3337}&0.3561\\
(40,800)&0.1542&\textbf{0.2466}&0.2480\\
(40,900)&0.1546&\textbf{0.2454}&0.2596\\
(40,1000)&0.1550&\textbf{0.1619}&0.2737\\
(40,1200)&0.1555&\textbf{0.2745}&0.2988\\
(50,1000)&0.1379&\textbf{0.2229}&0.2368\\
\hline
\end{tabular}
\end{center}
\end{table}

\begin{table*}[!h]
\small
\centering
\caption{Comparison of run time (seconds) of the algorithms to construct complex frames of small and medium dimensions.}
\label{time}
\begin{tabular}{|p{1.3cm}|p{1.7cm}|p{1.7cm}|p{1.7cm}|p{1.7cm}|p{1.7cm}|p{1.7cm}|}
\hline
$(d,N)$  &${\textrm{Frame}}_{\textrm{TELET}}$&${\textrm{Frame}}_{\textrm{ICBP}}$&${\textrm{Frame}}_{\textrm{CBGC}}$&${\textrm{Frame}}_{\textrm{BCASC}}$&${\textrm{Frame}}_{\textrm{Xiong}}$&${\textrm{Frame}}_{\textrm{Tropp}}$\\        
 \hline
$(2,8)$   & $0.8408$ & $0.16$ &$9.68$&$5749.9$&$0.034$ &$\textbf{0.014}$\\
$(3,16)$ & 13.4313&3.58&496.23&7017.9&\textbf{0.049}&0.061\\
(4,5)&0.2082&0.0734&0.0080&0.1288&0.0075&\textbf{0.0066}\\
(4,6)&40.8189&0.1344&6.6730&58.18&\textbf{0.0025}&0.2492\\
(4,7)&1.6661&0.3182&0.054&1.3887&0.1556&\textbf{0.0053}\\
(4,8)&2.8156&0.9512&11.754&180.116&0.37&\textbf{0.23}\\
(4,9)&7.1192&1.7749&30.1607&161.78&\textbf{0.189}&0.3126\\
(4,10)&16.1572&1.54&16.2422&146.28&\textbf{0.4609}&0.4701\\
(4,19)&82.1215&4.3984&103.5152&2909.2&\textbf{0.1584}&3.5493\\
(4,20)&224.2072&4.3790&223.69&5871&0.2831&\textbf{0.0425}\\
(5,6)&0.3619&0.0839&0.0315&0.4355&0.0047&\textbf{0.0040}\\
(5,7)&0.2017&0.1359&16.8251&2354.6&\textbf{0.0265}&0.5062\\
(5,8)&21.5216&1.1905&21.3538&3226.7&\textbf{0.4841}&0.5052\\
(5,9)&20.6065&1.5550&118720&81271&\textbf{0.2590}&0.4304\\
(5,10)&17.6087&0.7675&48.6249&10.1563&\textbf{0.3976}&0.4967\\
(5,16)&158.11&3.7181&396.65&3390.4&1.0742&\textbf{0.0345}\\
(5,26)&124.627&6.2891&1927.3&9932.3&6.3617&\textbf{0.2744}\\
(6,37)&548.18&1.7595&1760.1&31121&7.708&\textbf{1.5603}\\
(20,30)&1024.7&\textbf{36.54}&5443.2&122950&-&-\\
(20,50)&1083.8&\textbf{61.1703}&6321.6&343680&-&-\\
(20,100)&7246.7&\textbf{99.9807}&35599&-&-&-\\
(30,40)&1749.9&\textbf{53.6401}&9260.3&-&-&-\\
(30,50)&2778.5&\textbf{116.5102}&3928&-&-&-\\
(30,100)&6397.8&\textbf{286.835}&43210&-&-&-\\
\hline
\end{tabular}
\end{table*}

\begin{table}[!h]
\small
\centering
\caption{Comparison of run time (seconds) of the proposed and state-of-the-art algorithms to construct large dimensional complex frames.}
\label{timec}
\begin{tabular}{|p{1.3cm}|p{1.7cm}|p{1.7cm}|}
\hline
$(d,N)$  &${\textrm{Frame}}_{\textrm{TELET}}$&${\textrm{Frame}}_{\textrm{ICBP}}$\\        
 \hline
 (23,500)&9627.5&\textbf{1528}\\
(23,600)&7053.6&\textbf{1268.1}\\
(23,700)&15235&\textbf{3498.4}\\
(23,800)&15307&\textbf{2668.4}\\
(23,900)&22147&\textbf{3269.4}\\
(23,1000)&42103&\textbf{9255.4}\\
(23,1200)&48966&\textbf{7868.4}\\
 (25,500)&6778&\textbf{1597.1}\\
(25,600)&13701&\textbf{2832.3}\\
(25,700)&10573&\textbf{2014.5}\\
(25,800)&20653&\textbf{7385}\\
(25,900)&28069&\textbf{9803.2}\\
(25,1000)&45089&\textbf{7142.4}\\
(25,1200)&86220&\textbf{11260}\\
(27,500)&9658.9&\textbf{1581.8}\\
(27,600)&\textbf{1390.7}&{6952}\\
(27,700)&24756&\textbf{3900.3}\\
(27,800)&14244&\textbf{2777.8}\\
(27,900)&28770&\textbf{10211}\\
(27,1000)&37806&\textbf{13187}\\
 (27,1200)&80170&\textbf{19889}\\
 (30,800)&31733&\textbf{5093.1}\\
(30,900)&47413&\textbf{7525.5}\\
(30,1000)&43566&\textbf{7661.9}\\
(30,1200)&89700&\textbf{11891}\\
(40,800)&16350&\textbf{3732}\\
(40,900)&35033&\textbf{7978.6}\\
(40,1000)&31342&\textbf{5591.0}\\
(40,1200)&55046&\textbf{11676}\\
(50,1000)&46612&\textbf{9186.1}\\
\hline
\end{tabular}
\end{table}

\begin{figure} [!h]
\centering
\begin{tabular}{ccccccc}
\includegraphics[height=0.15\textwidth, width=0.15\textwidth]{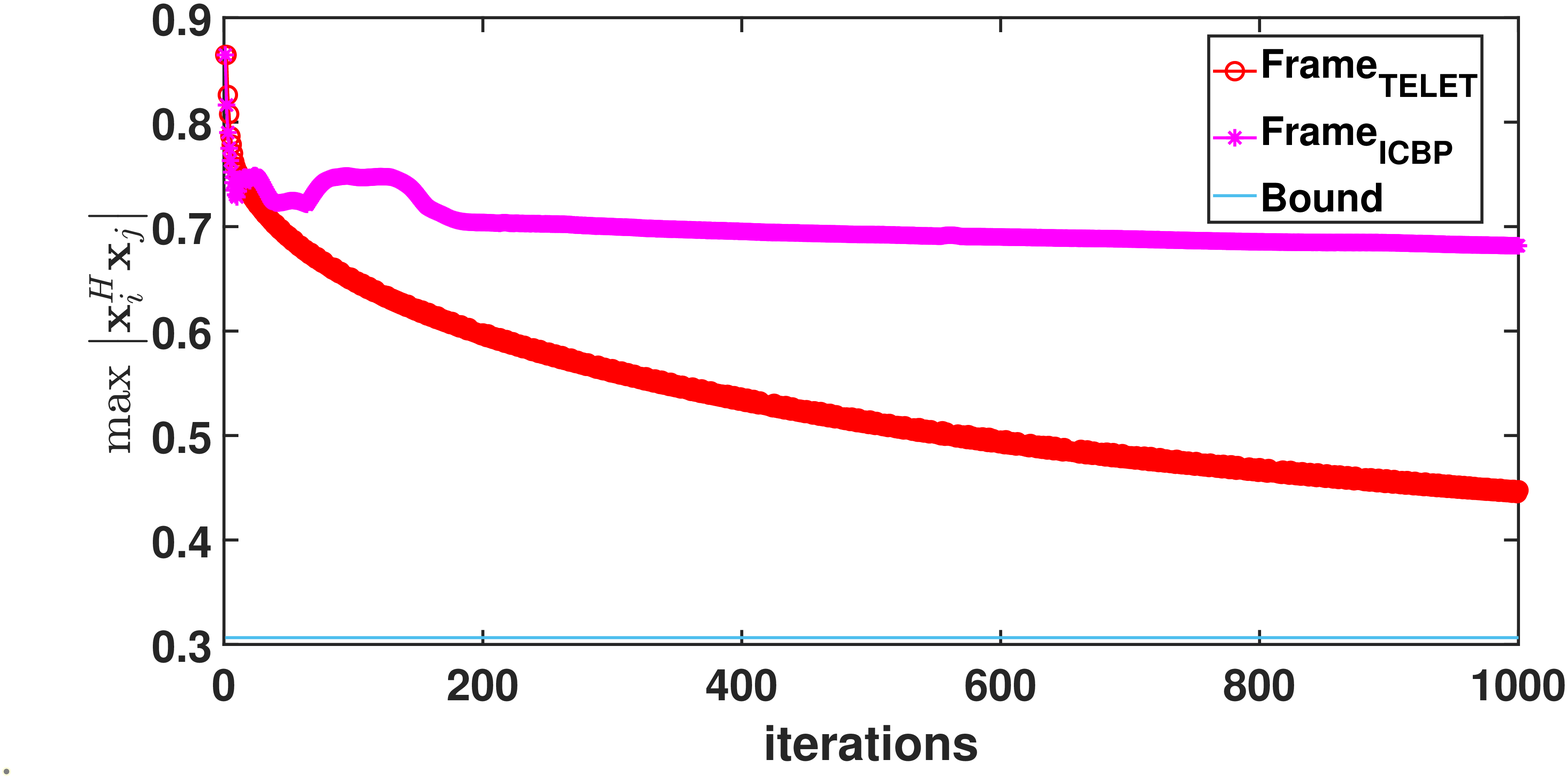} &
\includegraphics[height=0.15\textwidth,width=0.15\textwidth]{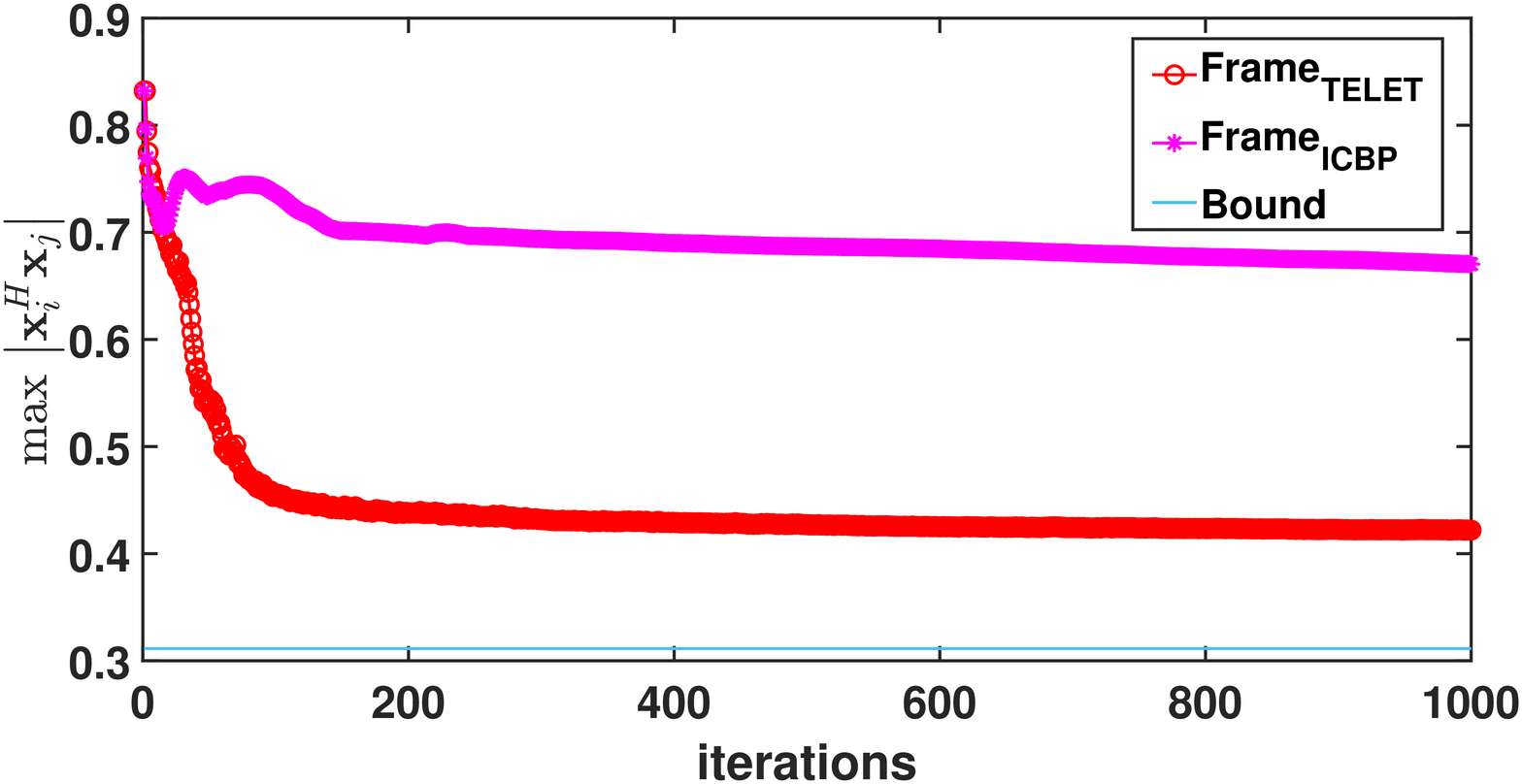} &
\includegraphics[height=0.15\textwidth,width=0.15\textwidth]{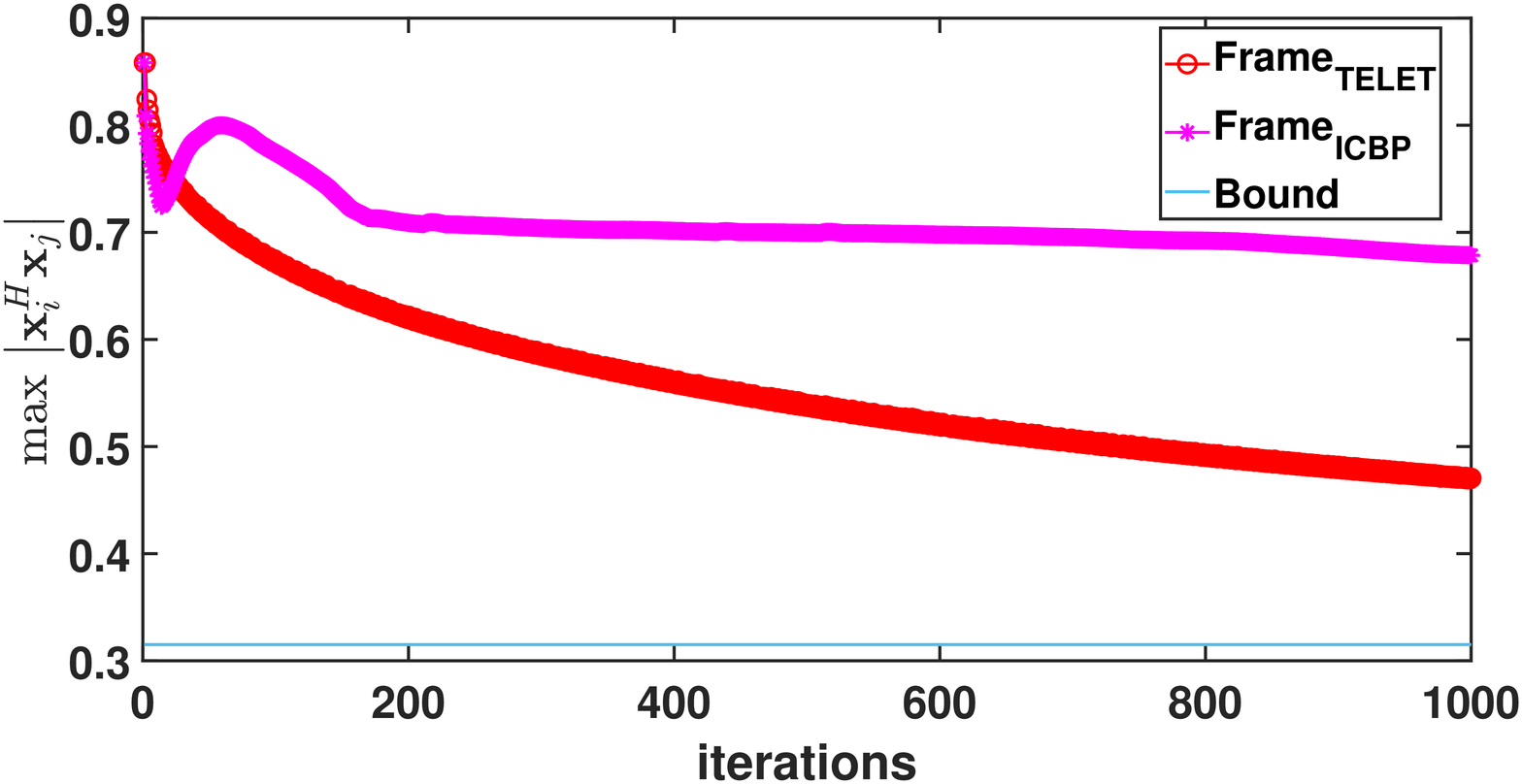} &
\includegraphics[height=0.15\textwidth, width=0.15\textwidth]{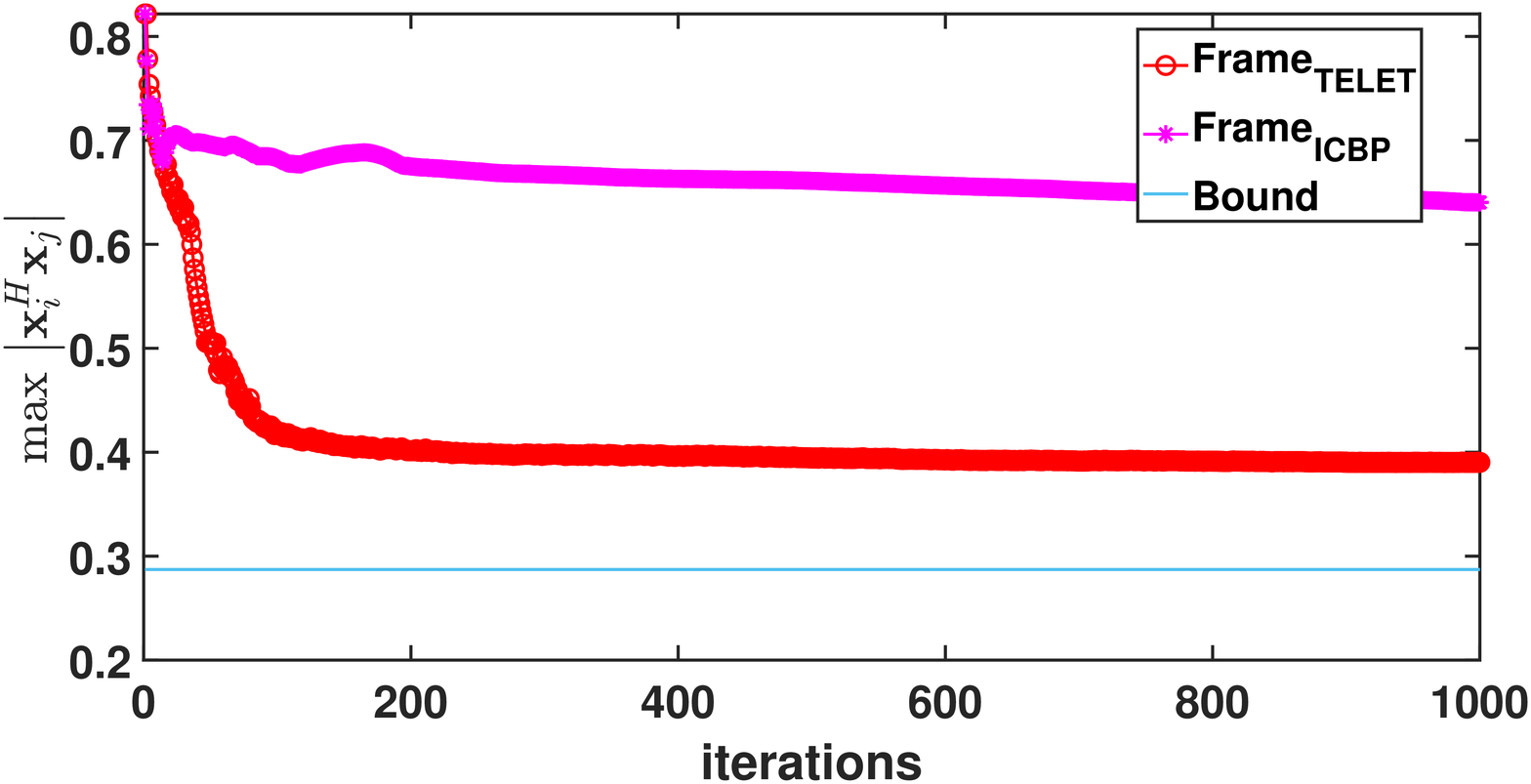} &
\includegraphics[height=0.15\textwidth, width=0.15\textwidth]{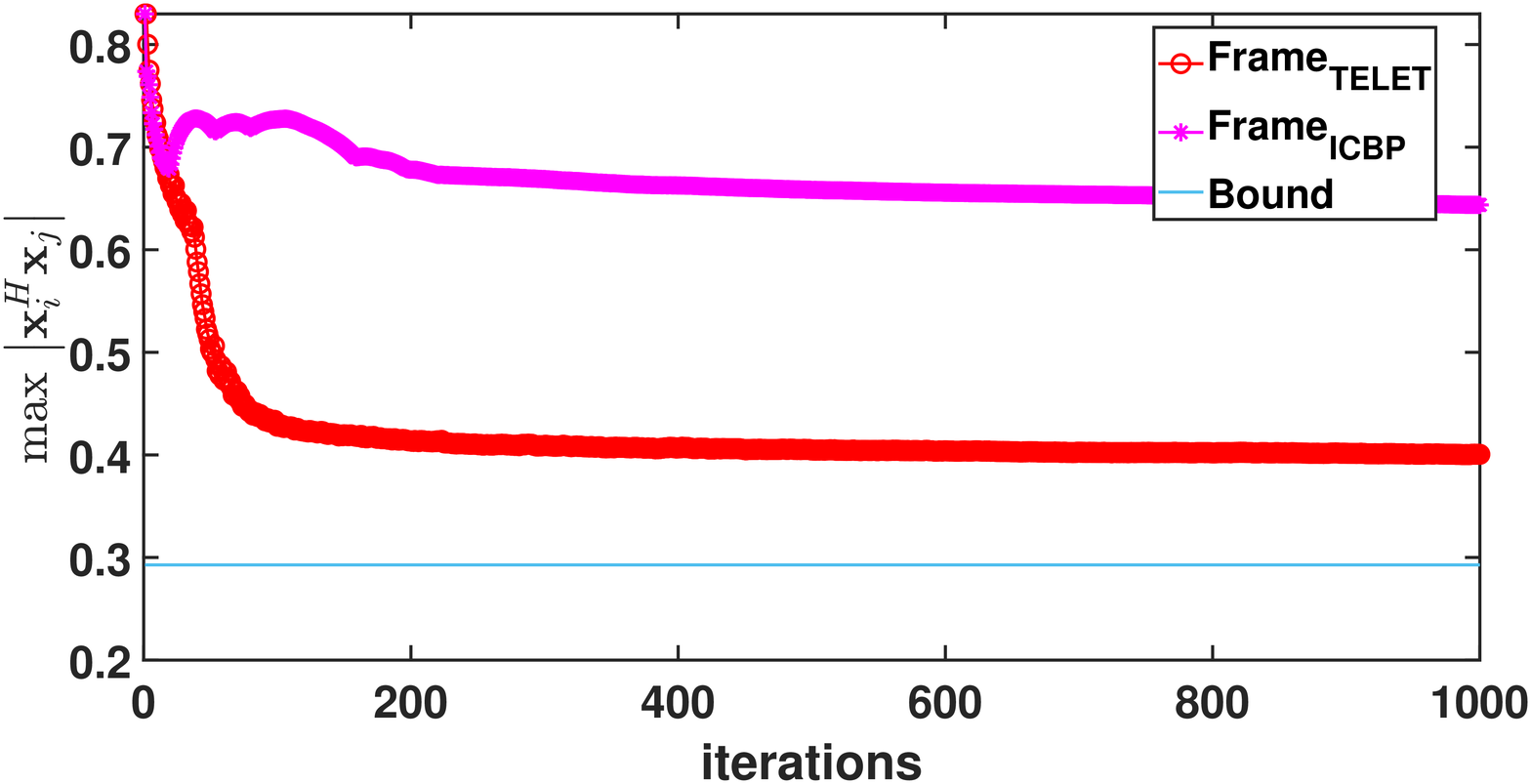} &
\includegraphics[height=0.15\textwidth, width=0.15\textwidth]{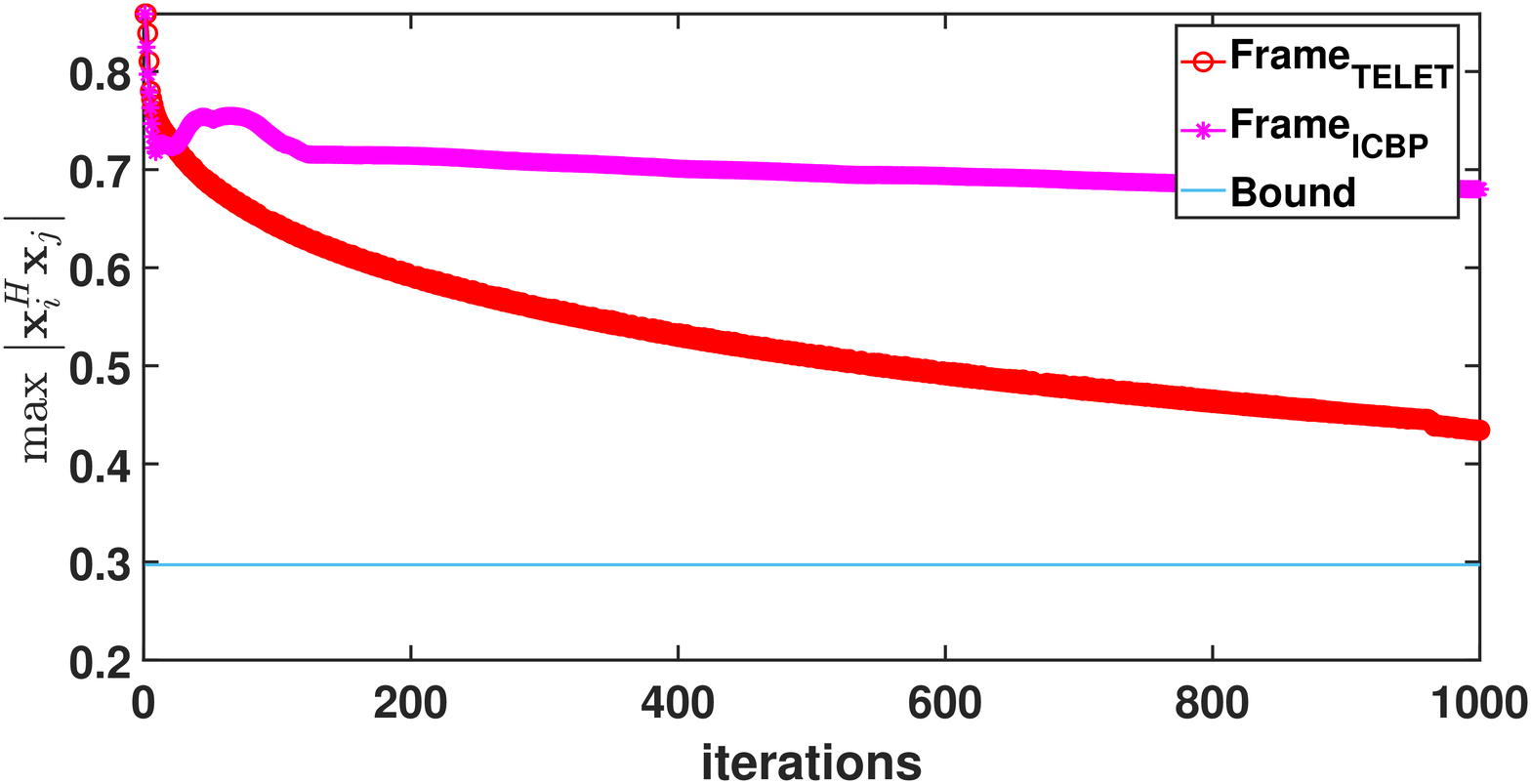} \\
\boldmath{$23\times 800$}  & \boldmath{$23\times900$} & \boldmath{$23\times 1000$} &\boldmath{$25\times 800$}  & \boldmath{$25\times900$} & \boldmath{$25\times 1000$} \\[6pt]
\end{tabular}
\begin{tabular}{ccccccc}
\includegraphics[height=0.15\textwidth, width=0.15\textwidth]{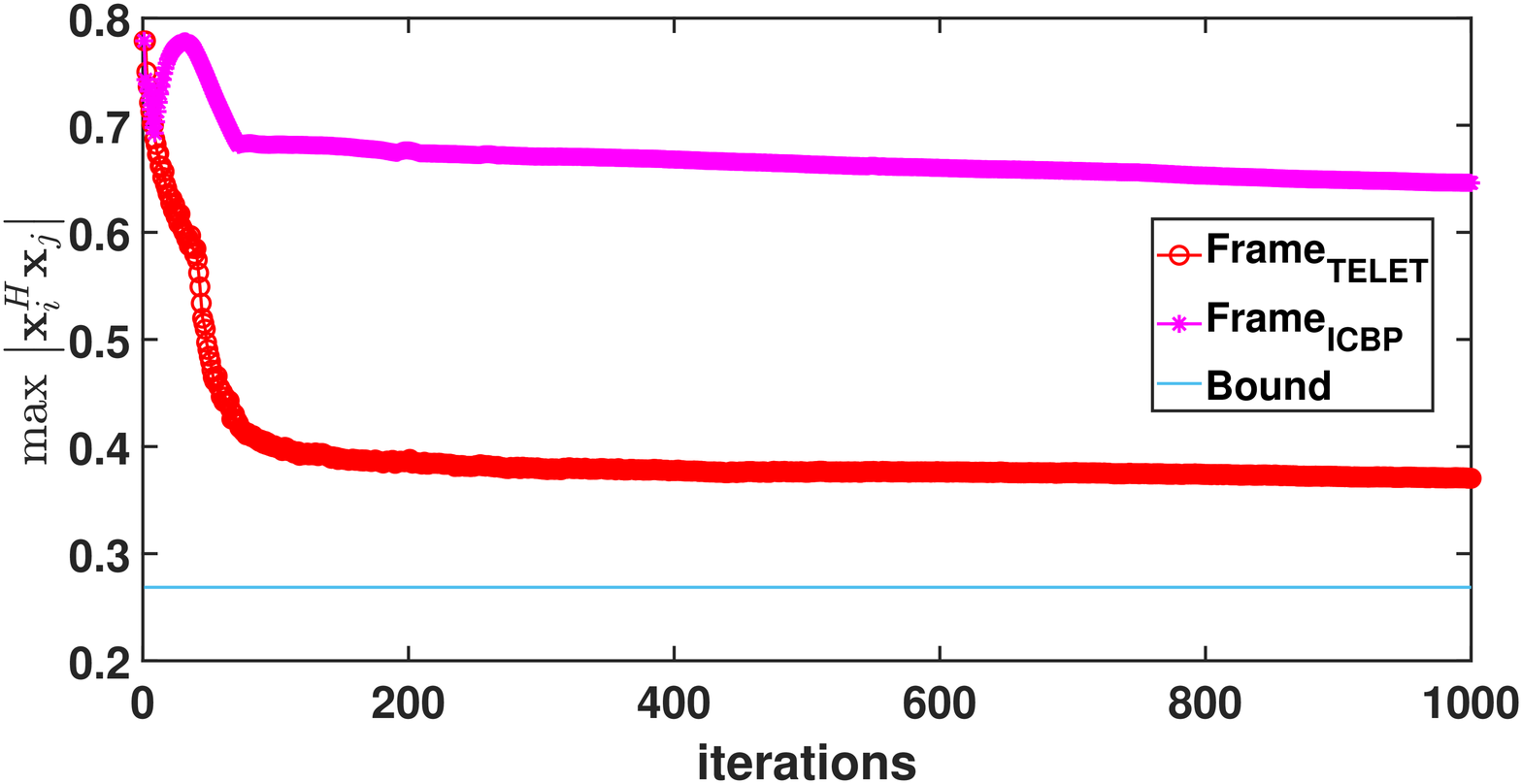} &
\includegraphics[height=0.15\textwidth, width=0.15\textwidth]{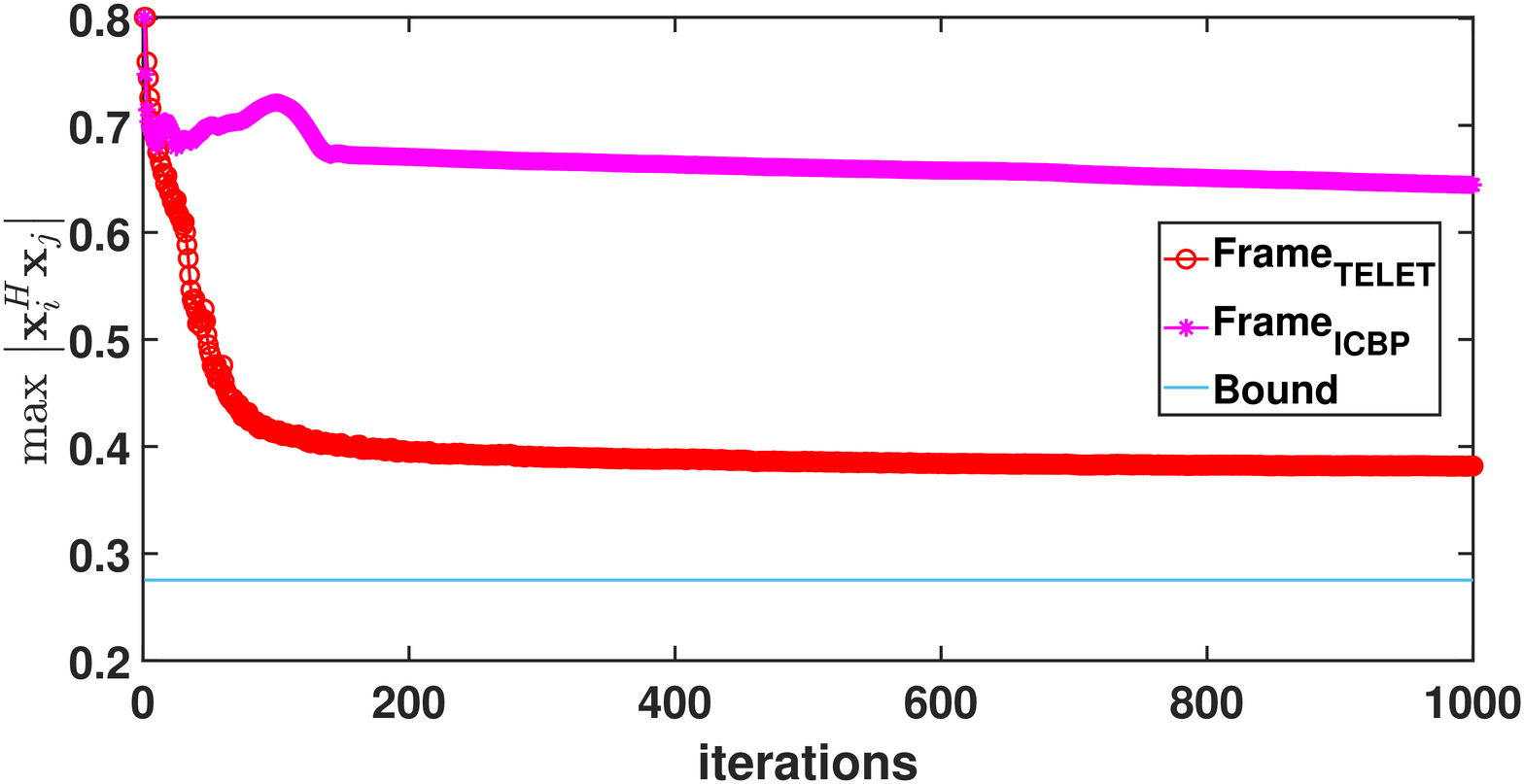} &
\includegraphics[height=0.15\textwidth, width=0.15\textwidth]{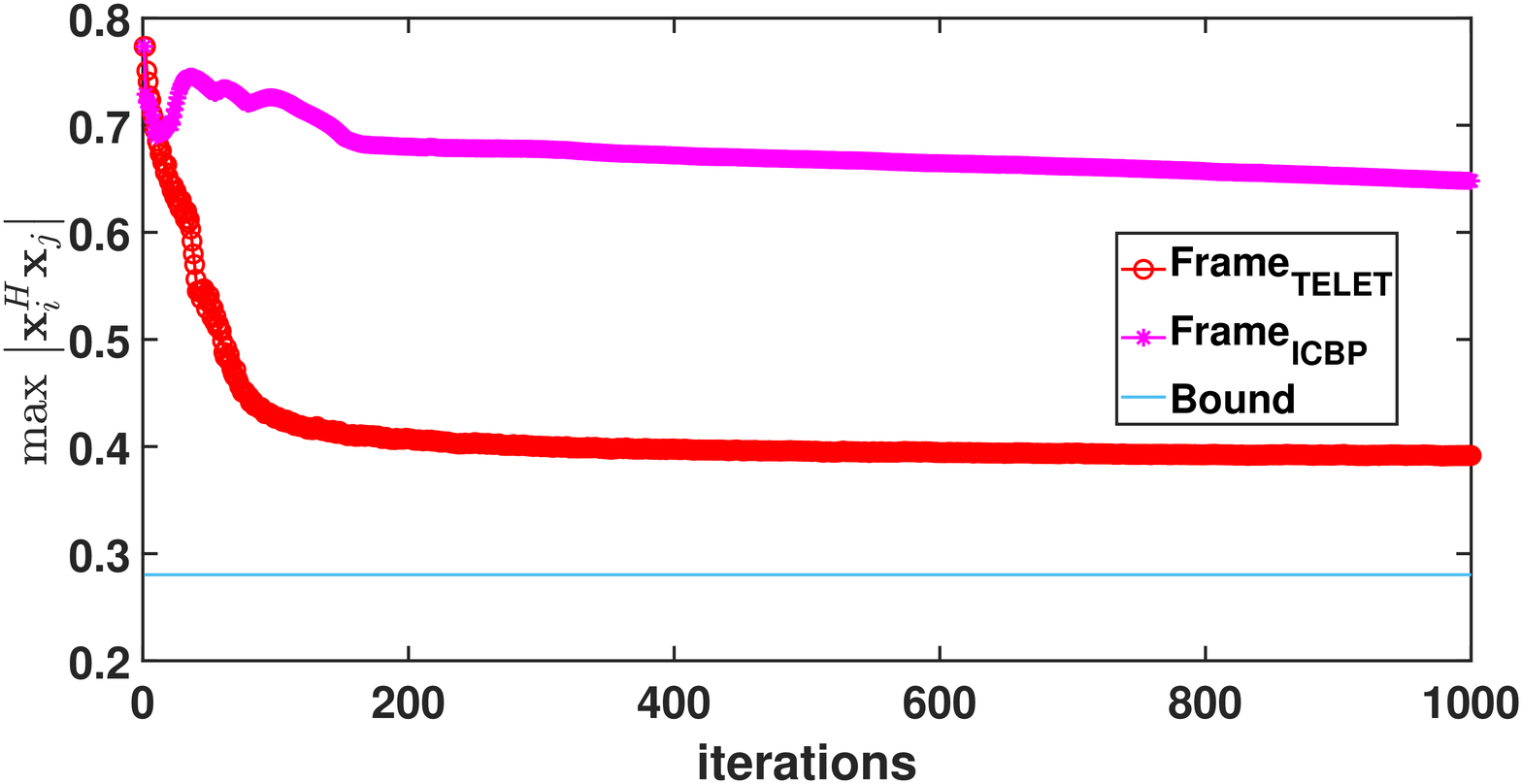} &
\includegraphics[height=0.15\textwidth, width=0.15\textwidth]{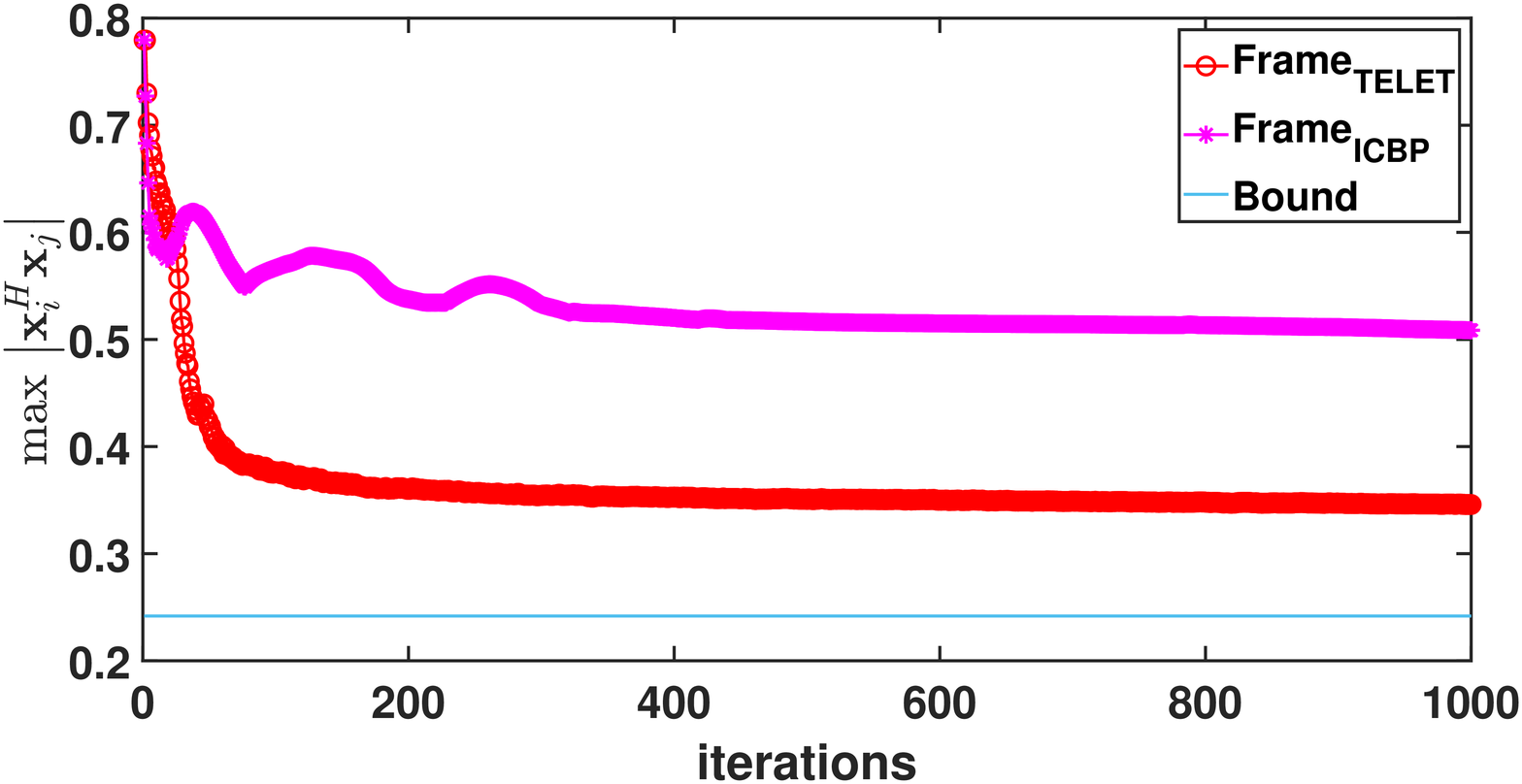} &
\includegraphics[height=0.15\textwidth, width=0.15\textwidth]{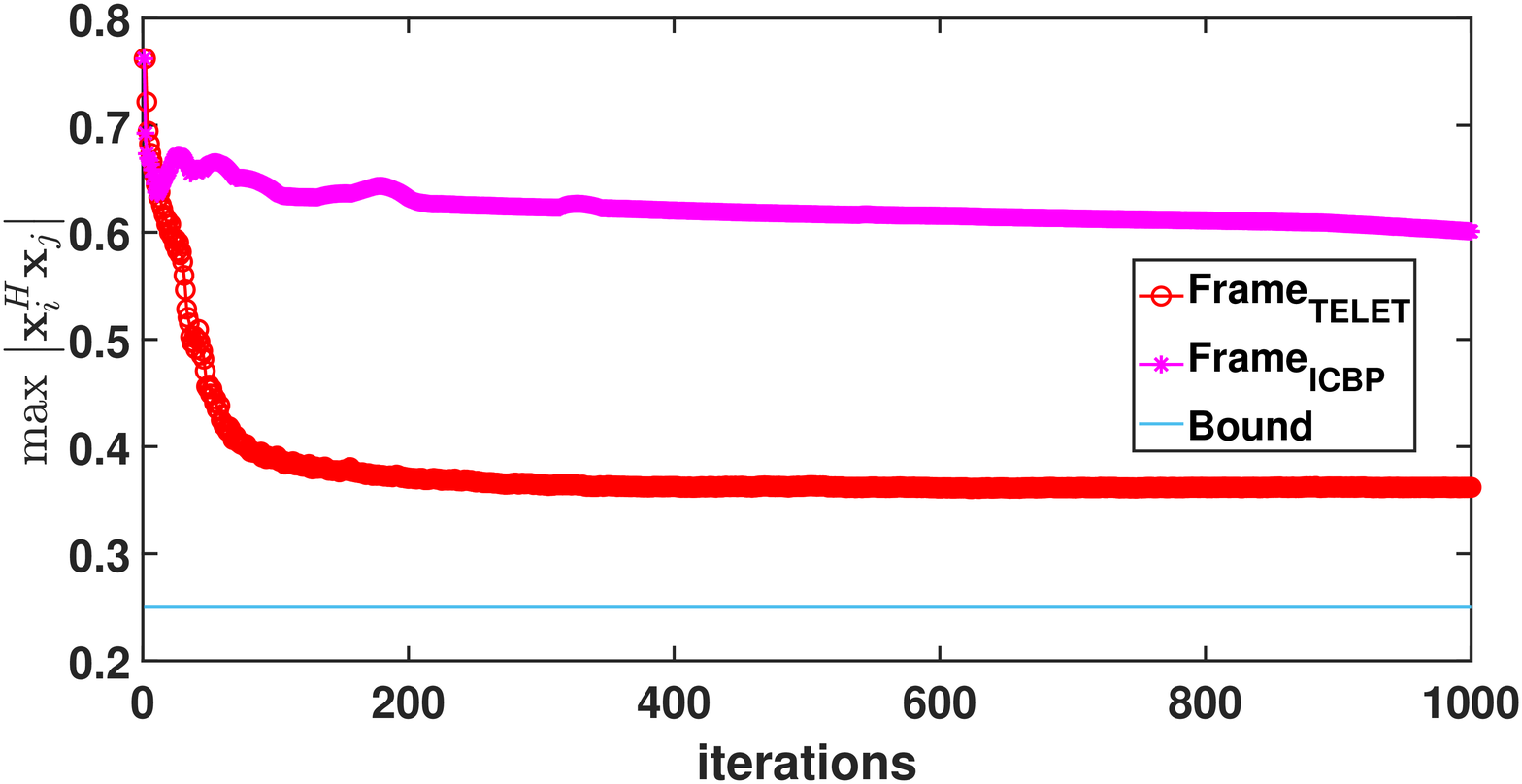} &
\includegraphics[height=0.15\textwidth, width=0.15\textwidth]{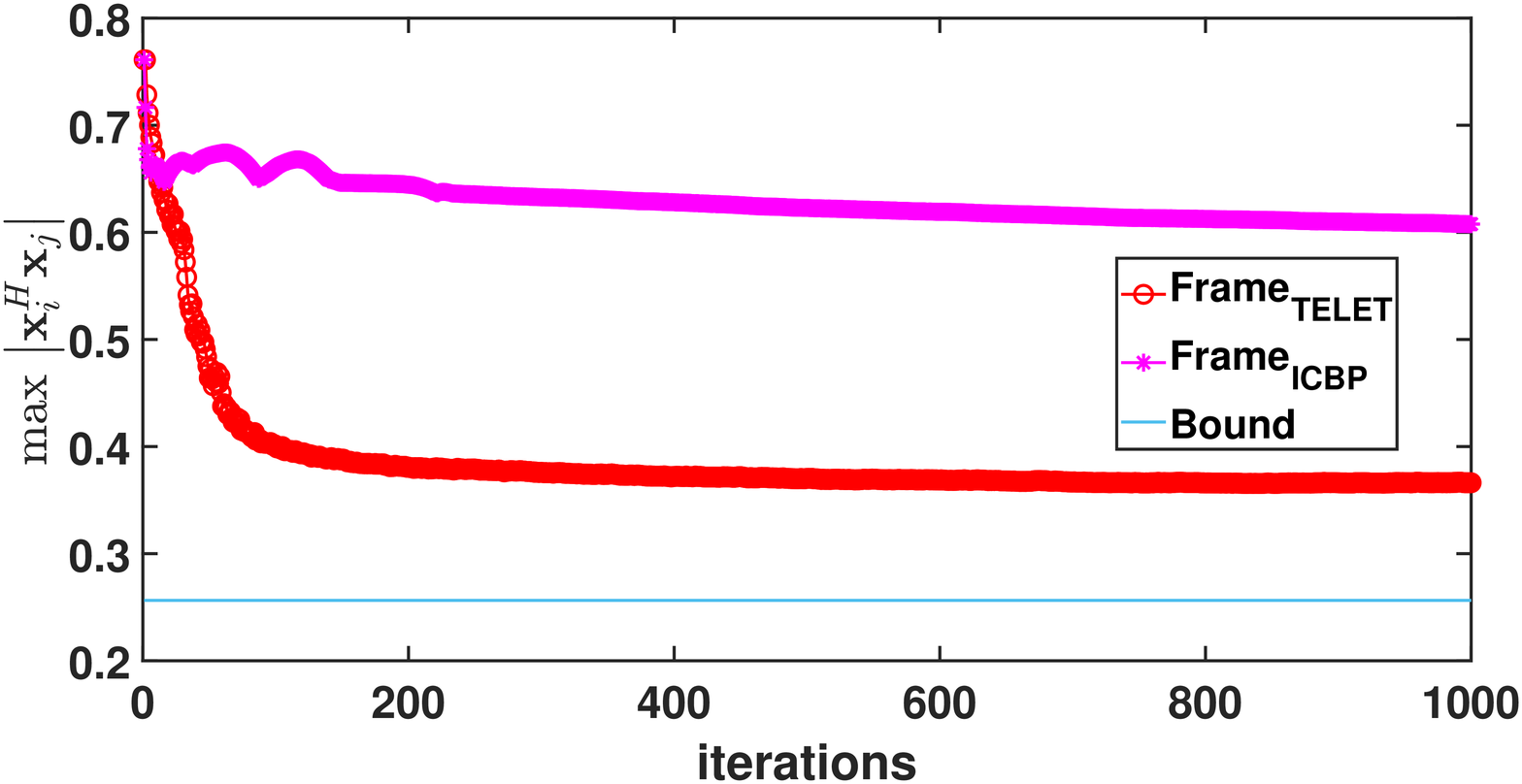} \\
\boldmath{$27\times 800$}  & \boldmath{$27\times900$} & \boldmath{$27\times 1000$} &\boldmath{$30\times 800$}  & \boldmath{$30\times900$}  & \boldmath{$30\times1000$} \\[6pt]
\end{tabular}
\caption{Convergence plot: ${\rm max}\:  \left| \textbf{x}_{i}^{H}\textbf{x}_{j}\right|$ vs. iteration of real frames of large dimensions.Red line - converge of ${\textrm{Frame}}_{\textrm{TELET}}$; Magenta line - convergence of ${\textrm{Frame}}_{\textrm{ICBP}}$ and Blue line - composite bound.}\vspace{-7mm}
\label{realbig}
\end{figure}

\noindent{3}. We now repeat the above experiments for real frames. All the algorithms were initialized with the same random vectors with the initialization scheme as discussed in the previous simulation. Similar to the previous simulation, the maximum number of iterations was set equal to $10^{4}$ in the case of small and medium frame dimensions and $10^{3}$ in the case of large frame dimensions. Table. \ref{coh_real} - Table. \ref{timel_real} compares the mutual coherence and run time of the algorithms for frames of different dimensions, respectively. In Table. \ref{coh_real} - Table. \ref{timel_real} we have highlighted the algorithms which performs the best with respect to mutual coherence value and run time. Similar to the observations made for complex case, from Table. \ref{coh_real} it can seen that ${\textrm{Frame}}_{\textrm{Xiong}}$ and ${\textrm{Frame}}_{\textrm{Tropp}}$ have higher mutual coherence for the considered small frame dimensions and from Table. \ref{time_real} it can be seen that CBGC algorithm takes more time to converge. Also, from Table. \ref{cohl_real} it can be seen that the proposed algorithm always achieves the lowest coherence value for the large frame dimensions when compared to the other state-of-the-art algorithms. Moreover, from Table. \ref{coh_real} and Table. \ref{cohl_real} it can be seen that out of the $54$ frames considered, the proposed algorithm achieves the least mutual coherence value $43$ times. Also, from Table. \ref{time_real} and Table. \ref{timel_real}, it can be seen that ${\textrm{Frame}}_{\textrm{ICBP}}$ takes lesser time for medium and large frame dimensions. In Fig. \ref{realbig} we also show the convergence plot of some of the large frame dimensions considered in Table. \ref{cohl_real}. Similar to the observation made for complex case in Fig. \ref{complexbig}, it can be seen in Fig. \ref{realbig} that for some of the frame dimensions (such as $23 \times 1000$ and $25 \times 1000$), the proposed algorithm even though achieves lower coherence value, has not converged and is stopped due to the maximum iteration limit.\\
We now conclude the results of the simulations conducted for various dimensions of complex and real frames. As can be observed from the simulations, in the case of complex frames, the two best algorithms (among the competing algorithms) are the proposed algorithm  ${\textrm{Frame}}_{\textrm{TELET}}$ and  ${\textrm{Frame}}_{\textrm{ICBP}}$ algorithm. The mutual coherence value attained by the two  best algorithms are comparable for small and medium dimension frames. However, with the increase in the value of the frame dimensions, it can be seen that  the proposed algorithm almost always achieves lower mutual coherence value when compared to ${\textrm{Frame}}_{\textrm{ICBP}}$ algorithm. We also calculated the percentage decrease in the mutual coherence value achieved by the proposed algorithm with respect to  coherence value achieved by the ${\textrm{Frame}}_{\textrm{ICBP}}$ using the formula:
\begin{equation}\label{pd}
 {\textrm{Mutual coherence decrease in}}\: \% = \left(\dfrac{\mu_{\textrm{ICBP}}-\mu_{\textrm{TELET}}}{\mu_{\textrm{ICBP}}}\right)\times 100
\end{equation}
where $\mu_{\textrm{ICBP}}$ and $\mu_{\textrm{TELET}}$ are the mutual coherence values obtained by ${\textrm{Frame}}_{\textrm{ICBP}}$ and ${\textrm{Frame}}_{\textrm{TELET}}$ algorithms, respectively  for the large frame dimensions mentioned in Table. \ref{cohc}. We found that the average percentage decrease in the mutual coherence value achieved by the proposed algorithm with respect to ${\textrm{Frame}}_{\textrm{ICBP}}$ to be equal to $11.58 \%$. Also, for some large frame dimensions, it was noticed that while the ${\textrm{Frame}}_{\textrm{ICBP}}$ algorithm had converged, the proposed algorithm had not converged and still achieved lower coherence value. However, for large dimensional frames, the average run time of ${\textrm{Frame}}_{\textrm{ICBP}}$ algorithm was found to be about $5$ times computationally faster than the proposed algorithm. Similarly, in the case of real frames, for small and medium dimensions, the mutual coherence value attained by ${\textrm{Frame}}_{\textrm{TELET}}$ and  ${\textrm{Frame}}_{\textrm{ICBP}}$ algorithm were comparable. However, with the increase in the value of the frame dimensions, the ${\textrm{Frame}}_{\textrm{TELET}}$ algorithm always achieving lower coherence value when compared to ${\textrm{Frame}}_{\textrm{ICBP}}$. For large dimensions, the average percentage decrease in the mutual coherence value achieved by the proposed algorithm with respect to ${\textrm{Frame}}_{\textrm{ICBP}}$ was found to be equal to $33.12\%$.  Also, similar to complex case, for some of the large dimensional frames, it was found that while ${\textrm{Frame}}_{\textrm{ICBP}}$ algorithm had converged to a larger coherence value, the proposed algorithm had not converged and yet achieved a lower coherence value. However, with respect to average run time, ${\textrm{Frame}}_{\textrm{ICBP}}$ was found to be faster than ${\textrm{Frame}}_{\textrm{TELET}}$ for large frames. 
\begin{table*}[!h]
\small
\begin{center}
\caption{Comparison of Mutual coherence of real frames constructed by different algorithms for small and medium  dimensions}
\label{coh_real}
\begin{tabular}{|p{1.2cm}|p{1.7cm}|p{1.7cm}|p{1.7cm}|p{1.7cm}|p{1.7cm}|p{1.7cm}|}
\hline
$(d,N)$ &$\mu_{\textrm{CB}}$&${\textrm{Frame}}_{\textrm{TELET}}$&${\textrm{Frame}}_{\textrm{ICBP}}$&${\textrm{Frame}}_{\textrm{CBGC}}$&${\textrm{Frame}}_{\textrm{Xiong}}$&${\textrm{Frame}}_{\textrm{Tropp}}$\\       
 \hline
$(2,8)$ & $0.8165$&\textbf{0.9239}& \textbf{0.9239} &\textbf{0.9239}&$0.99$&$0.9602$ \\
$(3,16)$&0.7125&\textbf{0.7947}&\textbf{0.7947}&\textbf{0.7947}&0.9799&0.9778\\
(4,5)&0.25&\textbf{0.25}&\textbf{0.25}&\textbf{0.25}&\textbf{0.25}&\textbf{0.25}\\
(4,6)&0.3162&\textbf{0.3333}&0.3334&0.3338&0.4998&0.3536\\
(4,7)&0.3536&0.3909&\textbf{0.3904}&0.3905&0.5&0.4438\\
(4,8)&0.3780&\textbf{0.3780}&0.3952&\textbf{0.3780}&0.5&\textbf{0.3780}\\
(4,9)&0.3953&\textbf{0.4343}&{0.4344}&{0.4344}&0.5&0.5192\\
(4,10)&0.4082&\textbf{0.4343}&\textbf{0.4343}&0.4344&0.5&0.4407\\
(4,19)&0.6055&\textbf{0.6467}&0.6774&0.65529&0.7149&0.7377\\
(4,20)&0.6124&0.6548&\textbf{0.6546}&0.6553&0.7114&0.9065\\
(5,6)&0.2&\textbf{0.2}&\textbf{0.2}&\textbf{0.2}&\textbf{0.2}&\textbf{0.2}\\
(5,7)&0.2582&0.2863&\textbf{0.2862}&0.2865&0.3960&0.3861\\
(5,8)&0.2928&\textbf{03291}&0.3304&0.3338&0.4472&0.4015\\
(5,9)&0.3162&\textbf{0.3334}&\textbf{0.3334}&0.3336&0.4472&0.4148\\
(5,10)&0.3333&\textbf{0.3333}&0.3855&\textbf{0.3333}&0.4472&\textbf{0.3333}\\
(5,16)&0.4109&\textbf{0.4472}&\textbf{0.4472}&\textbf{0.4472}&\textbf{0.4472}&0.4636\\
(5,26)&0.5408&0.5950&0.6316&\textbf{0.5910}&0.7150&0.8141\\
(6,37)&0.5040&0.5583&\textbf{0.5345}&0.5354&0.5764&0.8225\\
(20,30)&0.1313&0.1406&0.141&\textbf{0.1402}&-&-\\
(20,50)&0.1750&0.1970&0.1949&\textbf{0.1937}&-&-\\
(20,100)&0.2010&0.2609&\textbf{0.2465}&0.2471&-&-\\
(30,40)&0.0925&0.0994&0.1006&\textbf{0.0984}&-&-\\
(30,50)&0.1166&0.1266&0.1258&\textbf{0.1243}&-&-\\
(30,100)&0.1535&0.1919&0.1768&\textbf{0.1764}&-&-\\
\hline
\end{tabular}
\end{center}
\end{table*}

\begin{table}[!h]
\small
\begin{center}
\caption{Comparison of Mutual coherence of large dimension real frames constructed by different algorithms}
\label{cohl_real}
\begin{tabular}{|p{1.2cm}|p{1.2cm}|p{1.4cm}|p{1.4cm}|p{1.4cm}|}
\hline
$(d,N)$ &$\mu_{\textrm{CB}}$&${\textrm{Frame}}_{\textrm{TELET}}$&${\textrm{Frame}}_{\textrm{ICBP}}$\\       
 \hline
 (23,500)&0.2785&\textbf{0.3703}&0.5850\\
(23,600)&0.2914&\textbf{0.3869}&0.6139\\
(23,700)&0.3002&\textbf{0.4006}&0.6027\\
(23,800)&0.3065&\textbf{0.4449}&0.6817\\
(23,900)&0.3113&\textbf{0.4220}&0.6703\\
(23,1000)&0.3151&\textbf{0.4697}&0.6785\\
(23,1200)&0.3206&\textbf{0.4476}&0.6736\\
 (25,500)&0.2536&\textbf{0.3468}&0.4878\\
(25,600)&0.2692&\textbf{0.3641}&0.5772\\
(25,700)&0.2796&\textbf{0.4028}&0.6091\\
(25,800)&0.2871&\textbf{0.3900}&0.6403\\
(25,900)&0.2928&\textbf{0.4003}&0.6437\\
(25,1000)&0.2972&\textbf{0.4340}&0.6799\\
(25,1200)&0.3036&\textbf{0.4486}&0.6926\\
(27,500)&0.2286&\textbf{0.3275}&0.4336\\
(27,600)&0.2474&\textbf{0.3440}&0.5119\\
(27,700)&0.2598&\textbf{0.3597}&0.5715\\
(27,800)&0.2686&\textbf{0.3704}&0.6461\\
(27,900)&0.2752&\textbf{0.3816}&0.6444\\
(27,1000)&0.2803&\textbf{0.3911}&0.6480\\
(27,1200)&0.2878&\textbf{0.4070}&0.6604\\
(30,800)&0.2417&\textbf{0.3458}&0.5086\\
(30,900)&0.25&\textbf{0.3607}&0.6008\\
(30,1000)&0.2564&\textbf{0.3655}&0.6076\\
(30,1200)&0.2655&\textbf{0.3851}&0.6549\\
(40,800)&0.1542&\textbf{0.2909}&0.3357\\
(40,900)&0.1680&\textbf{0.3015}&0.3773\\
(40,1000)&0.1809&\textbf{0.3119}&0.3954\\
(40,1200)&0.1985&\textbf{0.3275}&0.4488\\
(50,1000)&0.1379&\textbf{0.2788}&0.3217\\
\hline
\end{tabular}
\end{center}
\end{table}

\begin{table*}[!h]
\small
\begin{center}
\caption{Comparison of run time (seconds) of the proposed and state-of-the-art algorithms to construct real frames of small, medium and large dimensions}
\label{time_real}
\begin{tabular}{|p{1.2cm}|p{1.7cm}|p{1.7cm}|p{1.7cm}|p{1.7cm}|p{1.7cm}|}
\hline
$(d,N)$&${\textrm{Frame}}_{\textrm{TELET}}$&${\textrm{Frame}}_{\textrm{ICBP}}$&${\textrm{Frame}}_{\textrm{CBGC}}$&${\textrm{Frame}}_{\textrm{Xiong}}$&${\textrm{Frame}}_{\textrm{Tropp}}$\\       
 \hline
$(2,8)$ & $0.1920$&$0.0287$& $1.9849$ &$0.0009$&$\textbf{0.0004}$ \\
(3,16)&11.46&0.0915&15.4802&\textbf{0.0159}&0.059\\
(4,5)&0.3121&0.0039&62&0.0031&\textbf{0.00003}\\
(4,6)&0.5354&\textbf{0.0288}&0.1023&0.2349&0.3027\\
(4,7)&6.8206&\textbf{0.1115}&2.1150&0.1972&0.3523\\
(4,8)&0.5354&0.4750&11.9228&0.2812&0.2473\\
(4,9)&1.43&0.0758&1.7694&0.2693&0.1309\\
(4,10)&7.8786&0.079&1.1954&0.2651&0.176\\
(4,19)&12.218&5.4175&818.0925&\textbf{0.0324}&0.66\\
(4,20)&24.2077&0.4647&49.6893&0.0070&\textbf{0.0025}\\
(5,6)&0.5059&0.0067&0.0078&0.0016&0.0012\\
(5,7)&1.8895&0.32&0.7191&0.007&\textbf{0.0003}\\
(5,8)&2.5685&0.7696&0.2503&0.2406&\textbf{0.0122}\\
(5,9)&43.7984&0.2205&0.8046&0.4533&\textbf{0.0005}\\
(5,10)&3.7073&1.3572&0.0341&0.2521&\textbf{0.0025}\\
(5,16)&19.586&1.0214&0.8949&0.159&\textbf{0.1139}\\
(5,26)&74.3653&8.9504&524.6116&0.0348&\textbf{0.0098}\\
(6,37)&95.505&0.8462&47.4128&2.5269&\textbf{0.0029}\\
(20,30)&127.97&6.128&206.79&-&-\\
(20,50)&56.78&11.7099&1700.7&-&-\\
(20,100)&1626.6&59.1218&12201&-&-\\
(30,40)&181.3425&4.4492&322.132&-&-\\
(30,50)&289.8671&10.9571&2221.2&-&-\\
(30,100)&2608&52.827&14339&-&-\\
\hline
\end{tabular}
\end{center}
\end{table*}

\begin{table}[!h]
\small
\begin{center}
\caption{Comparison of run time (seconds) of the proposed and state-of-the-art algorithms to construct large dimension real frames.}
\label{timel_real}
\begin{tabular}{|p{1.2cm}|p{1.2cm}|p{1.4cm}|}
\hline
$(d,N)$&${\textrm{Frame}}_{\textrm{TELET}}$&${\textrm{Frame}}_{\textrm{ICBP}}$\\       
 \hline
 (23,500)&7447.6&936.7474\\
(23,600)&12993&1871\\
(23,700)&20205&2764.8\\
(23,800)&18562&2523.5\\
(23,900)&19512&1878.8\\
(23,1000)&35523&4342.4\\
(23,1200)&57621&4086.1\\
(25,500)&9089.4&745.7\\
(25,600)&16677&1420.4\\
(25,700)&13243&1769.4\\
(25,800)&15117&1202.5\\
(25,900)&19670&1937.4\\
(25,1000)&34711&4371.5\\
(25,1200)&81627&7430.7\\
(27,500)&7142.5&1029.8\\
(27,600)&15357&1505\\
(27,700)&22302&2247.6\\
(27,800)&29570&4348.4\\
(27,900)&25403&3374.8\\
(27,1000)&33407&6421\\
(27,1200)&54222&4083\\
(30,800)&33546&2915.4\\
(30,900)&13589&1816\\
(30,1000)&22993&2778.5\\
(30,1200)&57005&4112.9\\
(40,800)&19930&2856.1\\
(40,900)&46389&4809.8\\
(40,1000)&34270&7634.8\\
(40,1200)&77984&6932.5\\
(50,1000)&26612&3332.7\\
\hline
\end{tabular}
\end{center}
\end{table}

\subsection{Compressed sensing application}
\begin{figure}[h]
\centering
\begin{subfigure}{0.49\textwidth}
\centering
\captionsetup{justification=centering}
\includegraphics[height=1.6in,width=2.5in]{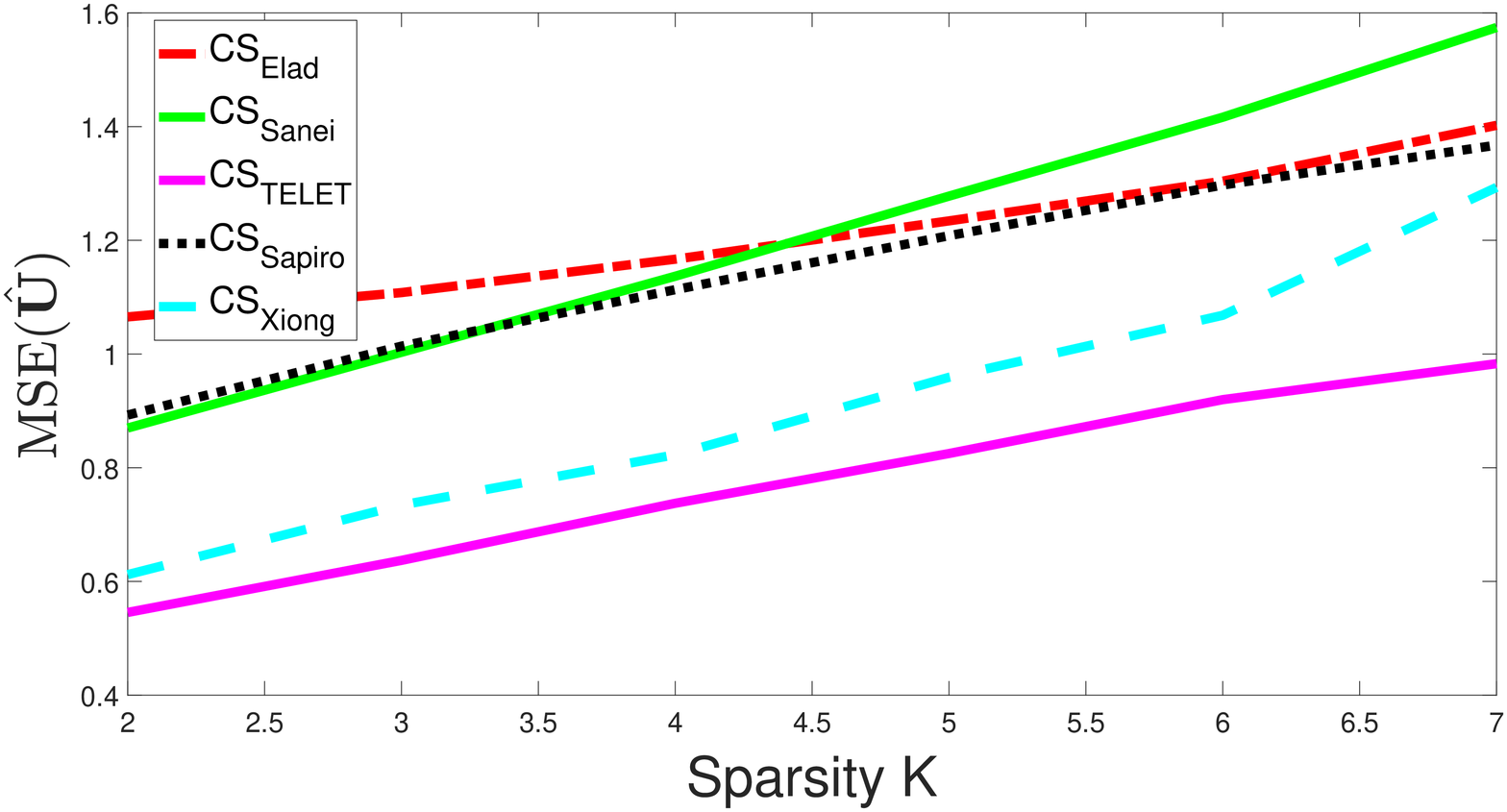}
\caption{ MSE ($\hat{\bU}$)  vs. Sparsity $K$  for Haar wavelet dictionary}
\end{subfigure}
\begin{subfigure}{0.49\textwidth}
\centering
\captionsetup{justification=centering}
\includegraphics[height=1.6in,width=2.5in]{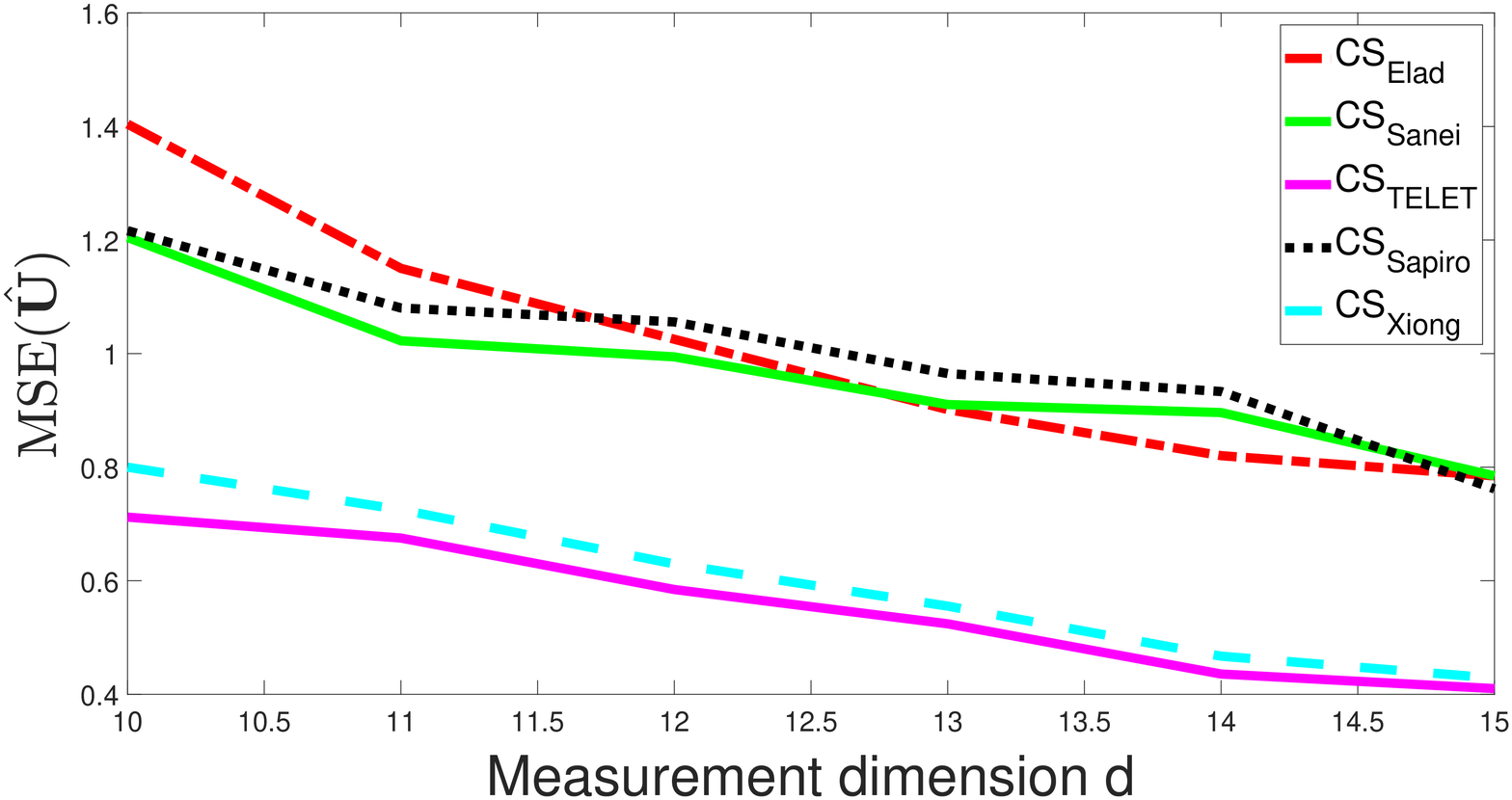}
\caption{MSE ($\hat{\bU}$)  vs. measurement dimension $d$  for Haar wavelet dictionary}
\end{subfigure}
\caption{ MSE ($\hat{\bU}$)  vs. Sparsity $K$ and measurement dimension $d$ for Haar wavelet dictionary using BP reconstruction algorithm}\vspace{-1.8mm}
\label{mse}
\end{figure}

\begin{figure} [h]
\centering
\begin{tabular}{cccc}
\includegraphics[width=0.15\textwidth]{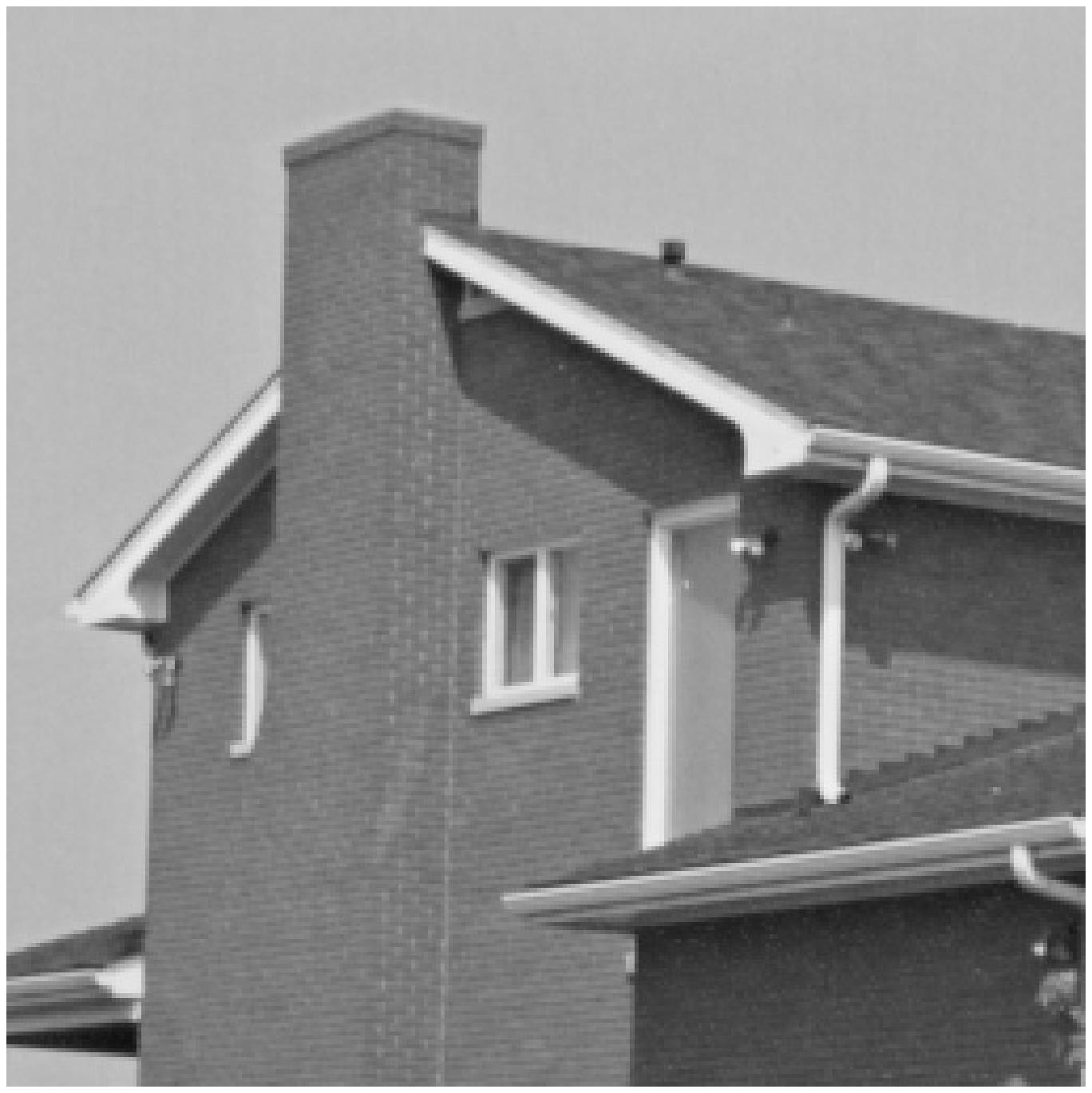} &
\includegraphics[width=0.15\textwidth]{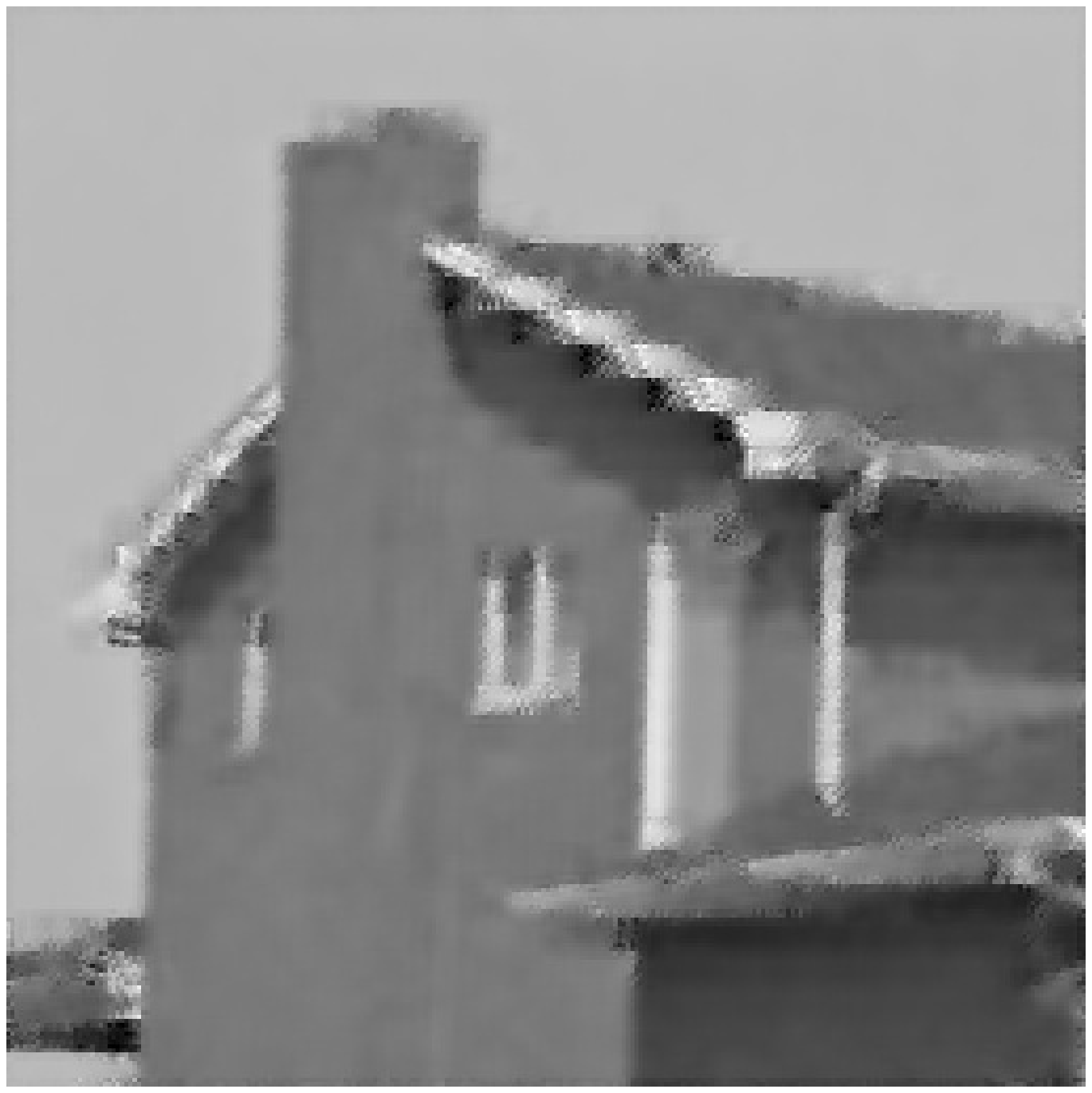} &
\includegraphics[width=0.15\textwidth]{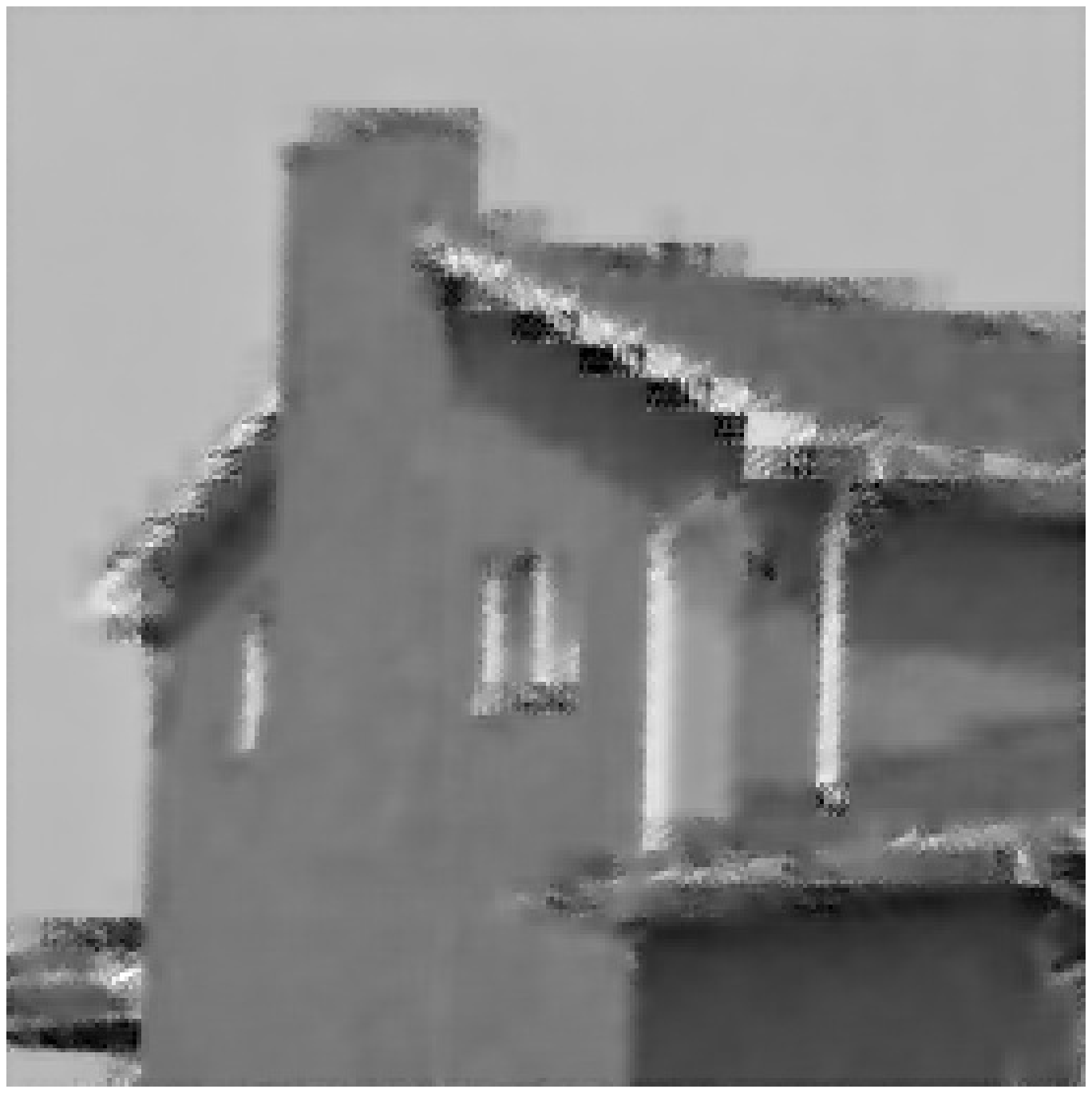} \\
\textbf{1.(a)}  & \textbf{1. (b)} & \textbf{1. (c)}\\[6pt]
\end{tabular}
\begin{tabular}{cccc}
\includegraphics[width=0.15\textwidth]{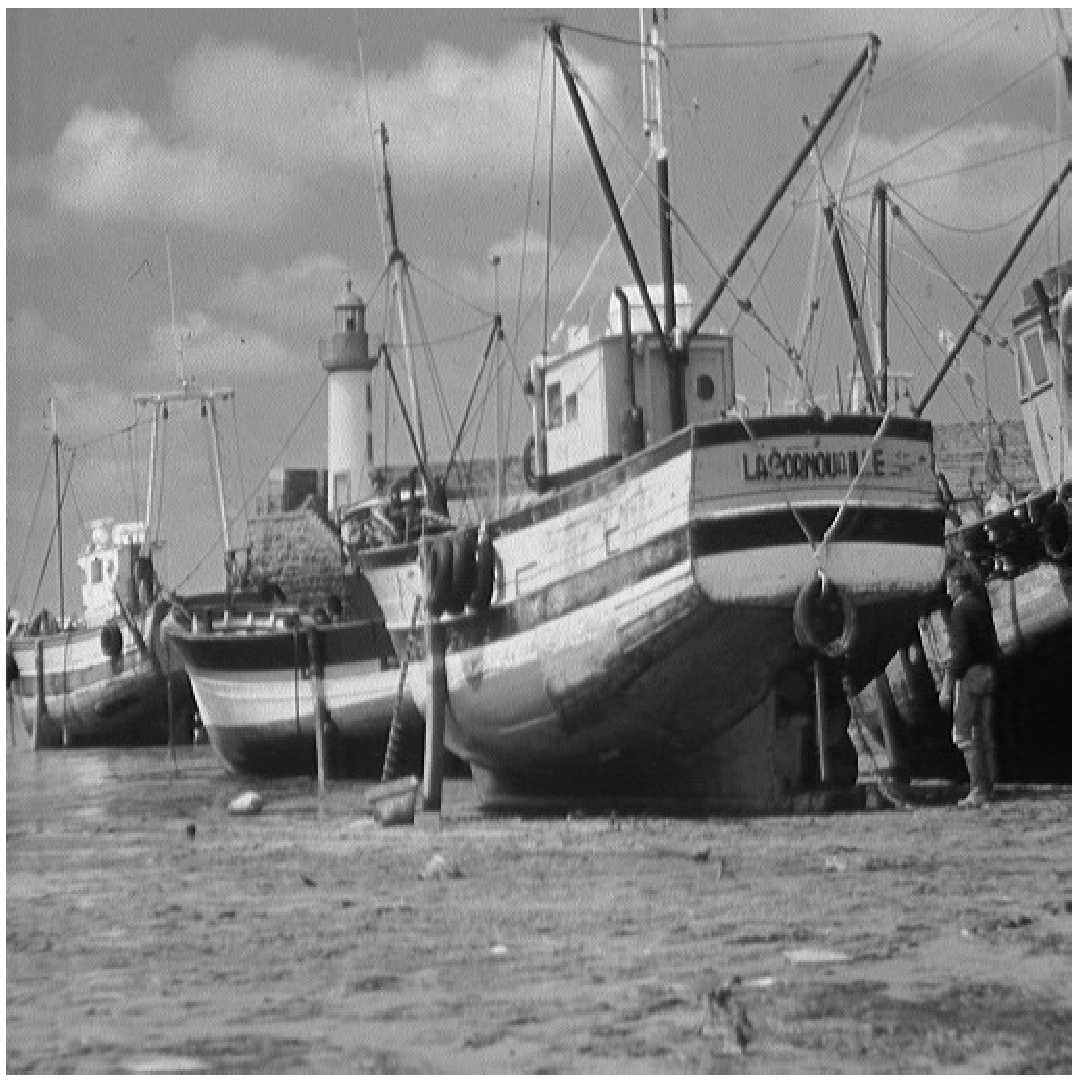} &
\includegraphics[width=0.15\textwidth]{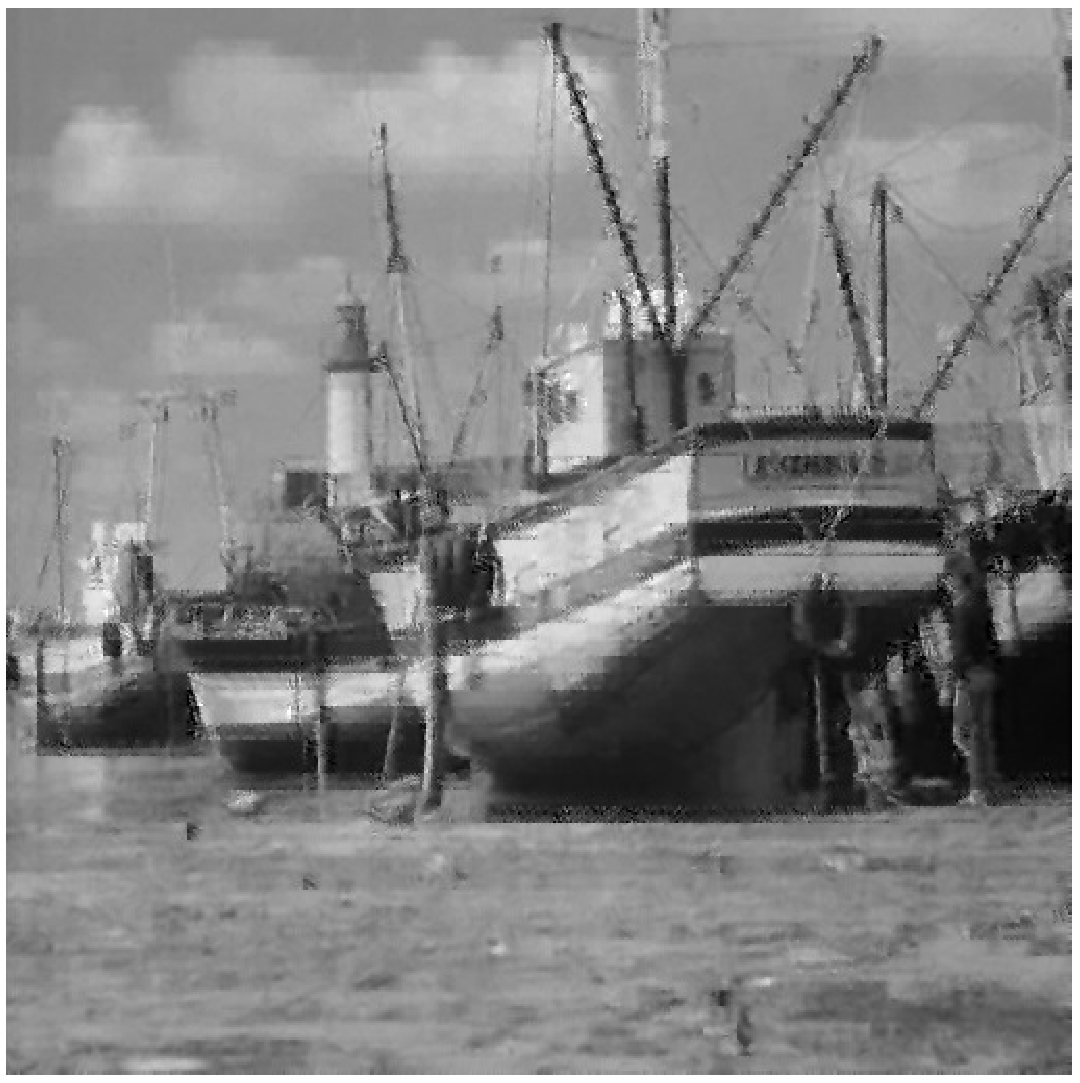} &
\includegraphics[width=0.15\textwidth]{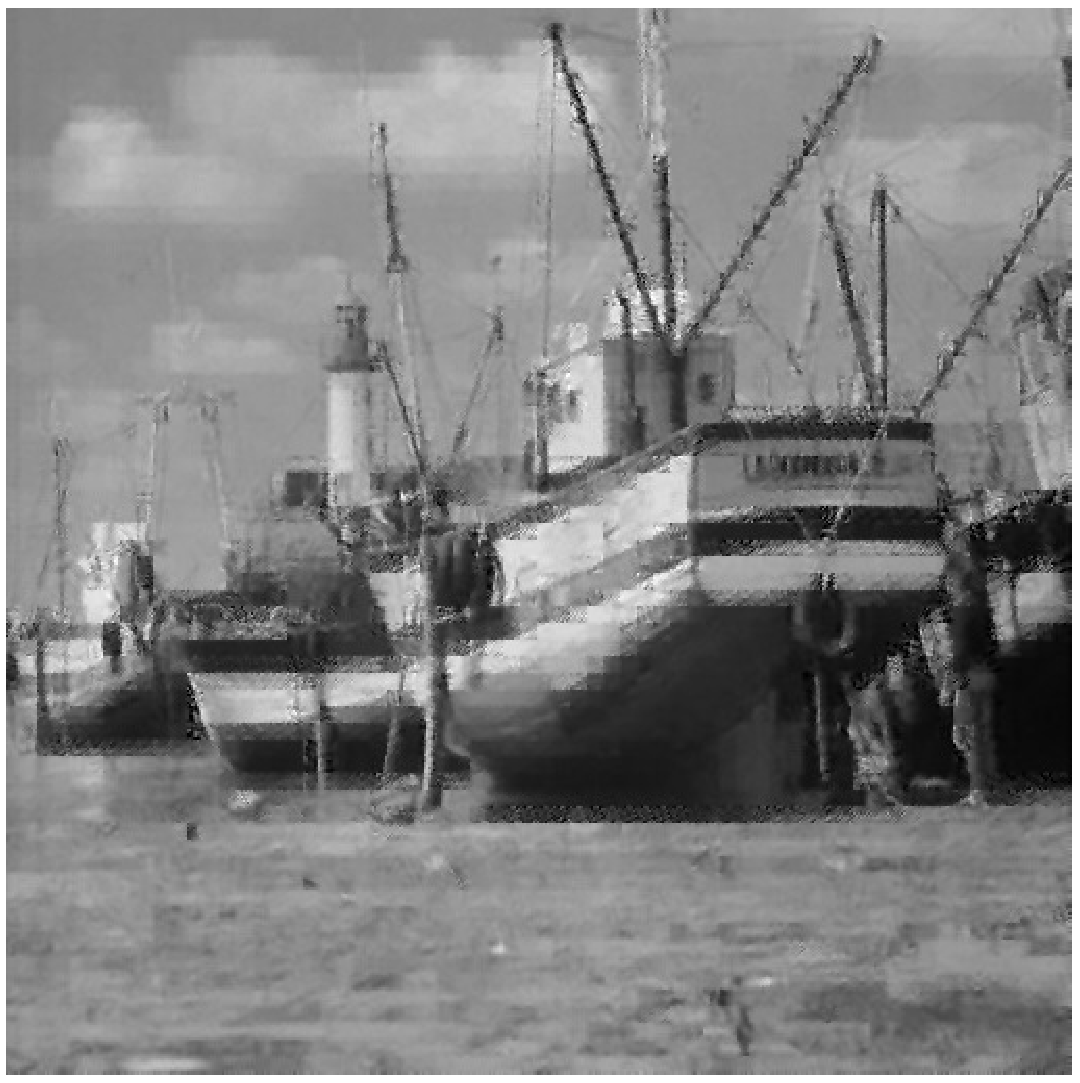} \\
\textbf{2. (a)}  & \textbf{2. (b)} & \textbf{2. (c)}  \\[6pt]
\end{tabular}
\begin{tabular}{cccc}
\includegraphics[width=0.15\textwidth]{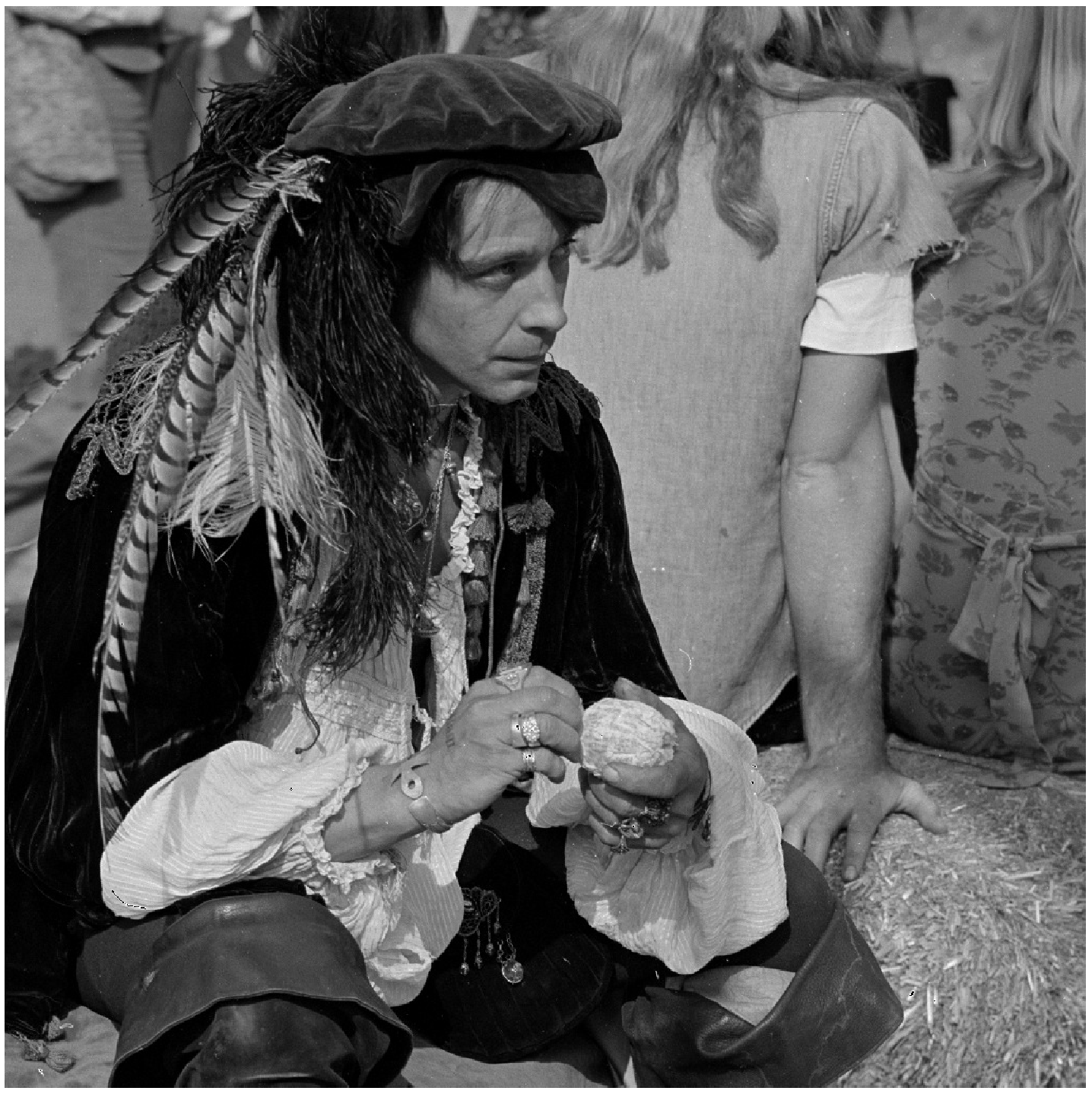} &
\includegraphics[width=0.15\textwidth]{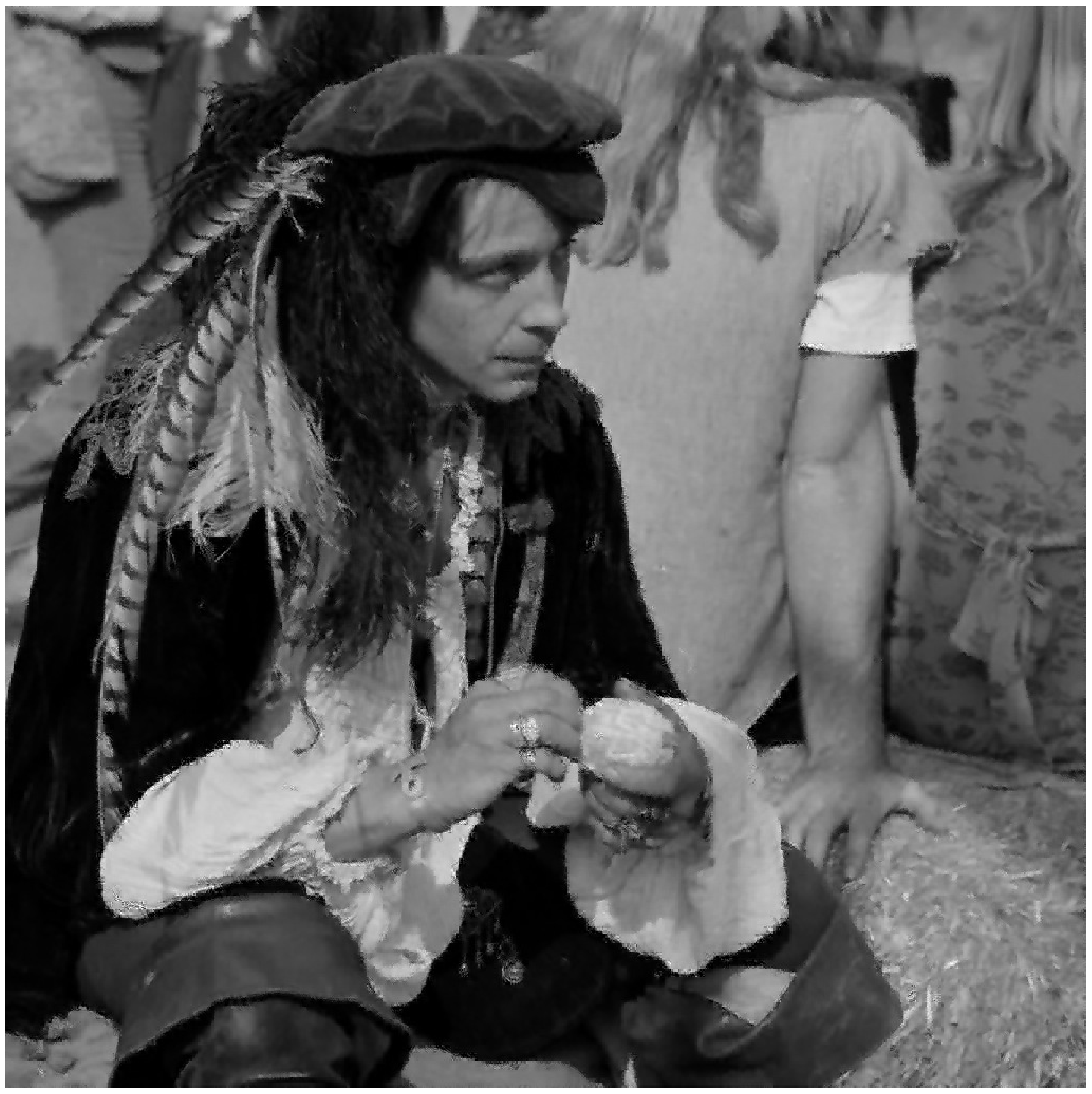} &
\includegraphics[width=0.15\textwidth]{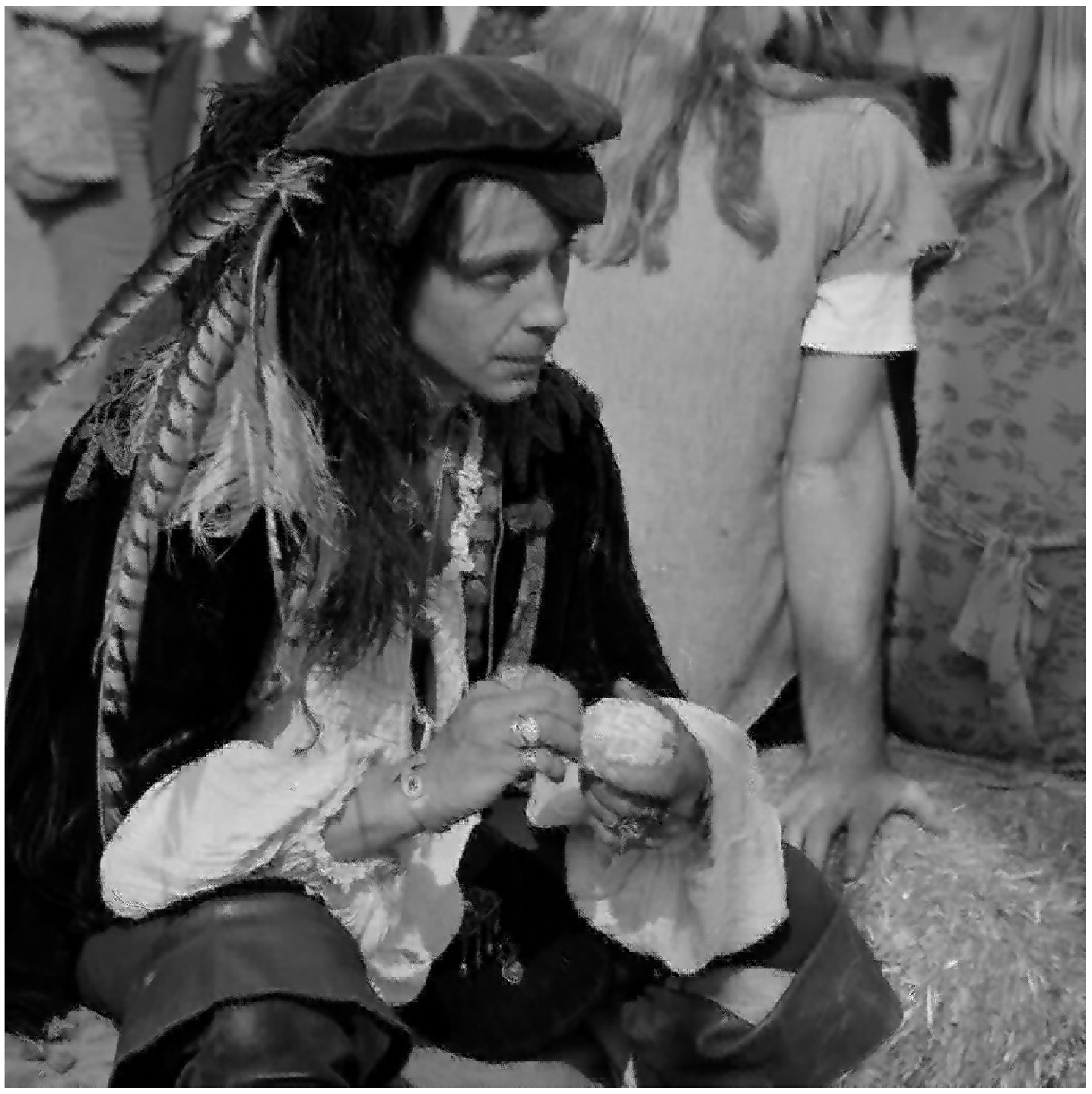} \\
\textbf{3. (a)}  & \textbf{3. (b)} & \textbf{3. (c)}\\[6pt]
\end{tabular}
\caption{\textbf{(a)} Original image
\textbf{(b)} Reconstructed image by the proposed algorithm
\textbf{(c)} Reconstructed image by the algorithm in \cite{xiong}}\vspace{-7mm}
\label{images}
\end{figure}

As discussed in the introduction section, an optimal sensing matrix must be constructed such that its equivalent dictionary $\bX_{{\rm{ED}}} = \bTh\bpsi$ has small mutual coherence and also by taking the sparse representation error into account.
Therefore, using the model proposed in \cite{xiong}, we solve the following problem to construct an optimal sensing matrix: 
\begin{equation}\label{ae}
\begin{array}{ll}
\underset{\bTh, \bX_{\rm{ED-Target}}}{\rm minimize}\:\:\omega\|\bX_{\rm{ED-Target}}- \bTh\bpsi\|_{F}^{2}  + (1-\omega)\|\bTh\bE\|_{F}^{2}
\end{array}
\end{equation}  
where $\omega$ is the trade-off factor which takes value in $[0,1]$, $\bX_{\rm{ED-Target}}$ is a target equivalent dictionary with properties of the ETF. The above problem is solved iteratively using the alternating minimization scheme wherein minimization is performed with respect to one matrix while keeping the other matrix fixed and vice-versa. At the $t^{th}$ iteration, the sensing matrix $\bTh^{t}$ is fixed and the equivalent dictionary $\bX_{\rm{ED-Target}}^{t+1}$ is updated using the TELET algorithm shown in Algorithm 2. Next, using the updated equivalent dictionary the problem in (\ref{ae}) can be solved - which has a closed-form solution and can be found leveraging Theorem $2$ in \cite{xiong}.\\
We now conduct simulations to demonstrate the performance of the CS system using the optimized sensing matrices constructed using the above described approach and the state-of-the-art algorithms (\cite{elad, gradient, sapiro, xiong}). We denote the CS system of the proposed algorithm as ${\rm{CS_{TELET}}}$ and the algorithms in  \cite{elad}, \cite{gradient}, \cite{sapiro} and \cite{xiong} as ${\rm{CS_{Elad}}}$, ${\rm{CS_{Sanei}}}$, ${\rm{CS_{Sapiro}}}$ and ${\rm{CS_{Xiong}}}$, respectively.\\ 
\emph{1. Experiments on synthetic data}\\
To evaluate the CS systems under various conditions, we first obtained a set of lower dimensional measurement vectors $\{\by_{t}\}_{t=1}^{R}$ ($R$ is the number of monte-carlo experiments) using the following experimental procedure. We first generated $R$ number of vectors $\{\bs_{t}\}_{t=1}^{R}$  with $K$ non-zero entries of length $N$. The non-zero elements of $\bs_{t}$ were randomly positioned and they were generated using Gaussian distribution with zero mean and unit variance. Then, the sparse signal vector $\bu^{*}_{t}$ was generated as $\bu^{*}_{t} = \bpsi\bs_{t}$, where the dictionary $\bpsi$ of dimension $N \times N$ dictionary was chosen as the Haar wavelet matrix. Next, we obtain the following contaminated signals: 
\begin{equation}\label{td}
\begin{array}{ll}
\bu_{t} = \bu^{*}_{t} +\epsilon_{t} =\bpsi\bs_{t}+\epsilon_{t}, \:\: t=1, 2 \cdots R
\end{array}
\end{equation}
where $\{\epsilon_{t}\}_{t=1}^{R}$ represents the sparse representation error and is assumed to Gaussian distributed noise with zero mean and variance $\sigma^{2}$. The lower dimensional measurement vectors $\{\by_{t}\}_{t=1}^{R}$ each of length $d$ were obtained using $\by_{t} = \bTh\bu_{t}$ where $\bTh$ is the optimized sensing matrix constructed using the proposed approach ${\rm{CS_{TELET}}}$ and the other state-of-the-art algorithms. From the obtained measurement vectors, the high dimensional signal vector $\hat{\bu}_{t}$ is reconstructed using $\hat{\bu}_{t} = \bpsi\hat{\bs}_{t}$ where $\hat{\bs}_{t}$ is obtained by the Basis Pursuit algorithm \cite{bp}. The reconstruction accuracy is quantified using the mean square error (MSE)  metric defined as:
\begin{equation}
\begin{array}{ll}
{\rm{MSE}(\hat{\bU})}= \dfrac{1}{d \times R}  \|\hat{\bU} - \bU^{*}\|_{F}^{2}
\end{array}
\end{equation}
where $\bU^{*}= [\bu_{1}^{*}, \bu_{2}^{*}, \cdots, \bu_{R}^{*}]$, $\hat{\bU} = [\hat{\bu}_{1},\hat{\bu}_{2},\cdots, \hat{\bu}_{R}]$, each $\hat{\bu}_{t}$ is the reconstructed version of $\bu_{t}^{*}$  and is equal to $\hat{\bu}_{t} = \bpsi\hat{\bs}_{t}$.
We evaluate the CS systems by varying the number of measurements $d$ and the sparsity level $K$. In the first experiment we fixed the number of measurements $d$ and varied sparsity level $K$ while in the second experiment we varied sparsity level $K$ for fixed number of measurements $d$. The parameters of the first experiment are as follows: Haar wavelet matrix of dimension $N=32$ was chosen, the number of measurements $d=10$, the number of monte-carlo experiments $R=50$, noise variance $\sigma^{2} = 0.25$, $\omega= 0.5$ \cite{xiong} and varied the sparsity of the vector $K$ from $2$ to $7$. In the second experiment,  we fixed the signal sparsity $K=4$ and varied the number of measurements $d$ from $10$ to $15$ with the same dimension of the dictionary $N$, noise variance $\sigma^{2}$ and the number of monte-carlo experiments $R$ as that of the first experiment.  Fig. \ref{mse}.a and Fig. \ref{mse}.b shows the  MSE ($\hat{\bU}$) vs. sparsity $K$ and MSE ($\hat{\bU}$) vs. measurement dimension $d$ of the proposed and the state-of-the-art algorithms, respectively. From Fig. \ref{mse} it can be seen that the proposed algorithm has lower  MSE ($\hat{\bU}$)  for all the values of sparsity $K$ and measurement dimension $d$.\\
Next, we compare the two competing algorithms ${\rm{CS_{Xiong}}}$ and ${\rm{CS_{TELET}}}$ in the context of image reconstruction from lower dimensional patches. \\
\emph{2. Experiments on image data}\\
Images from USC-SIPI database \cite{imagedata} were obtained to evaluate the performance of ${\rm{CS_{Xiong}}}$ and ${\rm{CS_{TELET}}}$ in the context of image reconstruction. This database is commonly used for conducting image processing research and contains images in TIFF format. We choose three different images of different sizes from the database - `House' image of size $256 \times 256$, `Boat' image of size $512 \times 512$ and a `Male' image of size $1024 \times 1024$.
Each image of size $Q \times Q$ is first  divided into non-overlapping patches of size $8\times 8$ and each patch is vectorized to get  $\bu_{t}$'s of size $64 \times 1$. Then from each patch, we obtained the compressed lower dimensional vector $\by_{t}$'s of size $d \times 1$ using the relation $\by_{t} = \bTh\bu_{t}$ where $\bTh$ is the optimized sensing matrix constructed using ${\rm{CS_{TELET}}}$ and ${\rm{CS_{Xiong}}}$ algorithm. For `House' image we obtained $31.25\: \%$ ($d=20$) compressed measurements from each patch. In the case of `Boat' image we took $35.9\: \%$ ($d=23$) compressed measurements from each patch and finally for `Male' image we obtained $40.62\: \%$ ($d=30$) compressed measurements. From the obtained compressed measurements, the high dimensional vector $\hat{\bu}_{t}$ is reconstructed using $\hat{\bu}_{t} = \bpsi\hat{\bs}_{t}$ where we have chosen $\bpsi$ as the DCT matrix, $\hat{\bs}_{t}$ is obtained by using the basis pursuit algorithm. Then, each reconstructed vector $\hat{\bu}_{t}$ is reshaped to get back a patch of size $p \times p$.
Fig. \ref{images} visually compares the original and the reconstructed images using ${\rm{CS_{Xiong}}}$ and ${\rm{CS_{TELET}}}$ algorithm. From Fig. \ref{images} it can be seen that ${\rm{CS_{TELET}}}$ algorithm is able to reconstruct the last two test images with high fidelity. In the case of `House' image, even though the reconstructed image using  ${\rm{CS_{TELET}}}$ looks better than the image reconstructed using ${\rm{CS_{Xiong}}}$, its fidelity is low. This could be because of the smaller size of the `House' image and also because of the higher compression from each patch. The quantitative accuracy of the reconstructed images is evaluated using PSNR metric as defined in \cite{psnr}. 
image of the proposed algorithm was found to be $25.910$
dB, $25.66$ dB and $28.244$ dB, respectively while that of the
algorithm in [9] was found to $24.457$ dB, $25.12$ dB and $27$
dB, respectively. Hence, ${\rm{CS_{TELET}}}$ has higher PSNR for all the images, with mean SNR of $26.604$ dB, when compared to ${\rm{CS_{Xiong}}}$ which has mean SNR equal to $25.526$ dB. 
\section{Conclusion}\label{sec:6}
\vspace{-5mm}
In this paper, we proposed an iterative algorithm TELET based on the Majorization Minimization principle to construct equiangular tight frames. Since the proposed algorithm is developed using Majorization Minimization procedure, it enjoys nice properties such as monotonicity and guaranteed convergence to a stationary point of the ETF problem. We also apply TELET algorithm to construct optimized sensing matrix for compressed sensing systems. We show through numerical simulations that when compared to the state-of-the-art algorithms TELET can construct complex and real frames with very low mutual coherence value, especially for large frame dimensions. However, with respect to run time, when compared to state-of-the-art algorithm, it is found that TELET is computationally slower. We also compared the performance of the optimized sensing matrix obtained using TELET algorithm with the state-of-the-algorithms used in compressed sensing systems using both synthetic and real images and found that the proposed algorithm performs better in terms of reconstruction accuracy. 
\section{Appendix}

1. {Proof of $\lambda_{\text{max}}({\boldsymbol{\boldsymbol{\Phi}}}_{ij})= d$}
\begin{IEEEproof}
We first recall that: 
\begin{equation}
{\boldsymbol{\boldsymbol{\boldsymbol{\Phi}}}_{ij}} =\text{vec}\left({{\bA}_{ij}}\right)\text{vec}^{H}\left(\left({{\bA}_{ij}}\right)^{H}\right)+ \text{vec}\left(\left({{\bA}_{ij}}\right)^{H}\right)\text{vec}^{H}\left({{\bA}_{ij}}\right)
\end{equation}
Notice that by construction of $\bA_{ij}= \bS_{j}^{H}\bS_{i}$, (where $\bS_{l} = [\bzero_{d \times (l-1)}, \bI_{d}, \bzero_{d \times (N-l)d}]$ is the selection matrix), it will have $d$  off-diagonal elements equal to one and zeros everywhere else. Let $\bU_{ij} = \text{vec}(\bA_{ij})\text{vec}^{H}({\bA_{ij}^{H}}) = \bq_{ij}\bs_{ij}^{H}$. Then it is easy to see that:
\begin{equation}\label{p1}
\text{Tr}(\bU_{ij}\bU_{ij}^{H}) = \|\bs_{ij}\|_{2}^{2}\|\bq_{ij}\|_{2}^{2} = d^{2}.
\end{equation} 
\begin{equation}\label{p2}
\text{Tr}(\bU_{ij}) = 0
\end{equation}
Since $\bPhi_{ij}$ is the sum of two rank one matrices, its  rank can be at most two. Let $\lambda_{1}$ and $\lambda_{2}$ be the two non-zero eigen values of $\bPhi_{ij}$.  Then using (\ref{p1}) and (\ref{p2}) we have:
\begin{equation}\label{eig:1}
\begin{array}{ll}
\lambda_{1}+ \lambda_{2}  = \rm{Tr}(\bU_{ij}+ \bU_{ij}^{H}) = 0
\end{array}
\end{equation}
\begin{equation}\label{eig:2}
\begin{array}{ll}
\lambda_{1}^{2} + \lambda_{2}^{2} = \rm{Tr}(\bU_{ij} + \bU_{ij}^{H})^{2} = 2 \rm{Tr}({\bU_{ij}\bU_{ij}^{H}}) = 2d^{2}
\end{array}
\end{equation}
Using (\ref{eig:1}) and (\ref{eig:2}) we get:
\begin{equation}
\begin{array}{ll}
\lambda_{1}\lambda_{2} = \dfrac{1}{2} \left(\left(\lambda_{1}+ \lambda_{2}\right)^{2} - \left(\lambda_{1}^{2} + \lambda_{2}^{2}\right)\right) = -d^{2}
\end{array}
\end{equation}
Then, we can find the maximum eigen value by solving the following equation: 
\begin{equation}
\begin{array}{ll}
\lambda^{2} - d^{2} = 0 
\end{array}
\end{equation}
whose solution is equal to $\lambda = \pm d$, choosing its maximum gives $\lambda_{max}(\bPhi_{ij}) = d$
\end{IEEEproof}
2. {Proof of $\lambda_{\text{max}}({\bB}_{ij}) = |c_{ij}|= |\bx_{i}^{H}\bx_{j}|$}
\begin{IEEEproof}
To prove the above, we first define $\bar{\bA}_{ij}$ and $\bar{\bB}_{ij}$ matrices as follows: 
\[  \begin{array}{cc}
\bar{\bA}_{ij}=
\end{array}   \left( \begin{array}{cc}
0 & c_{ij} \\
c_{ij}^{*} & 0 
\end{array} \right),\: \bar{\bB}_{ij}= \left( \begin{array}{cc}
\bzero_{d\times d} & c_{ij} \bI_{d\times d} \\
c_{ij}^{*} \bI_{d\times d} & \bzero_{d\times d}
\end{array} \right)\] 
One can easily see that the eigenvalues of $\bar{\bA}_{ij}$ is equal to $\lambda_{\bar{\bA}_{ij}} = \pm |c_{ij}|$. 
Next, the eigenvalues of $\bar{\bB}_{ij}$ can be obtained by solving the following equation (which we obtained by cofactor expansion): 
\begin{equation}
{\lambda_{\bar{\bB}_{ij}}}= (-{\lambda}^{2} + |c_{ij}|^{2})^{d-1} \rm{det}(\bar{\bA}_{ij}-\lambda\bI)=0
\end{equation}  
Hence, the maximum eigenvalue of $\bar{\bB}_{ij}$ will be equal to $|c_{ij}|$. Also, note that by inserting zero vector columns and rows between the blocks of $\bar{\bB}_{ij}$ wont change its maximum eigenvalue. We use these observations to prove that $\lambda_{\text{max}}({\bB}_{ij}) = |c_{ij}|$ where $\bB_{ij} =  \bA_{ij} (c_{ij}^{*})+\bA_{ij}^{H}(c_{ij})$. By construction, the matrix $\bB_{ij}$ will be a block diagonal matrix with a main block equal to $\bar{\bB}_{ij}$ (with or without the zero vectors inserted between its blocks) and the remaining main blocks equal to zero square matrices. Sometimes, depending on the value of $(i,j)$, the entire matrix $\bB_{ij}$ will be equal to $\bar{\bB}_{ij}$  with zero vectors inserted between the blocks. For the later case, as discussed earlier, the addition of zero vectors do not change the maximum eigenvalue of $\bar{\bB}_{ij}$, hence the maximum eigenvalue of $\bB_{ij} = |c_{ij}|$. For the former case, we make use of the property that the eigenvalues of a block diagonal matrix will be the eigenvalues of its blocks which implies the maximum eigenvalue of $\bB_{ij} = |c_{ij}|$. 
\end{IEEEproof}
\bibliographystyle{IEEEtran} 
\bibliography{ref_frame}

\begin{thebibliography}{10}
\providecommand{\url}[1]{#1}
\csname url@samestyle\endcsname
\providecommand{\newblock}{\relax}
\providecommand{\bibinfo}[2]{#2}
\providecommand{\BIBentrySTDinterwordspacing}{\spaceskip=0pt\relax}
\providecommand{\BIBentryALTinterwordstretchfactor}{4}
\providecommand{\BIBentryALTinterwordspacing}{\spaceskip=\fontdimen2\font plus
\BIBentryALTinterwordstretchfactor\fontdimen3\font minus
  \fontdimen4\font\relax}
\providecommand{\BIBforeignlanguage}[2]{{%
\expandafter\ifx\csname l@#1\endcsname\relax
\typeout{** WARNING: IEEEtran.bst: No hyphenation pattern has been}%
\typeout{** loaded for the language `#1'. Using the pattern for}%
\typeout{** the default language instead.}%
\else
\language=\csname l@#1\endcsname
\fi
#2}}
\providecommand{\BIBdecl}{\relax}
\BIBdecl

\bibitem{fourier}
G.~Bachmann, L.~Narici, and E.~Beckenstein, ``Fourier and wavelet analysis,''
  \emph{Springer Science \& Business Media}, 2012.

\bibitem{gabor}
T.~Strohmer and R.~Heath, ``Grassmannian frames with applications to coding and
  communication,'' \emph{arXiv preprint math/0301135}, 2003.

\bibitem{comm}
P.~Viswanath and V.~Anantharam, ``Optimal sequences and sum capacity of
  synchronous cdma systems,'' \emph{IEEE Transactions on information theory},
  vol.~45, no.~6, pp. 1984--1991, 1999.

\bibitem{sparse2}
D.~Donoho and M.~Elad, ``Maximal sparsity representation via $\ell_{1}$
  minimization,'' \emph{Proceedings of National Academy of Sciences}, vol. 100,
  pp. 2197--2202, 2003.

\bibitem{rt}
M.~Fickus and D.~G. Mixon, ``Numerically erasure-robust frames,'' \emph{arXiv
  preprint arXiv:1202.4525}, 2012.

\bibitem{qt}
Y.~C. Eldar and G.~D. Forney, ``Optimal tight frames and quantum measurement,''
  \emph{IEEE Transactions on Information Theory}, vol.~48, no.~3, pp. 599--610,
  2002.

\bibitem{upperbound}
J.~A. Tropp, ``Complex equiangular tight frames,'' \emph{International Society
  for Optics and Photonics}, vol. 5914, p. 591401, 2005.

\bibitem{tropp}
J.~A. Tropp, I.~S. Dhillon, R.~W. Heath, and T.~Strohmer, ``Designing
  structured tight frames via an alternating projection method,'' \emph{IEEE
  Transactions on information theory}, vol.~51, no.~1, pp. 188--209, 2005.

\bibitem{xiong}
H.~Bai, S.~Li, and X.~He, ``Sensing matrix optimization based on equiangular
  tight frames with consideration of sparse representation error,'' \emph{IEEE
  Transactions on Multimedia}, vol.~18, no.~10, pp. 2040--2053, 2016.

\bibitem{agelos}
E.~V. Tsiligianni, L.~P. Kondi, and A.~K. Katsaggelos, ``Construction of
  incoherent unit norm tight frames with application to compressed sensing,''
  \emph{IEEE Transactions on Information Theory}, vol.~60, no.~4, pp.
  2319--2330, 2014.

\bibitem{alternatingprojection}
K.~Jaganathan and B.~Hassibi, ``Reconstruction of integers from pairwise
  distances,'' \emph{International Conference on Acoustics, Speech and Signal
  Processing}, 2012.

\bibitem{tahir}
B.~Tahir, S.~Schwarz, and M.~Rupp, ``Constructing grassmannian frames by an
  iterative collision-based packing,'' \emph{IEEE Signal Processing Letters},
  vol.~26, no.~7, pp. 1056--1060, 2019.

\bibitem{codebook}
H.~E.~A. Laue and W.~P. Du~Plessis, ``A coherence-based algorithm for
  optimizing rank-1 grassmannian codebooks,'' \emph{IEEE Signal Processing
  Letters}, vol.~24, no.~6, pp. 823--827, 2017.

\bibitem{preclic}
C.~Rusu and N.~Gonzalez-Prelcic, ``Designing incoherent frames through convex
  techniques for optimized compressed sensing,'' \emph{IEEE Transactions on
  Signal Processing}, vol.~64, no.~9, pp. 2334--2344, 2016.

\bibitem{rusu}
C.~Rusu, ``Design of incoherent frames via convex optimization,'' \emph{IEEE
  Signal Processing Letters}, vol.~20, no.~7, pp. 673--676, 2013.

\bibitem{bcasc}
H.~Z{\"o}rlein and M.~Bossert, ``Coherence optimization and best complex
  antipodal spherical codes,'' \emph{IEEE Transactions on Signal Processing},
  vol.~63, no.~24, pp. 6606--6615, 2015.

\bibitem{ub}
D.~L. Donoho and M.~Elad, ``Optimally sparse representation in general
  (nonorthogonal) dictionaries via $\ell_{1}$ minimization,'' \emph{Proceedings
  of the National Academy of Sciences}, vol. 100, no.~5, pp. 2197--2202, 2003.

\bibitem{r2}
J.~A. Tropp, ``Greed is good: Algorithmic results for sparse approximation,''
  \emph{IEEE Transactions on Information theory}, vol.~50, no.~10, pp.
  2231--2242, 2004.

\bibitem{o1}
J.~A. Tropp and A.~C. Gilbert, ``Signal recovery from random measurements via
  orthogonal matching pursuit,'' \emph{IEEE Transactions on information
  theory}, vol.~53, no.~12, pp. 4655--4666, 2007.

\bibitem{bp}
S.~S. Chen, D.~L. Donoho, and M.~A. Saunders, ``Atomic decomposition by basis
  pursuit,'' \emph{SIAM review}, vol.~43, no.~1, pp. 129--159, 2001.

\bibitem{elad}
M.~Elad, ``Optimized projections for compressed sensing,'' \emph{IEEE
  Transactions on Signal Processing}, vol.~55, no.~12, pp. 5695--5702, 2007.

\bibitem{sapiro}
J.~M. Duarte-Carvajalino and G.~Sapiro, ``Learning to sense sparse signals:
  Simultaneous sensing matrix and sparsifying dictionary optimization,''
  \emph{IEEE Transactions on Image Processing}, vol.~18, no.~7, pp. 1395--1408,
  2009.

\bibitem{Li}
G.~Li, Z.~Zhu, D.~Yang, L.~Chang, and H.~Bai, ``On projection matrix
  optimization for compressive sensing systems,'' \emph{IEEE Transactions on
  Signal Processing}, vol.~61, no.~11, pp. 2887--2898, 2013.

\bibitem{cao}
J.~Xu, Y.~Pi, and Z.~Cao, ``Optimized projection matrix for compressive
  sensing,'' \emph{EURASIP Journal on Advances in Signal Processing}, vol.
  2010, no.~1, p. 560349, 2010.

\bibitem{gradient}
V.~Abolghasemi, S.~Ferdowsi, and S.~Sanei, ``A gradient-based alternating
  minimization approach for optimization of the measurement matrix in
  compressive sensing,'' \emph{Signal Processing}, vol.~92, no.~4, pp.
  999--1009, 2012.

\bibitem{hunter}
D.~R. Hunter and K.~Lange, ``A tutorial on mm algorithms,'' \emph{The American
  Statistician}, vol.~58, no.~1, pp. 30--37, 2004.

\bibitem{sir}
Y.~Sun, P.~Babu, and D.~P. Palomar, ``Majorization-minimization algorithms in
  signal processing, communications, and machine learning,'' \emph{IEEE
  Transactions on Signal Processing}, vol.~65, no.~3, pp. 794--816, 2016.

\bibitem{convergence}
M.~Razaviyayn, M.~Hong, and Z.-Q. Luo, ``A unified convergence analysis of
  block successive minimization methods for nonsmooth optimization,''
  \emph{SIAM Journal on Optimization}, vol.~23, no.~2, pp. 1126--1153, 2013.

\bibitem{boyd}
S.~Boyd, S.~P. Boyd, and L.~Vandenberghe, \emph{Convex optimization}.\hskip 1em
  plus 0.5em minus 0.4em\relax Cambridge university press, 2004.

\bibitem{minmax}
J.~v. Neumann, ``Zur theorie der gesellschaftsspiele,'' \emph{Mathematische
  annalen}, vol. 100, no.~1, pp. 295--320, 1928.

\bibitem{MDA1}
A.~Beck and M.~Teboulle, ``Mirror descent and nonlinear projected subgradient
  methods for convex optimization,'' \emph{Operations Research Letters},
  vol.~31, no.~3, pp. 167--175, 2003.

\bibitem{squarem}
R.~Varadhan and C.~Roland, ``Simple and globally convergent methods for
  accelerating the convergence of any em algorithm,'' \emph{Scandinavian
  Journal of Statistics}, vol.~35, no.~2, pp. 335--353, 2008.

\bibitem{imagedata}
\BIBentryALTinterwordspacing
U.~SIPI, ``The usc-sipi image database,'' 2016. [Online]. Available:
  \url{http://sipi.usc.edu/services/database/data-base.html}
\BIBentrySTDinterwordspacing

\bibitem{psnr}
R.~Gonzalez and R.~Woods, ``Digital image processing: Pearson prentice hall,''
  \emph{Upper Saddle River, NJ}, vol.~1, pp. 376--376, 2008.

\end{thebibliography}
\end{document}